# LiDAL: Light Detection and Localization

Aubida A. Al-Hameed[1], Safwan Hafeedh Younus[1], Ahmed Taha Hussein , Mohammed T. Alresheedi[2] and Jaafar M. H. Elmirghani[1]
[1]School of Electronic and Electrical Engineering, University of Leeds, LS2 9JT, United Kingdom
[2]Department of Electrical Engineering, King Saud University, Riyadh, Saudi Arabia
{elaawj, elshy @leeds.ac.uk, asdftaha@yahoo.com , malresheedi@ksu.edu.sa, j.m.h.elmirghani@leeds.ac.uk}

*Abstract*—**In this paper, we present the first indoor light-based detection and localization system that builds on concepts from radio detection and ranging (radar) making use of the expected growth in the use and adoption of visible light communication (VLC), which can provide the infrastructure for our LiDAL system. Our system enables active detection, counting and localization of people, in addition to being fully compatible with existing VLC systems. In order to detect human (targets), LiDAL uses the visible light spectrum, it sends pulses using a VLC transmitter and analyses the reflected signal collected by a photodetector receiver. Although we examine the use of the visible spectrum here, LiDAL can be used in the infrared spectrum and other parts of the light spectrum. We introduce LiDAL with different transmitter-receiver configurations and optimum detectors considering the fluctuation of the received reflected signal from the target in the presence of Gaussian noise. We design an efficient multiple input multiple output (MIMO) LiDAL system with wide field of view (FOV) single photodetector receiver, and also design a multiple input single output (MISO) LiDAL system with an imaging receiver to eliminate the ambiguity in target detection and localization. We develop models for the human body and its reflections and consider the impact of the colour and texture of the cloth used as well as the impact of target mobility. A number of detection and localization methods are developed for our LiDAL system including cross correaltion and a background subtraction method. These methods are considered to distinguish a mobile target from the ambient reflections due to background obstacles (furniture) in a realistic indoor environment.**

## I. INTRODUCTION

Visible Light Communication (VLC) systems are used to provide illumination and data communications. VLC uses light emitting diodes (LEDs) or lasers to encode data into light intensity in the visible spectrum [1-5]. VLC systems have many advantages such as cost-effectiveness using the existing lighting infrastructure, operating on a broad, unlicensed bandwidth, secure (light signals do not penetrate walls) and there is no interference with Radio Frequency (RF) signals [4], [6-8].

People counting has become an emerging and attractive area in the past decade [9, 10]. Many approaches have been developed for counting in public places such as subways, bus stations and supermarkets [10, 11] . The outcome of these techniques can be used for public security, resources allocation and marketing decisions. Passive infrared (PIR) imaging systems have been employed to detect and count people, however, the PIR system is temperature dependent, thus leading to a vast number of detection failures [12], [11]. Ultra-wideband (UWB) radar has been utilized to effectively detect and track outdoor pedestrians. However, for an indoor environment, the effects of signal scattering and absorption by obstacles significantly impairs the performance of UWB indoor radar [11, 13]. IR Laser detection and ranging (LADAR) has been used to detect people by monitoring the reflected signal patterns of people legs [11]. Counting systems based on computer vision and digital image processing are becoming meaningful and useful. Video cameras with image processing algorithms have been widely used to count people indoor and count pedestrians outdoor [13], [14], [10]. It should be noted however that acquiring images of people poses in many cases privacy concerns, whereas our LiDAL system uses light reflections from people and therefore no images of people are acquired, stored or transmitted.

In this paper, we introduce for the first time indoor light-based detection, counting and localization of people based on the use of radar-like reflections. This can significantly expand the utility of indoor VLC systems. The key concept behind our LiDAL system is the use of the (visible) light reflected from targets (people) where the light reflectivity is a function of the material type and colour of the target's surface. The reflected light signal is captured by a photodetector which monitors the change in the light intensity in the time domain. LiDAL can be a system embedded in the VLC system to provide additional functionality to detect, count and localize people. In addition, LiDAL reduces the complexity and cost associated with the acquisition and digital processing of images to detect the presence of people.

To the authors' best knowledge, the proposed system is the first to employ an indoor optical radar for people detection and localization. It uses the visible light spectrum of VLC systems, and can potentially use other parts of the light spectrum. It is worth noting that the use of the infrared spectrum for example can eliminate issues with light dimming and switching off light sources. The concept of LiDAL has the benefits of active radio waves radar systems while avoiding, as mentioned, the issues associated with UWB (and other radio) radar signal propagation indoor. It also makes use of the existing lighting / illumination systems and potentially the existing VLC systems infrastructure.

Due to the fact that (visible) light is reflected from opaque objects, the major critical issue in LiDAL is how to distinguish the people (targets) from other background objects, i.e., furniture. In order to overcome this problem, we have considered the mobility of people as a key distinguishing feature between humans and furniture. Even in the case of nomadic users, people exhibit movement of body parts while stationary, for example while sat working in an office environment. They may also standup from time to time. We introduced a background subtraction method and a cross-correlation method targets when they are mobile. The contributions of this work can be summarized as follows:

1) We proposed for the first time an indoor (visible) light pulsed radar-like system which utilises the VLC system transmitters to detect, count and localize multiple targets.
2) We developed a model for the human body and its reflections and the impact of the colour and texture of the clothing used, which are all important attributes of the target of interest.

3) We considered a range of different mobility models for humans and used these as an important input to our LiDAL human detection and localization system.
4) We designed optimum receivers and algorithms for the proposed LiDAL systems.
5) We designed MIMO-LiDAL and MISO-Imaging-LiDAL systems which are compatible with VLC and light fidelity (Li-Fi) systems.

This paper is divided into sections as follows: Section II introduces the modelling of the environment and the target. Section III presents the analysis of LiDAL system range, resolution and optimum receiver design. Section IV presents target distinguishing approaches and mobility modelling in a realistic environment. Section V develops the design of MIMO-LiDAL systems. Section VI considers the design of the MISO-IMG-LiDAL system. Section VII presents the simulation setup and results. Finally, conclusions are drawn in Section VIII.

## II. Realistic Environment and Target Modelling

To study the performance of the proposed LiDAL system, simulations were performed in a typical office consisting of a furnished room, with dimensions of 4 m (width) × 8 m (length) × 3 m (height) as shown in Fig.1a. The walls, furniture and floor were segmented into small reflective elements. The reflective elements were represented as small secondary emitters that diffuse the received signal in the shape of a Lambertian pattern, with a reflectivity of 0.8 for the walls and ceiling and 0.3 for the floor [15], [16]. In addition, the reflection elements can be treated as small secondary transmitters that diffuse the incident rays back into space from their centre [16], [17]. The accuracy of the received impulse response profile was controlled by the size of the reflective elements, which were 5 cm × 5 cm and 20 cm × 20 cm for the first and second order reflections, respectively [15], [16], [18]. Eight light units were placed at a height of 3m above the floor and were used to satisfy ISO and European illumination standards. Each unit had 9 RGB laser diodes (LDs), and the total transmitted power from each RGB-LDs light unit was 18 W [19], [20], [21]. It is worth mentioning that, each light unit consists of red, green and blue laser diodes which are driven by different modulation currents to meet the illumination standards [19], [20].

The average target (person) dimensions considered were 15 cm × 48 cm × 170 cm (depth × width × height) [22] as shown in Fig.1b and coloured polyester fabric was considered as the target coating material. The fabric reflection model used was based on the work in [23], which analysed the reflections from different types of fabric including silk, cotton, polyester, acetate and glass fibre. We also made use of the work in [24] which examined the combination of fabric colour and material and their impact on light reflection. The resulting reflections in [23] were observed to be a combination of diffuse (Lambertian) and specular reflections. In [23], the distribution of the reflected visible light of several cloth materials was experimentally studied. In particular, cotton reflectance was about 9% specular and 91% diffuse, while polyester reflectance was 10% specular, 26% diffuse and 63% internal multiple reflections which are treated as diffuse reflections as can be seen in Table I. It should be noted that 1% of the polyester reflections are internal reflections which occur inside the fabric layers [23]. Therefore, in our simulation, we only considered a Lambertian pattern (ie diffuse) as the model for the target's surface material. The reflectivity factor of different dyed polyester fabric ranges between 0.25 and 0.72 [24]. Moreover, the reflectivity of dark and white human skin is 0.04-0.35 and 0.16-0.86 respectively [25]. Regarding furniture, office desks (1.54 m (width) × 0.76 m (length) × 0.75 m (height)) and a bookshelf (3 m × 0.8 m × 2 m) are considered, and are located in the room as shown in Fig.1a, where the office desks and bookshelf materials were finished-wood with a reflectivity factor of 0.55 and diffuse reflections [26]. The Lambertain diffuse reflections order for the furniture and target is assumed to be 1.

TABLE I
REFLECTION MODEL FOR A DIFFERENT TARGET COATING MATERIALS [16].

| | Specular reflector (%) | Diffuse reflector (%) |
|---|---|---|
| Cotton | 9% | 91% |
| Polyester | 10% | 89% |

## III. LiDAL System

In this section, we analyse the LiDAL system maximum range which is related to the receiver's field of view. We also pay attention to the received reflected signal in two LiDAL configurations that relate to the colocation or separation of transmitter and receiver in space. Furthermore, we analyse the resolution and the ambiguity of target detection which are related to the transmitted pulse width. In addition, we examine the optical receiver design for LiDAL and consider the receiver bandwidth and thermal and ambient noises. In our LiDAL system, the sources of randomness are attributed to the target colour of cloth, the target orientation and the receiver noise. Note that in terms of indoor optical wireless channel, we consider the channel at the target's maximum range dictated by the receiver field of view (and the receiver sensitivity). The fluctuation of the received reflected signal attributed to the different colours worn by the target is modelled leading to a pdf of the target reflection factor. The target (human) random orientation and its impact on reflections was determined through extensive simulations, leading a pdf of the effective target cross-section. The optimum LiDAL receiver is then formulated using Bayes structures and signal space theory for single and multiple targets in the presence of the impairments outlined above.

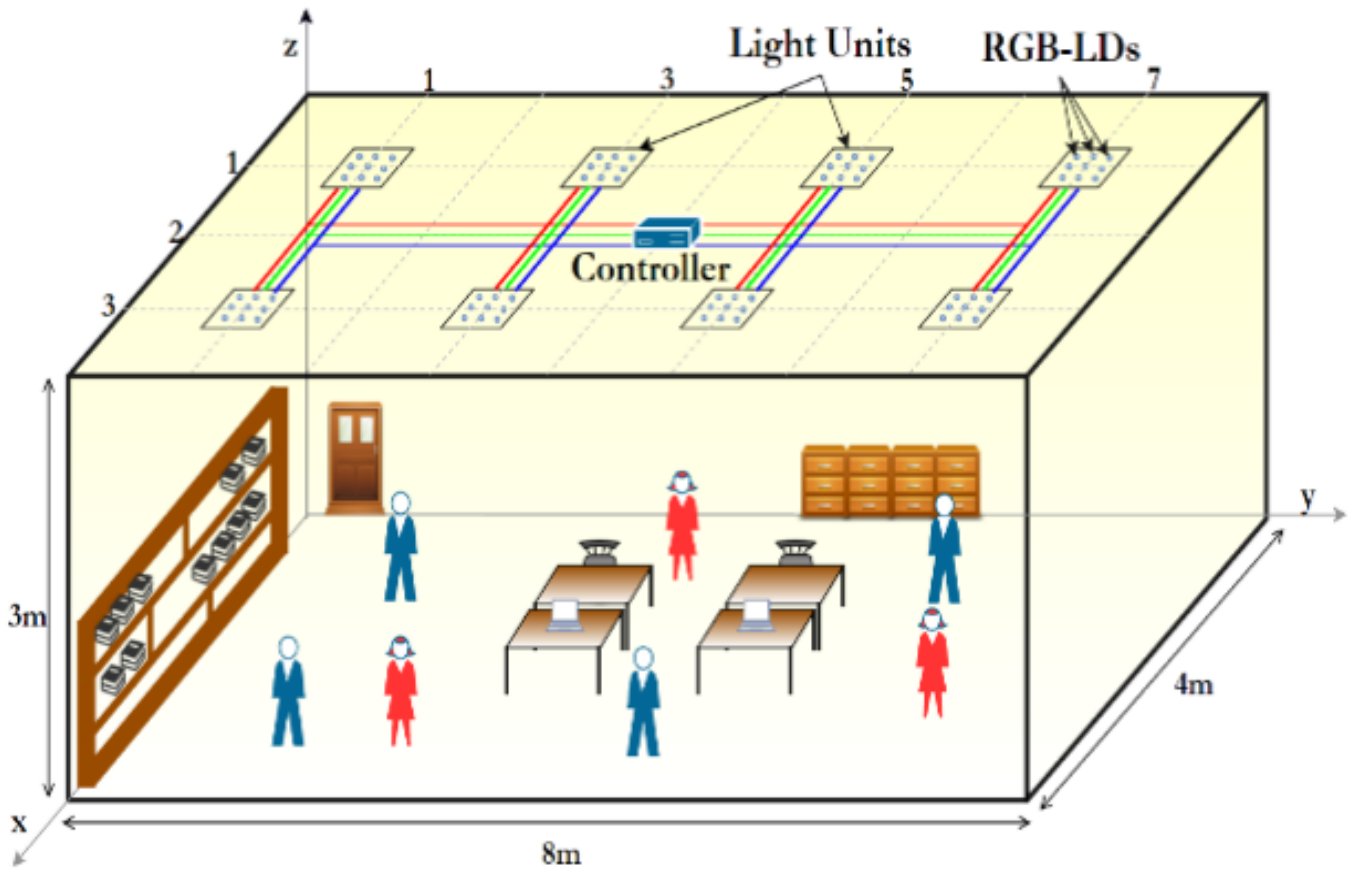

(a)

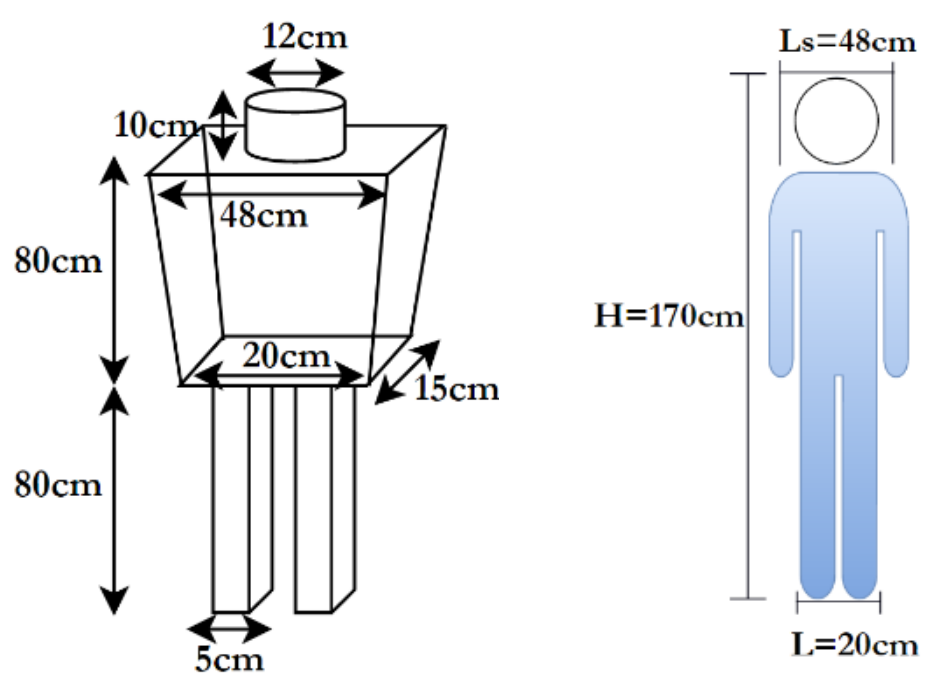

(b)

Fig. 1: (a) Realistic simulation environment setup and (b) basic 3D target model.

### *A. Range Analysis*

The light unit emits a narrow pulse in a wide optical beam (Lambertian radiation pattern) directed towards the floor. An optical receiver, collocated or separated from the transmitter, collects the received reflected pulses. The received signal is a superposition of the reflected pulses from the target(s), static environment obstacles (furniture) and noise. Note that, in this section we assumed the target(s) are located in an ideal environment (i.e. an empty room with zero reflectively from walls, ceiling and floor). Therefore, the received reflected signal randomness is only due to target(s) colours and effective cross-section and is corrupted by noise. In Section III we deal with the presence of furniture (reflections) and reflections from the walls.

The maximum range of LiDAL can be determined depending on the receiver's photodetector FOV. The maximum range $R_{Max}^{FOV}$ for a certain receiver concentrator FOV ($\Psi_c$) is given as (see Fig. 2):

$$R_{Max}^{FOV} = tan(\Psi_c)\ (d_o - h) \tag{1}$$

where $\Psi_c$ is the semi-angle of photodetector's concentrator, $d_o$ is the perpendicular distance between the $i^{th}$ receiver location $L_{Rx}^i(x_{Rx}^i, y_{Rx}^i, z_{Rx}^i)$ and the ground reference point $L_o^i(x_o^i, y_o^i, 0)$ as shown in Fig. 2 and $h$ is the target height.

Fig. 2 present two arrangements of different possible transmitter and receiver configurations with a target located inside the receiver optical footprint (i.e. receiver FOV). We refer to collocated transmitter-receiver configuration as 'monostatic LiDAL' and refer to the spaced transmitter-receiver configuration as 'bistatic LIDAL'.

The received reflected optical power ($P_{r_{R_{Max}^{FOV}}}^B$) from a target at maximum range, ie located in the receiver optical footprint at a radius of $R_{Max}^{FOV}$, for a bistatic LIDAL (see Fig.2a, b) is derived as:

$$P_{r_{R_{Max}^{FOV}}}^B = \frac{(n+1)(n_{ele}+1)}{4\pi^2 R_1^2 R_2^2} T_f(\Psi_c) G_c(\Psi_c) P_t\, d_A \rho A_R \\ cos^n(\theta)\ cos\,(\varphi)\, cos^{n_{ele}}(\varphi_1)\, cos(\Psi_c) \tag{2}$$

and for monostatic LIDAL (see Fig. 2c, d), the $P_{r_{R_{Max}^{FOV}}}^M$ is written as:

$$P_{r_{R_{Max}^{FOV}}}^M = \frac{(n+1)(n_{ele}+1)}{4\pi^2\left((d_o-h)^2+R_{Max}^{FOV\,2}\right)^2} T_f(\Psi_c) G_c(\Psi_c) P_t\, d_A \rho A_R cos^{n+3}(\Psi_c) \tag{3}$$

where $R_1$ is the distance between the transmitter and target, $R_2$ is the distance between the taregt and receiver, $R_2 = \left((d_o-h)^2 + R_{Max}^{FOV\,2}\right)^{\frac{1}{2}}$, $T_f(\Psi_c)$ is the optical filter transmission factor, $G_c(\Psi_c)$ is the gain of the concentrator, $P_t$ is the transmitted power, $d_A$ is target cross section area (top and/or the sides), $A_R$ is the photodetector physical area, $\rho$ is the target reflection coefficient, $\theta$ and $\varphi$ are the angles of irradiance and incidence respectively, $n_{ele}$ is Lambertian order for the target diffuse reflector and $n$ is the Lambertian emission factor of LD defined as [27]:

$$n = -\frac{ln(2)}{ln(cos(\Phi))}. \tag{4}$$

The gain of the concentrator $G_c(\Psi_c)$ is given as [28] :

$$G_c(\Psi) = \frac{N^2}{sin^2(\Psi_c)} \tag{5}$$

where $\Phi$ is the semi-angle at half power of LD ($\Phi > \Psi_c$) and $N$ is the concentrator refractive index.

It should be noted that the transmitter has a broad radiation pattern ($n$=0.52 for illumination purposes [19] [20]) and the target assumed has a diffuse emission factor of $n_{ele}$=1. Therefore, the target has a narrow radiation pattern compared to the transmitter's radiation pattern. With such narrow radiation pattern, the target delivers maximum power to the receiver if it is directly under or near the receiver. As such, the weakest received reflected signal from a target occurs when the target is at the edge of the receiver FOV (i.e. target located at $R_{Max}^{FOV}$).

The photodetector area ($A_R$) and the concentrator's FOV and gain are among the receiver's key parameters that determine the LiDAL detection performance. The values of these parameters have to satisfy the LiDAL (radar) design requirements. We analyze their impacts later in this paper. In addition, the transmitted power $P_t$ is set at the maximum power needed for normal illumination in the room. (i.e. $P_t$ =18W according to the design in [19]). We therefore do not consider in this paper the impact of dimming on our LiDAL system, and in cases where dimming is an issue, infrared sources and detectors can be used for LiDAL.

### *B. Receiver Bandwidth*

To determine the maximum receiver bandwidth needed, we selected the LiDAL configurations that result in the largest channel bandwidths which the receiver has to deal with. The largest channel bandwidths occur when the target is under the receiver. We have also evaluated the channel bandwidths at a large number of target locations. Figs. 3a and b show a target located underneath the receiver for the LiDAL bistatic and monostatic scenarios respectively. We have simulated the pulse dispersion associated with the bistatic and monostatic LiDAL channels due to target presence at different target locations. The target's locations have been generated uniformly inside the receiver optical footprint (see Figs.3a and b) to calculate the channel impulse response and then to obtain the 3dB channel bandwidth for each location. It should be noted that we considered an ideal indoor environment without furniture or background obstacles, and we treated the room's floor as a non-reflective surface (i.e. zero reflection factor). In addition, the simulation and calculations of the received reflected signal were carried out using MATLAB. Our simulation tool is similar to the one developed by Barry [28] in terms of the indoor channel impulse response calculation method. Figs.3c and 3d depict the probability distribution of the channel bandwidth ($Bw_{ch}$) for the bistatic and monostatic LiDALs respectively. As can be seen in Figs. 3c and d, the bistatic LiDAL channel is more dispersive than the monostatic

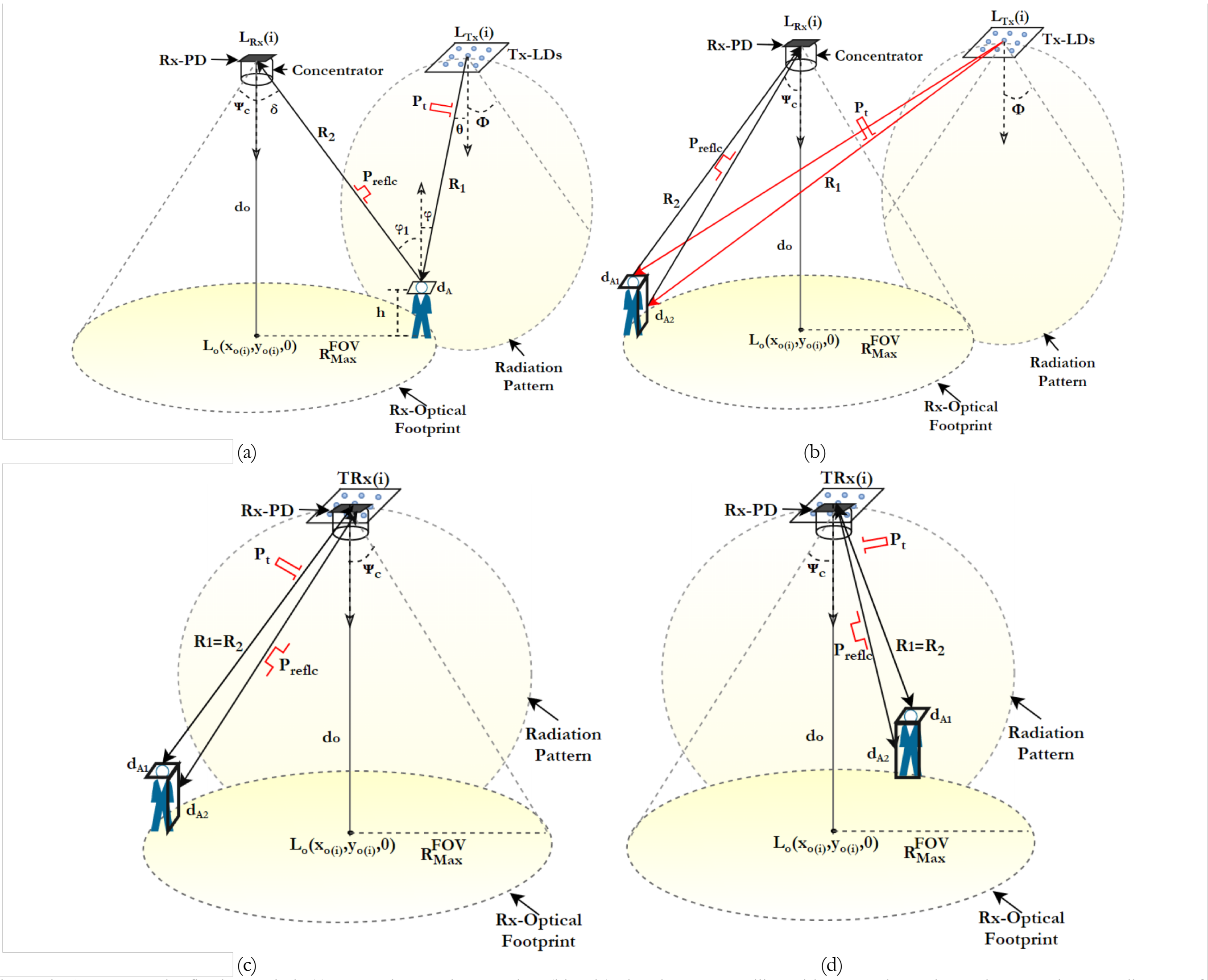

Fig. 2: LiDAL range and reflection analysis (a) a spaced transmitter-receiver (bistatic) placed on room ceiling with a target located near the transmitter at a distance of $R_{Max}^{FOV}$ from the receiver (b) a spaced transmitter-receiver placed on room ceiling with a target located away from the transmitter at a distance of $R_{Max}^{FOV}$ from the receiver (c) and (d) collocated transmitter-receiver (monostatic) placed on room ceiling with a target at two locations.

LiDAL channel due to the large distance between the transmitter, target and receiver. Table II summarizes the bistatic and monostatic LiDAL channels characteristics.

We calculated the channel bandwidth for the monostatic and bistatic LiDAL as follows:

1) An input pulse $pl(\tau)$ with time duration $\tau$ of 0.01ns (equal to the time bin duration used in simulation [29]) is presented to the input of a transmitter unit, RGB-LDs, with impulse response $h_{tx}(t)$ followed by calculation of $H_{tx}(f) = \mathcal{F}(h_{tx}(t))\,\mathcal{F}(pl(\tau))$. It is worth mentioning that, the RGB-LDs have a large bandwidth (few GHz) [19] and therefore, given a channel with few hundred MHz bandwidth, we ignored the laser transfer function.
2) We set the following simulation parameters for the monostatic and bistatic LiDAL systems: The room has dimensions of 8m × 4m × 3m and the illumination requirements were met using 8 light units distributed as shown in Fig. 1a. These light units also represent the LiDAL receiver locations. To provide overlapping LiDAL coverage zones, the receiver FOV was set to 43°. The transmitter beamwidth was set 75° for illumination purposes [20] and the impulse response was calculated with a time bin of 0.01ns. The monostatic transmitter-receiver pair was located at (2m, 4m, 3m) at the centre of the room in Fig.1a. The bistatic transmitter was located at (2m, 5m, 3m) and the receiver was located at (2m, 4m, 3m).
3) We calculated the LiDAL channel impulse response $h_{ch}(t)$ (i.e. the LiDAL system configuration with the target present) using the ray tracing propagation model in [28, 30], [27], [31]. In this paper, we considered the first and second order reflection components in the simulation of the impulse response of the LiDAL channel. We then determined the 3dB channel bandwidth, $Bw_{ch}$, using $h_{ch}(t)$.
4) The required 3dB receiver bandwidth is determined as

$$Bw_{Rx} = max\left(H_{tx}(f)|_{3dB}\,, H_{ch}(f)|_{3dB}\right) \tag{6}$$

TABLEII
CHARACTERISTICS OF LiDAL CHANNEL.

| | Min. $Bw_{ch}$ (MHz) | Max. $Bw_{ch}$ (MHz) | Mean. $Bw_{ch}$ (MHz) |
|---|---|---|---|
| Bistatic LiDAL | 65 | 260 | 125 |
| Monostatic LiDAL | 140 | 315 | 230 |

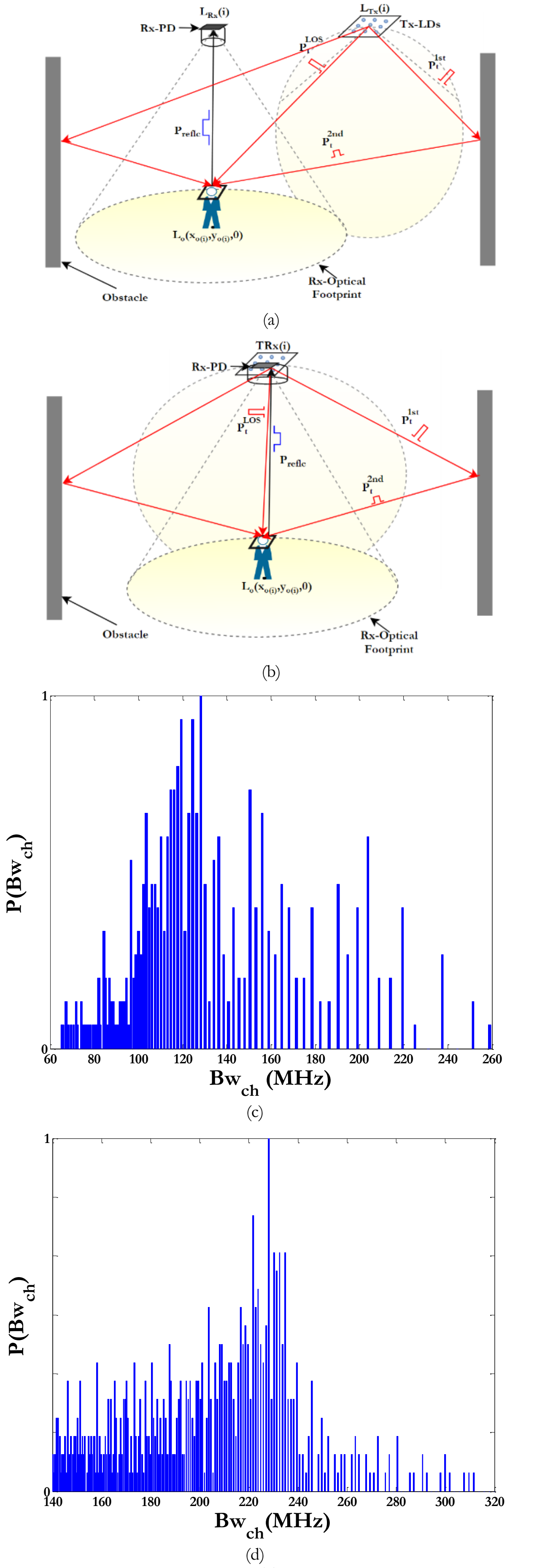

Fig. 3: LiDAL channel configurations (a) Tx and Rx placed in different locations (bistatic) with a target located at the centre of the optical footprint, (b) Tx and Rx placed in the same location (monostatic ) and a target located at the centre of the optical footprint, (c) the histogram of the received power in bistatic LiDAL versus bandwidth and (d) histogram of the received power in monostatic LiDAL versus bandwidth.

### *C. LiDAL Resolution and Ambiguity in Target Detection Analysis*

The distance ($R_1$) between the monostatic LiDAL transceiver unit ($T_{Rx}$) and the target is calculated based on the round trip time (i.e. time taken by the pulse from the transmitter to the target plus the time taken by the reflected pulse back from the target to the receiver), $t_{trip}$, and the speed of light, $c$, as:

$$R_1 = \frac{c\, t_{trip}}{2}. \tag{7}$$

The range resolution of LiDAL is defined as the minimum separation distance $(\Delta R)$ at which two or more targets can be reliably detected as illustrated in Fig.4. The range resolution is related to the pulse width of the transmitted signal. The LiDAL resolution $(\Delta R)$ is given as [32]:

$$\Delta R = R_{1,1} - R_{1,2} = \frac{c\,\tau}{2} \tag{8}$$

where $\tau$ is the transmitted pulse width. The separation distance $\Delta xy$ between two targets, as can be seen in Fig.4, is given as:

$$\Delta xy = R_{1,1} \sin \theta_{1,1} - R_{1,2} \sin \theta_{1,2} \tag{9}$$

and if $\theta_{1,1} \cong \theta_{1,2} = \theta$, then

$$\Delta xy = \left(R_{1,1} - R_{1,2}\right) \sin \theta = \Delta R \sin \theta\,. \tag{10}$$

Therefore, $\Delta xy \leq \Delta R$, and in a typical room such as that in Fig. 1a, we determined that $\theta = 43^0$, hence here $\Delta xy \leq 0.68\Delta R$. Fig.5 shows an example of the received pulse response attributed to the reflected signal, as received by a transceiver ($T_{Rx}$) unit which covers an optical footprint that includes two targets in the presence of noise. In this work, we considered typical room layouts, where for example in a meeting room (closest separation between people in a business setting), the designers recommend an inter-chair-distance more than 60cm as in [33] and 75cm as in [34], and the typical justifiable distance between two people having a conversation is 30 cm. Therefore, we selected a minimum LiDAL resolution of $\Delta R$=30cm and therefore given (10), $\Delta xy \leq$ 30cm which is the required minimum separation between two targets (i.e. the required $\tau$ is 2ns from (8)). Optical transmitters and optical receivers that support this bandwidth are readily available, and the optical wireless channel is able to provide such bandwidth [35], [36]. The analysis of the channel bandwidth for the bistatic and monostatic LiDAL systems (see fig.3c, d) showed high channel dispersion and low channel bandwidth which cannot accommodate a transmitted pulse of 2ns without pulse spreading in the receiver. Thus, an equilzer is required to mitigate the imperfections of the LiDAL channel.

Let us first assume an ideal indoor environment (i.e. no reflected signal from the room's background). Here ambiguity in multiple targets detection occurrs when the distance between targets is less than the LiDAL (radar) resolution $\Delta R$. In other words, when the difference of the targets' round trip times is less than the transmitted pulse width$\left(\left|t_{trip}(1) - t_{trip}(2)\right| < \tau\right)$, this leads to ambiguity. Furthermore, the ambiguity in target detection is affected by the configurations of the LiDAL system.

Table III provides a comparison between conventional radar and LiDAL when the only available information is range. Note that the angle of arrival in LiDAL can be determined through coherent optical detection, but this is too complex, and is not considered here. As Table III shows, complete localization is only achieved when three or more anchor points are available to provide range estimations.

Table III: LiDAL localization compared to traditional radar localization.

| Radar localization in literature | Localization in LiDAL | LiDAL Design Comments |
|---|---|---|
| **Single Transmitter Single receiver (SISO) 1x1** | | *One monostatic LiDAL* (a transmitter receiver pair are monostatic if they are collocated and the target is within the FOV of the receiver)<br>- Can only detect target presence.<br>- Cannot determine number of targets at same range.<br>- Cannot determine the exact target location.<br>- Therefore target ambiguity is very high. |
| - Only range is known, hence the target can be located on the surface of a sphere.<br>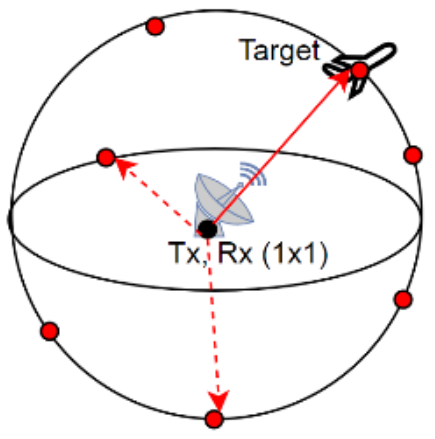<br> | - Only range is known, hence the target can be located on a circle.<br>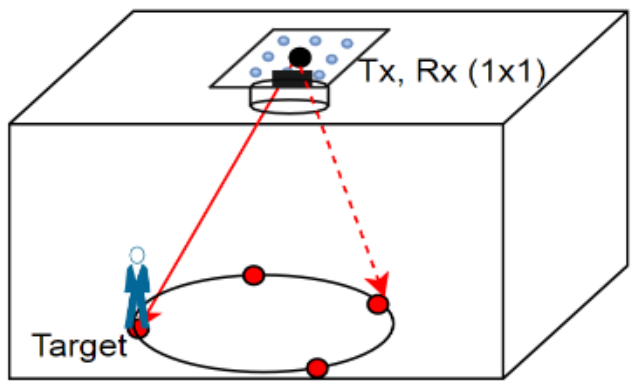<br> | |
| **Multiple Transmitters Multiple receivers (MIMO) 2x2** | | - *One Bistatic and one monostatic LiDAL*, (a transmitter receiver pair are bistatic if they are not collocated and the target is within the FOV of the receiver).<br>- Ambiguity in target detection is less due to use of both monostatic and bistatic LiDAL systems together.<br>- Scene localization can be implemented, but exact location is not known.<br>- Cannot determine the exact target location. |
| - Only two ranges are known, hence the target can be located on a circle.<br>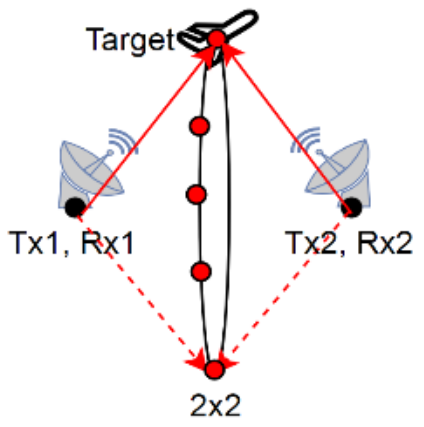<br> | - Only two ranges are known, hence the target can be located in one of two locations.<br>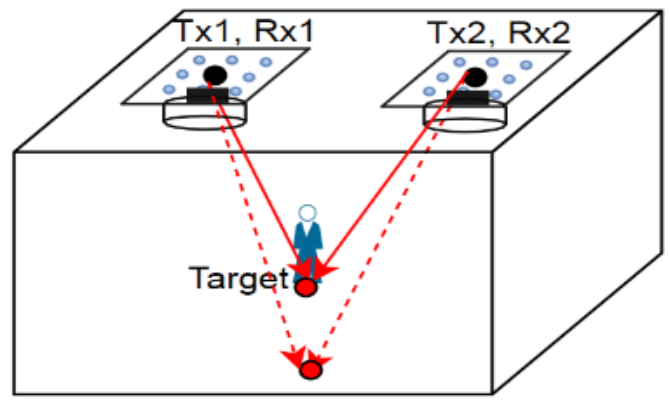<br> | |
| **Multiple Transmitters Multiple receivers (MIMO) 3x3** | | - *Two Bistatic LiDAL systems and One monostatic LiDAL system*; the target is within the FOV of one receiver only, and that receiver – transmitter pair act as monostatic LiDAL.<br>- Can detect, count and determine the exact location of multiple targets.<br>- The mean and standard deviation of the signals received from the three anchors can be different. |
| -Exact target location can be determined.<br>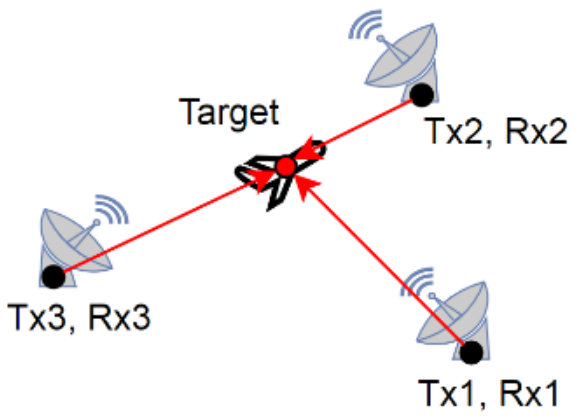<br> | -Exact target location can be determined.<br>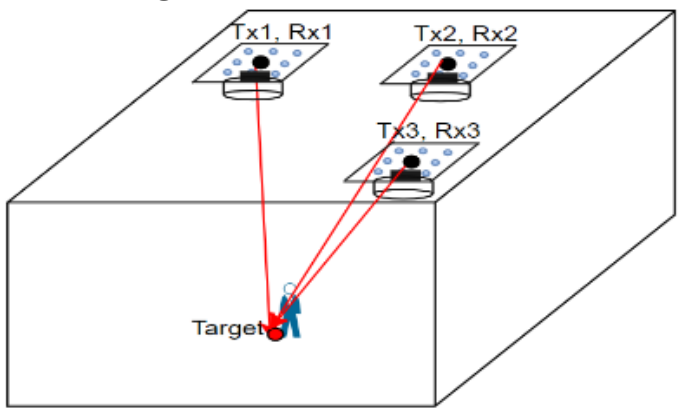<br> | |

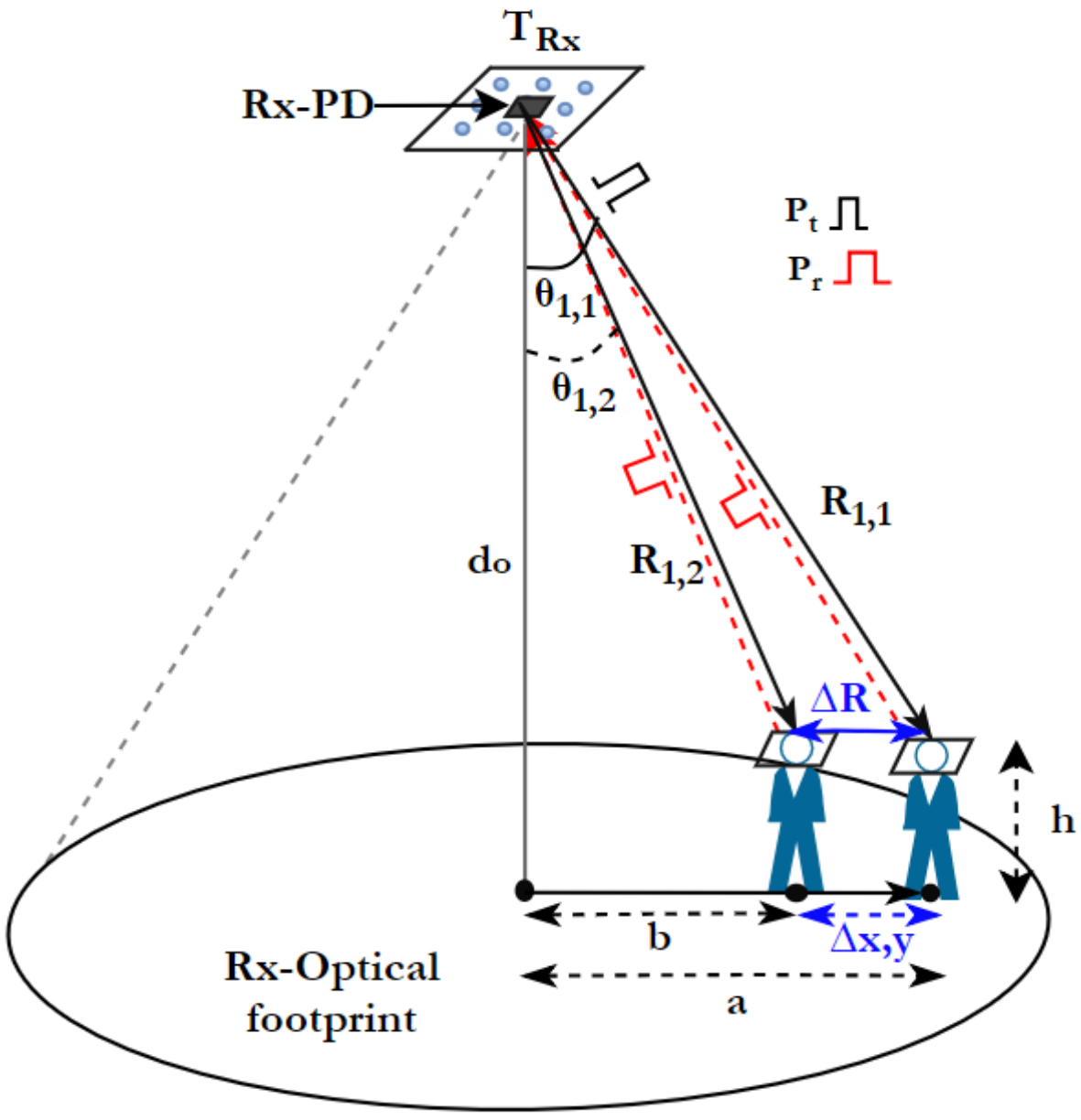

Fig. 4: The LiDAL resolution needed to distinguish two targets.

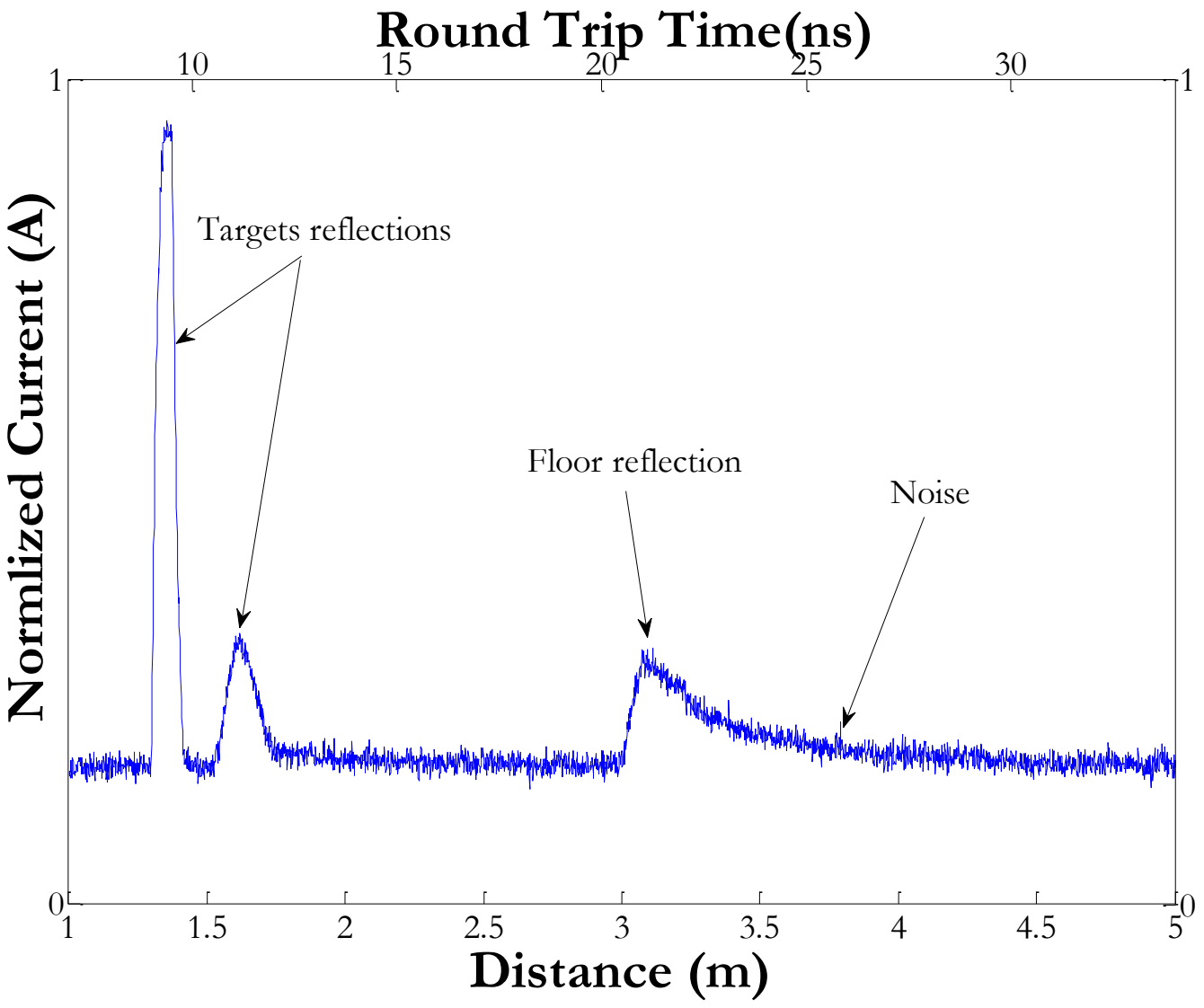

Fig. 5: The reflected received signal current from two targets located in an empty room in a monostatic LiDAL configuration.

*D. Receiver Noise*

We considered the receiver bandwidth needed in Section B, here we consider receiver noise in LiDAL.

In optical wireless (OW) systems, the noise can be divided into two components, a shot noise ($\sigma_{shot}^2$) component and a thermal noise component ($\sigma_{thermal}^2$). The total noise variance $\sigma_t{}^2$ is given by [16], [27]:

$$\sigma_t{}^2 = \sigma_{thermal}^2 + \sigma_{shot}^2. \quad (11)$$

The shot noise variance is defined as the sum of contributions from the ambient lights (direct sunlight, desk lamps etc.) and the noise from the received signal. The shot noise, $\sigma_{shot}^2$, is written as [37]:

$$\sigma_{shot}^2 = 2qBw_{Rx}\left(I_b + R_{esp}P_r\right) \quad (12)$$

where $q$ is the electronic charge, $Bw_{Rx}$ is the receiver bandwidth, $R_{esp}$ is the photodiode responsivity and $I_b$ is the background current due to ambient lights. We considered the effects of shot noise due to desk-lamps. For the four office desk-lamps shown in Fig. 1a, we considered Philips light bulbs where each light bulb has an optical power of 13w [38]. The background current measured in [38] was $I_b$ = 8.8μA (without optical filter) and corresponded to a typical setup, with a 0.85 cm$^2$ photodetector area at a distance of 2.2m from the light source with a line of sight path (worst case induced shot noise) between the light source and the receiver. The setup in [38] is comparable to the realistic environment setup used in LiDAL in terms of the distance between the desk-lamp and LiDAL receiver (distance of 2.25m in LiDAL). The background current was scaled by a factor that accounts for the difference in area between the photodetector in [38] and the photodetector we used, where our photodetector had an area of 20 mm$^2$ to provide sufficient bandwidth [27]. An optical bandpass filter (OBPF) can be used to supress the effect of the ambient noise. For example the background current in [38] was reduced from 8.8μA to 0.48μA when an OBPF is used. It is worth mentioning that the measurements in [38] included the infrared part of the optical spectrum, while this work focuses on the visible spectrum, however, an optical bandpass filter within the visible spectrum can be used to reduce the background noise to comparable levels. In addition, an electrical high pass filter can be implemented to reduce the DC component of the ambient noise. However, these solutions may increase the cost and the complexity of the LiDAL receiver.

In this work, the optical receiver was a silicon *p-i-n* photodetector with a transimpedance amplifier (TIA) to achieve high sensitivity and a good dynamic range [39], [40]. The receiver considered in this work had high speed and low input noise, designed by Texas Instruments® [41].The TIA with a $Bw_{Rx}$ of 300 MHz had a thermal input noise current of about 2.5 pA/√Hz [41].

*E. Received Signal Fluctuation and Target Reflectivity Modelling*

The fluctuation of the received optical power reflected from a target is related to the target coating reflection factor ($\rho$) (i.e. colour, material type and, reflection type) and the target effective cross section area ($A_e$). The target effective cross section area is the size of the target surface area illumined by the transmitted pulse (which reflects light) and depends on the target position, LiDAL transmitter and receiver configurations and LiDAL field of view. It should be noted that, the fluctuation of the received signal due to target reflection factor (colour of clothing and type of clothing worn) is independent of the target position and the target orientation (i.e., independent of the target effective cross section area).

Table IV presents a range of people favourite colours with their weights and reflection factors for dyed cotton coating material [42], [43]. The people favourite colours weights show features of a Gaussian distribution, Fig. 6. This can be explained by observing that the received reflected signals over an extended period of operation of the LiDAL system will be due to multiple colours and coating materials worn by the person (target) in every case; and this together with the large number of subjects allow the central limit theorem to be involved. Therefore, we fitted and optimized the survey data of the people favourite colours using a Gaussian distribution as can be seen in Fig. 6 where, the target reflection factor is the random variable of the distribution. The survey data of favourite colours [43], [42] was fitted to minimize the root mean square error (RMSE), and the minimum RMSE obtained was about 15%.

The probability distribution function (PDF) of the target reflection factor $p(\rho)$ is given as:

$$p(\rho) = \frac{1}{\sigma_\rho\sqrt{2\pi}} e^{-\left(\frac{(\rho-\mu_\rho)^2}{2\sigma_\rho^2}\right)} \quad (13)$$

where, $\mu_\rho$ and $\sigma_\rho$ are the mean and standard deviation of the target reflection factor respectively.

We determined the PDF of the effective target cross section area through simulation. The target shown in Fig. 1b (human body model) was placed at a large number of locations in the room and the ray tracing indoor propagation method of [16], [44], [45] was used to determine the power reflected by all the target surface area elements for the given target location and orientation, and the given LiDAL transmitter and receiver configurations. We then fitted the simulated data to a normalized Gaussian distribution as can be seen in Fig. 7 where the target is placed randomly in the receiver optical footprint edge with different locations and orientations. At each location, the target is rotated to eight directions with 45$^o$ angle randomly. The minimum RMSE of the effective target cross section area fitting obtained was 5%. The PDF of the effective target cross section area $p(A_e)$ is written as:

$$p(A_e) = \frac{1}{\sigma_{A_d}\sqrt{2\pi}} e^{-\left(\frac{\left(A_e-\mu_{A_e}\right)^2}{2\sigma_{A_e}^2}\right)} \quad (14)$$

where, $\mu_{A_e}$ and $\sigma_{A_e}$ are the mean and standard deviation of the target effective cross section area respectively. Observing the results in Fig. 7, it can be seen that the effective target cross section area variation is small with a $\sigma_{A_e}$=4 and a large mean $\mu_{A_e}$=50. Thus, the average value of target cross section area is used. In other words, the target effective cross section area is modelled as a random viable with mean ($\mu_{A_e}$) and very small variance, which is ignored.

The received reflected signal from target is given as:

$$P_r = A_o\,\rho \quad (15)$$

where, $A_o$ is the LiDAL channel gain for a target located at $R_{\text{Max}}^{\text{FOV}}$ as in equations (2) and (3) of bistatic and monastic LiDAL systems respectively; and $\rho$ is a Gaussian random variable described in equation (13). Thus, the PDF of the received reflected signal $p(P_r)$ without noise can be defined as:

$$p(P_r) = \frac{1}{\sigma_s\sqrt{2\pi}} e^{-\left(\frac{(P_r-\mu)^2}{2\,\sigma_s}\right)} \quad (16)$$

where, $(\mu = A_o\sigma_\rho)$ and $(\sigma_s = A_o\sigma_\rho)$ are the mean and standard deviation of the received reflected signal. Equation (16) represents a Gaussian random variable scaled by a positive constant representing the LiDAL channel gain for a target located at $R_{\text{Max}}^{\text{FOV}}$ .

In the OW channel, ambient light induces shot noise in the photodetector receiver in addition to the thermal noise of the receiver amplifier. This noise is modelled as a white Gaussian noise [27] with zero mean and variance of $\sigma_t{}^2$(see equation 6). The noise probability density is given as:

$$p(n) = \frac{1}{\sqrt{2\pi}\sigma_t} e^{-\left(\frac{n^2}{2\sigma_t{}^2}\right)} \quad (17)$$

where $n$ is the total detected noise current in the receiver and $\sigma_t$ is the noise current standard deviation.

The noise is statistically independent and additive to the received reflected signal from the target. The shot noise due to the signal presence may be neglected compared to the thermal and shot ambient noises. Therefore, the joint probability density of the received signal in the presence of noise $p(p_{r_n})$ is written as:

$$p(p_{r_n}) = \frac{1}{\sqrt{(\sigma_s{}^2 + \sigma_t{}^2)}\sqrt{2\pi}} e^{-\left(\frac{(P_r - \mu)^2}{2(\sigma_s{}^2 + \sigma_t{}^2)}\right)} \tag{18}$$

where, $\mu$ and $\sqrt{(\sigma_s{}^2 + \sigma_t{}^2)}$ are the mean and standard deviation of the received reflected random signal in noise.

TABLE IV
PEOPLE FAVOURATIE COLOURS SURVEY DATA

| Favourite colour [30] | Weight [42] | Target coating reflection factor ($\rho$)[43] |
|---|---|---|
| Black | 7% | 0 |
| Yellow | 3% | 0.5 |
| White | 4% | 1 |
| Red | 8% | 0.9 |
| Purple | 14% | 0.78 |
| Orange | 5% | 0.4 |
| Green | 14% | 0.6 |
| Brown | 3% | 0.45 |
| Blue | 42% | 0.75 |

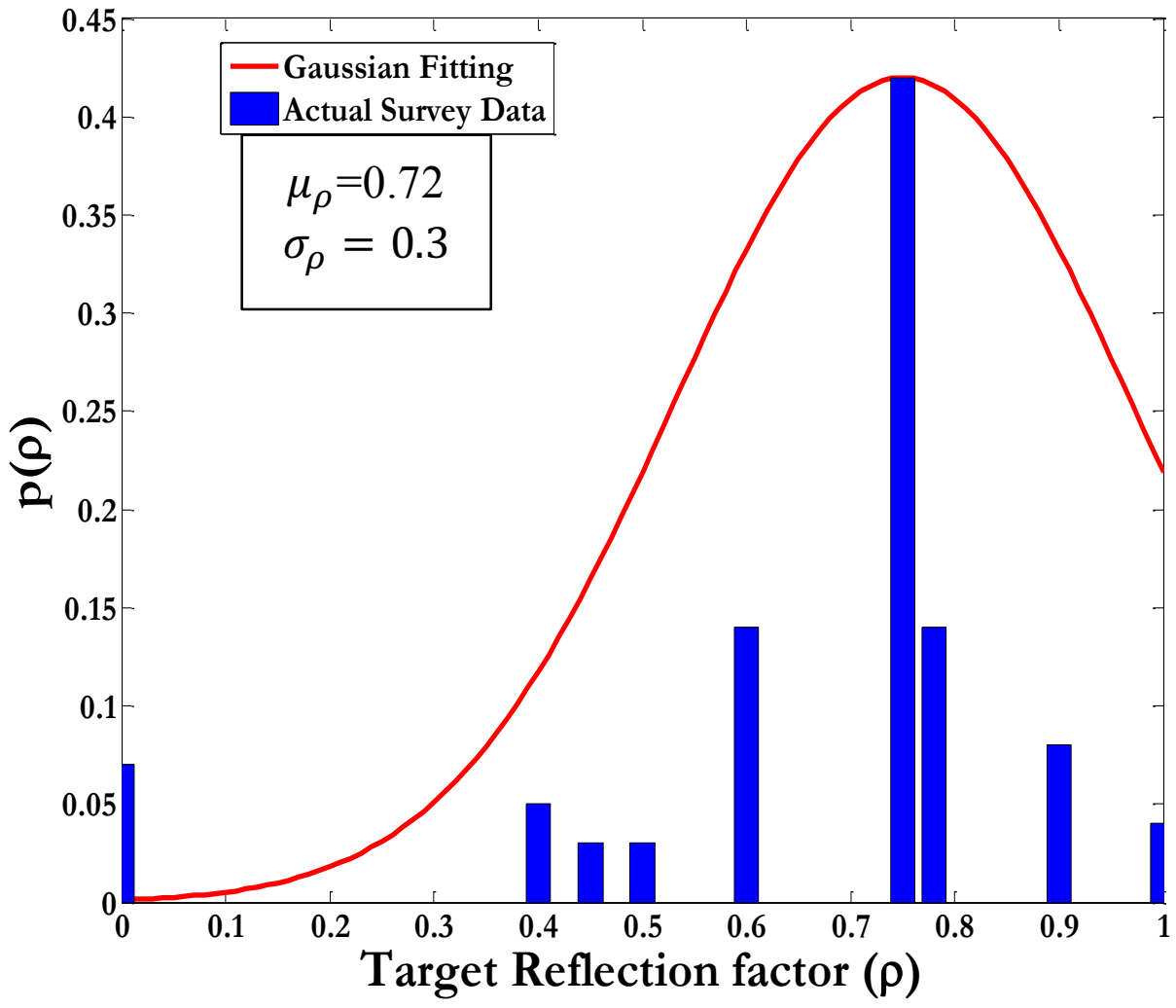

Fig. 6. The PDF of target reflection factor.

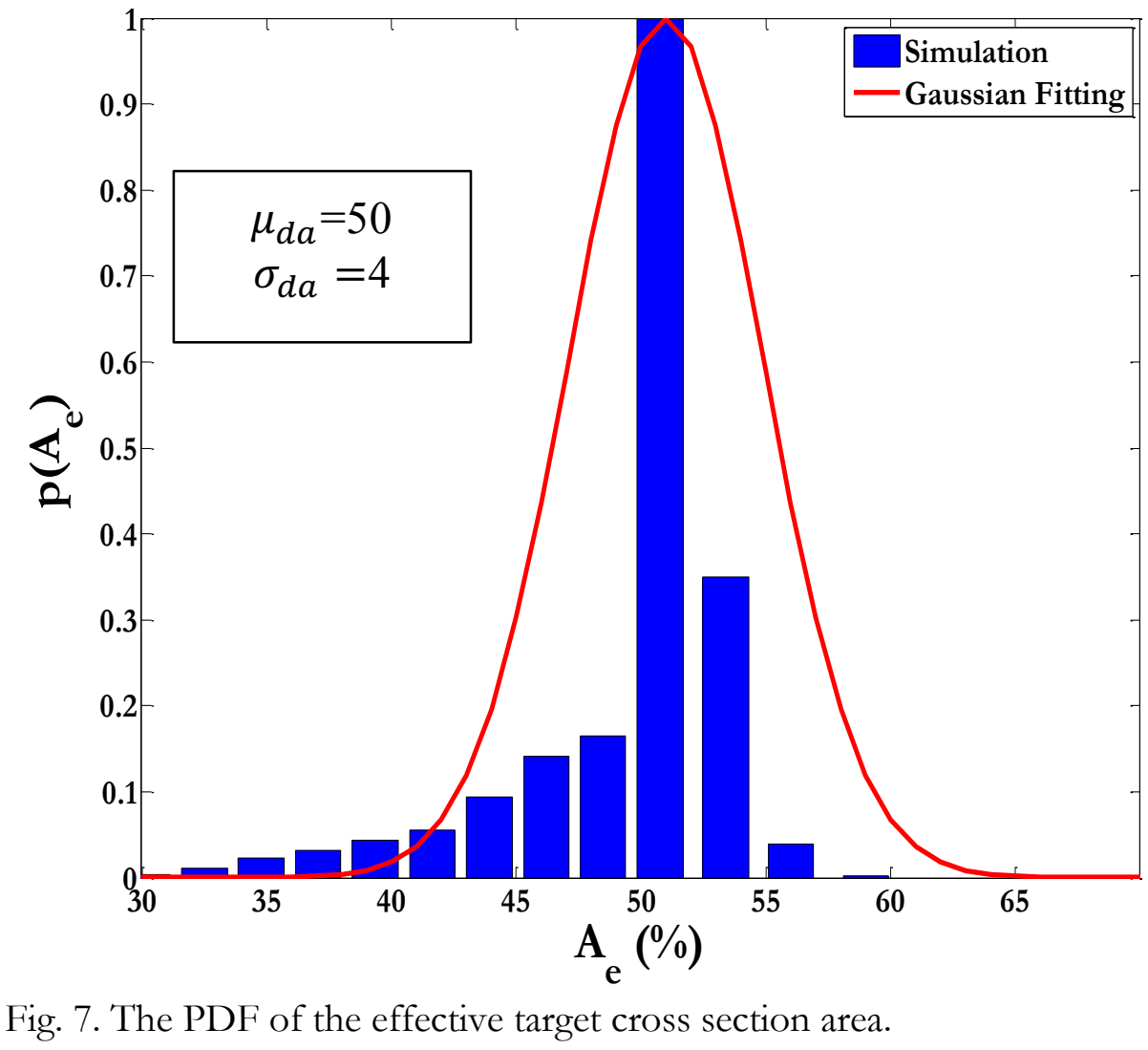

Fig. 7. The PDF of the effective target cross section area.

*F. LiDAL Optimum Receiver Design*

We used Bayes receivers and signal space theory to design an optimum receiver structure for LiDAL taking into account the minimization of the average cost of making decisions and the error in target detection. Bayes criterion takes into account the impact of the cost of making a wrong decision in different LiDAL applications by setting an optimum detection threshold. For instance, in a people counting application the cost of miss-detecting people may be low, however, for a LiDAL security application the cost of miss-detecting a target may be very high. We employed signal space techniques with a maximum posterior probability (MAP) decision rule to design an optimum LiDAL receiver based on minimum probability of error to detect target(s) for multiple cases as we discuss later in this paper. In addition, We evaluated the performance of the optimum detection threshold $D_{th}(z)$ where the random variable $z$ represents the received power in (18) . This was used to produce the receiver operating characteristics (ROC) in terms of the probability of false detection ($P_{FD}$) and the probability of detection ($P_D$).

*i. Optimum Detection Threshold Analysis (Hard Decision)*

We analyzed the optimum detection threshold for the LiDAL receiver considering the fluctuation of the received reflected signal and the cost of making a decision on LiDAL given the application considered. In LiDAL, the goal is to decide the presence or absence of a received reflected signal from a target in the presence of noise. This situation can be cast into two hypotheses. Let $H_1$ represent the hypothesis where noise is present and the reflected signal (from the target) is absent. Let $H_2$ represent the hypothesis where both the received signal (from target) and noise are present. The PDF of $H_1$ can be written as:

$$F_z(z|H_1) = \frac{1}{\sqrt{2\pi}\sigma_t} e^{-\left(\frac{z^2}{2\sigma_t{}^2}\right)} \tag{19}$$

and the PDF of $H_2$ is given as:

$$F_z(z|H_2) = \frac{1}{\sqrt{2\pi}\,\sigma} e^{-\left(\frac{(z-\mu)^2}{2\,\sigma^2}\right)} \tag{20}$$

where, $\sigma^2$ and $\mu$ are the variance and the mean of the received signal in $H_2$ with $\sigma^2 = (\sigma_s{}^2 + \sigma_t{}^2)$, see equation (18).

The Bayesian average cost of making decision $C(D)$ is given as [46], [47]:

$$C(D) = (p_o\alpha_{21} + q_o\alpha_{22}) + \int (q_o(\alpha_{12} - \alpha_{22})F_z(z|H_2) - (p_o(\alpha_{21} - \alpha_{11})F_z(z|H_1)d_z \tag{21}$$

where, $p_o$ and $q_o$ are the prior probabilities of $H_1$ and $H_2$ respectively. For LiDAL, we define the four prior costs as: $\alpha_{11}$which is the cost of deciding that the target is absent when it is true, $\alpha_{22}$ is the cost of deciding the target is present when it is true, $\alpha_{12}$ is the cost of deciding the target is absent when it is false and $\alpha_{21}$ is the cost of deciding the target is present when it is false. It should be observed that $p_o$ and $q_o$ were set to 0.5 which is a general case where it is equally likely to have a target or no target (for example in an indoor environment). In particular dense (user wise) indoor environments $q_o$ may be higher than $p_o$ and the converse is true in sparse indoor environments. Therefore, the parameters can be determined accordingly. We are interested in the costs of wrong decisions ($\alpha_{12}$ and $\alpha_{21}$), hence we assumed $\alpha_{11}$ and $\alpha_{22}$ (costs of correct decisions) are equal to zero. To clarify this, $\alpha_{12}$ is defined as the cost of missing a target, while $\alpha_{21}$ is defined as the cost of a false alarm. Note that, $\alpha_{12}$ should be set higher than $\alpha_{21}$, for security applications where missing a target is worse than a false alarm. However we are interested here in target counting applications, and therefore $\alpha_{12}$ was set equal to $\alpha_{21}$ where both wrong decisions equally contribute to wrong counting. Thus, the LiDAL average cost of making decision $C(D)_{VLP}$ can be written as:

$$C(D)_{VLP} = p_o\alpha_{21} + \left(q_o\alpha_{12}\int F_z(z|H_2)d_z - p_o\alpha_{21}\int F_z(z|H_1)\,d_z\right). \tag{22}$$

The first term of (22) represents the fixed cost while the second term represents the variable cost. We wish to minimize the second term of (22) by choosing the value of $z$. Mathematically (22) can be summarized by a pair of inequalities, and can thus be rewritten as:

$$q_o\alpha_{12}\,F_z(z|H_2)\ \underset{H_2}{\overset{H_1}{\lessgtr}}\ p_o\alpha_{21}\,F_z(z|H_1)\,. \tag{23}$$

For LiDAL, we define $\gamma_{FA}$ and $\gamma_{FP}$ as the cost factors of missing the target and false alarm respectively. Therefore, $\gamma_{FA}$ (FA is False Absence) is given as:

$$\gamma_{FA} = q_O\alpha_{12} \tag{24}$$

and the $\gamma_{FP}$ (FA is False Presence) is given as:

$$\gamma_{FP} = p_O\alpha_{21}. \tag{25}$$

Thus, we get:

$$\frac{F_z(z|H_2)}{F_z(z|H_1)}\ \underset{H_2}{\overset{H_1}{\lessgtr}}\ \frac{\gamma_{FP}}{\gamma_{FA}} \tag{26}$$

where $\frac{\gamma_{FP}}{\gamma_{FA}}$ is the LiDAL likelihood test threshold, and $\frac{F_z(z|H_2)}{F_z(z|H_1)}$ is the LiDAL likelihood test ratio.

Substituting equations (19) and (20) into equation (26), the optimum detection threshold $D_{th}(z)$ can be derived as:

$$D_{th}(z)\ \underset{H_2}{\overset{H_1}{\lessgtr}}\ \left(\sqrt{\frac{\mu^2}{(\beta_\sigma-1)^2}+\frac{\mu^2}{\beta_\sigma-1}+\frac{2(\sigma_s^2+\sigma_t^2)}{\beta_\sigma-1}\left(\ln\frac{\gamma_{FP}}{\gamma_{FA}}-\ln\frac{\sigma_t}{\sqrt{\sigma_s^2+\sigma_t^2}}\right)}\right) - \left(\frac{\mu}{\beta_\sigma-1}\right) \tag{27}$$

where, we define $\beta_\sigma = \left(\frac{{\sigma_s}^2+{\sigma_t}^2}{\sigma_t^2}\right)$ as a colour factor where $\beta_\sigma \geq 1$. The colour factor $\beta_\sigma$ is a measure of the variation in the received reflected signal due to the colour worn by the target, versus the variation in the received signal due to noise. For example, if all the targets wore the same colour, then $\sigma_s^2 = 0$ and $\beta_\sigma = 1$. At the other extreme, if the colours worn by the targets are very different and the receiver noise is very small, $\beta_\sigma \to \infty$. It is worth observing that in addition to colour, other optical properties of the target coating affect $\beta_\sigma$, such as the material used in the clothing (e.g. cotton verses polyester).

As can be noted in Fig. 8, when the weights of cost factors are equal ($\frac{\gamma_{FP}}{\gamma_{FA}} = 1$) and $\beta_\sigma \approx 1$ (i.e. the value of signal variance is very small $\sigma_s \approx 0$), the optimum $D_{th} \approx \frac{\mu}{2}$. This case is the classical scenario [47], which acts to validate our derivation of equation (27). Fig. 8 shows the main operating region for the LiDAL detection system. Firstly, the LiDAL system can be used for counting purposes only. In other words, to count the number of human pedestrians. Here the cost of missing a target and the cost of a false alarm are identical as they result in equal counting errors. This is represented by $\gamma_{FP} = \gamma_{FA}$. Secondly, if the application is such as that there is high cost associated with falsely identifying the presence of a target in the indoor environment, then the detection threshold is set high, represented for example by $\gamma_{FP} = 10$ and $\gamma_{FA} = 1$ in Fig. 8. Finally, if the cost of missing a human pedestrian target is very high (security or safety application), then the threshold should be set very low as shown in Fig. 8 where for example $\gamma_{FP} = 1$ and $\gamma_{FA} = 10$ and $\gamma_{FA} = 100$.

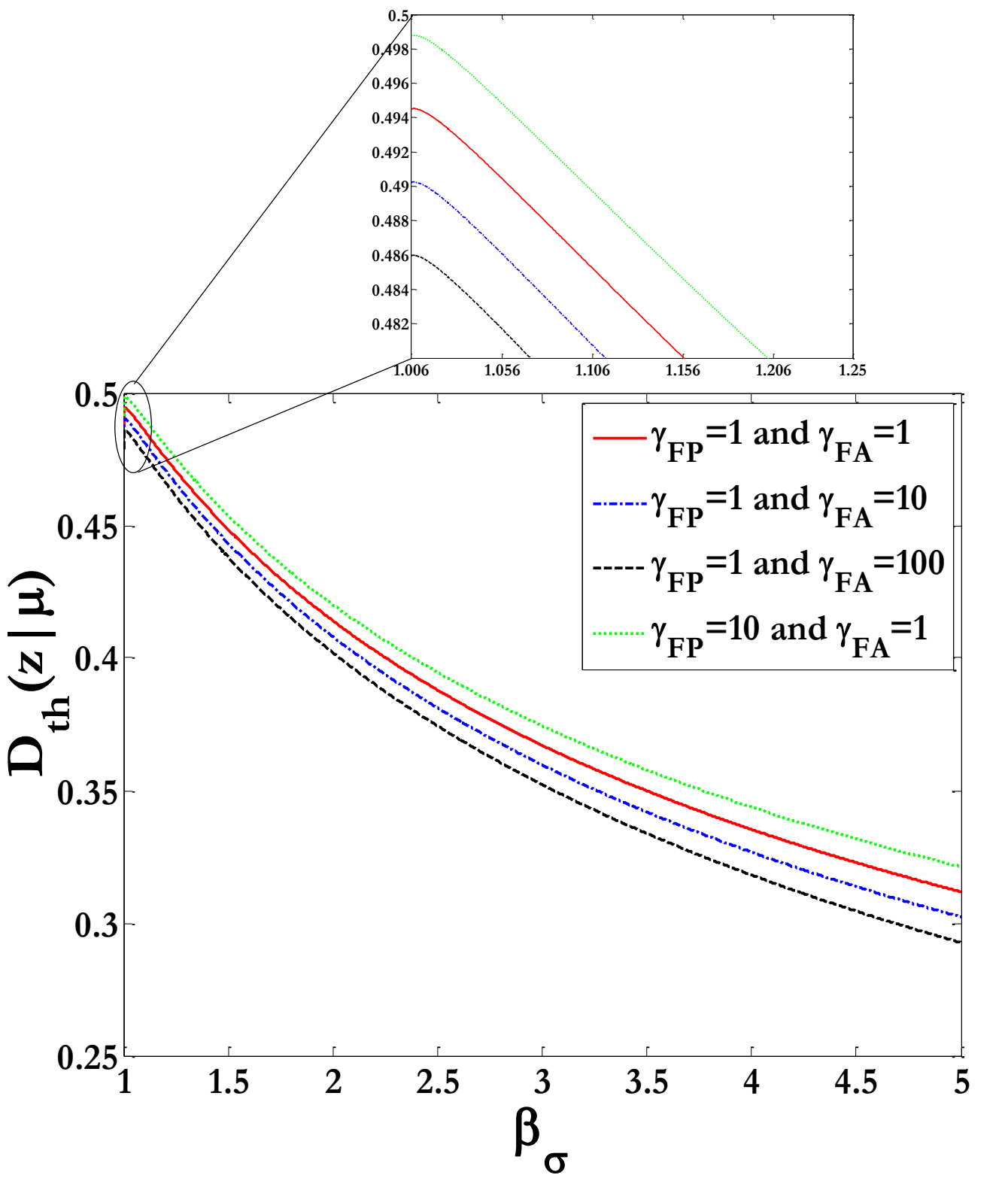

Fig. 8. The Optimum detection threshold with $\beta_\sigma$ and LiDAL cost factors.

*ii. Probability of False Detection ($P_{FD}$)*

In the absence of a target there is a chance that a noise signal from any ambient light source can exceed the detection threshold. This noise signal can thus be interpreted as a reflected signal from a target which causes false detection (earlier referred to as false presence or false alarm). The probability of false detection $(P_{FD})$ is therefore given as[32], [47]:

$$P_{FD} = \int_{D_{th}}^{\infty} F_z(z|H_1)\,d_z \tag{28}$$

By solving (28), $P_{FD}$ can be written as:

$$P_{FD} = \frac{1}{2}erfc\left(\frac{\left(\sqrt{\frac{\mu^2}{(\beta_\sigma-1)^2}+\frac{\mu^2}{\beta_\sigma-1}+\frac{2\sigma^2}{\beta_\sigma-1}\left(\ln\frac{\gamma_{FP}}{\gamma_{FA}}-\ln\frac{\sigma_t}{\sigma}\right)}\right)-\left(\frac{\mu}{\beta_\sigma-1}\right)}{\sqrt{2}\sigma_t}\right) \tag{29}$$

where $erfc$ is the error function complementary.

*iii. Probability of Detection ($P_D$)*

The probability of detecting a target relies on the received signal reflected by the target in the presence of noise. The probability of detection $(P_D)$ is therefore given as [32, 47]:

$$P_D = \int_{D_{th}}^{\infty} F_z(z|H_2)\,d_z \tag{30}$$

Solving equation (30) we get:

$$P_D = \frac{1}{2} erfc\left(\frac{\left(\left(\left(\sqrt{\frac{\mu^2}{(\beta_\sigma-1)^2}+\frac{\mu^2}{\beta_\sigma-1}+\frac{2\sigma^2}{\beta_\sigma-1}\left(\ln\frac{\gamma_{FP}}{\gamma_{FA}}-\ln\frac{\sigma_t}{\sigma}\right)}\right)-\left(\frac{\mu}{\beta_\sigma-1}\right)\right)-\mu\right)}{\sqrt{2}\sigma}\right) \quad (31)$$

As discussed in Section III.F, the ROC can now be evaluated for the proposed MIMO and Imaging LiDAL systems. This will be reported in Section IV, after considering the environments and systems of interest, hence the optimum threshold $D_{th}$ which in turn relies on the statistical characteristics of the reflected signal and noise in each LiDAL system.

### *iv. LiDAL Optimum Detector*

We use the term detector here to imply and include the initial signal detection by the optical receiver, followed by its optimum processing and finally decision making. We implemented a MAP detection approach in LiDAL to design an optimum receiver based on observation of the received reflected signal(s); and hence calculation of the posterior probability to minimize the probability of decision errors [46]. In LiDAL, a single transmitted pulse is sent and is reflected from the target(s) to the receiver where the receiver uses a finite listening time. The LiDAL receiver listening time ($T_s$) is divided into $N$ time slots. Two cases arise, the single target case and the multiple targets case. In the single target case, (i) if the target presence in all spatial locations is equally likely, then the time slots have equal prior probabilities for target reception; (ii) in the single target case, however, the reception of a pulse in a time slot implies that the remaining time slots (if any) will contain no pulses, hence the independence of the time slots does not hold. In the multiple targets case, condition (i) holds, and further in (ii) the reception of a pulse does not exclude the remaining time slots from having targets / pulses. Therefore, independence of the time slots can be assumed (ignoring instances where targets may walk in pairs for example). Therefore, we assume here equal prior probabilities for the time slots and assume the independence of the time slots, which is a general common case. The LiDAL receiver has to optimally determine (*i*) target presence, (*ii*) number of targets (number of time slots containing pulses) and (*iii*) identify the time slot (target's range).

The time slot width ($T_s$) is related to the desired LiDAL resolution and target ranging accuracy. Therefore, we select a time slot width equal to the transmitted pulse width ($T_s$=$\tau$) in order to obtain a $\Delta R$=30cm resolution. This 30cm resolution corresponds to the minimum typical separation of interest between humans in an indoor environment. Selecting narrower pulses can improve the resolution, however this is not needed and can lead to higher dispersion in the channel.

Here we analyse three cases of interest: Single target case, multiple targets case and multiple targets with channel dispersion.

1) Case I assumptions: Single target, noise present, no channel dispersion, the receiver's $N$ time slots are orthogonal (i.e. only one received reflected pulse), the received reflected pulse may fit into one time slot or overlap with a neighbour time slot (i.e. the received pulse is shifted in the listening frame depending on target location and may occur at the boundary of the time slot), and independent time slots. For the purpose of this case, the objectives of the designed receiver are detecting the target presence and its range.

Case I is similar to M-ary orthogonal signals (pulse position modulation (PPM)) [47], where a single transmitted pulse is reflected from one target and received by a time slot $T_{s_j}$. The MAP rule for minimum probability of error is given as [46], [47]:

$$P(H_i|z_1, \dots z_N) = \frac{f_Z(z_1, \dots z_N|H_i)P(H_i)}{f_Z(z_1, \dots z_N)} \quad (32)$$

where, $Z \in [z_1, .. z_N]$ is the observed received signal vector in $N$ time slots and $P(H_i)$ is the probability of receiving $H_i$, with $P(H_i) = \left(\frac{1}{N+1}\right)$, $i \in \{1 \dots, N+1\}$; $P(H_i)$ takes this values since the received reflected signal from a target can be present (equi-probably) in any of $N$ time slots depending on the target location. Note that, $P(H_i)$ and $f_Z(z_1, \dots z_N)$ do not depend on $H_i$ [46]. Therefore, we require a receiver to calculate $f_Z(z_1, \dots z_N|H_i)$ and choose the $H_i$ associated with the largest probability [47]. The orthonormal expansion $Z$ of the received signal can be written as [47]:

$$Z_j = \int_0^{T_s} \big(p_r\,(t) + n\,(t)\big)\phi_j(t)\,dt \qquad j \in \{1, \dots N\} \quad (33)$$

where, $p_r(t)$ is the received signal, $n\,(t)$ is the noise and $\phi_j(t)$ is the orthonormal basis function chosen as:

$$\int_0^{T_s} \phi_u(t)\,\phi_j(t) = \begin{cases} 1, & u = j \\ 0, & u \neq j \end{cases} \quad (34)$$

where, $\phi_j(t) = \prod(t - jT_s)$. It should be noted that $z_1, \dots z_N$ are uncorrledetd and statistically independent, therefore their joint probability is given as:

$$f_Z(z_1, \dots z_N|H_i) = \prod_{j=1}^{N} F_z(z_j|H_i) \qquad i \in \{1 \dots, N+1\} \quad (35)$$

The mean and variance of hypothesis $H_i$ are given as:

$$E\{Z_j|H_i\} = A_{i_j} \quad (36)$$

$$var\{Z_j|H_i\} = \sigma^2 \quad (37)$$

where $A_{i_j}$ is the orthonormal coefficient given as [47]:

$$A_{i_j} = \int_0^{T_s} p_r(t)\phi_j(t)dt \quad (38)$$

Equation (35) can be rewritten as:

$$f_Z(z_1, \dots z_N|H_i) = \prod_{j=1}^{N} \frac{e^{-\frac{\left(z_j - A_{i_j}\right)^2}{2\sigma^2}}}{\sigma\sqrt{2\pi}} \quad (39)$$

$$f_Z(z_1, \dots z_N|H_i) = \frac{e^{-\left(\sum_{j=1}^{N}\frac{\left(z_j - A_{i_j}\right)^2}{2\sigma^2}\right)}}{(\sigma 2\pi)^{N/2}}\,. \quad (40)$$

Thus,

$$f_Z(z_1, \dots z_N|H_i) = \frac{e^{-\left(\frac{||z_j - s_i||^2}{2\sigma^2}\right)}}{(\sigma 2\pi)^{N/2}} \quad (41)$$

where:

$$s_i(t) = \sum_{j=1}^{N} A_{i_j}\,\phi_j(t) \quad (42)$$

Therefore, as equation (41) shows, the optimum receiver that maximizes the likelihood is one that minimizes the distance between $z$ and $s_i$. In other words, it is a receiver that chooses the minimum distance to the orthonormal coefficient coordinates.

For instance when $N = 2$, we have three hypotheses: (*i*) $H_0$ no target and both time slots contain only noise (note equation 19 for $F_z(z)$), (*ii*) $H_1$ time slot $T_{s_1}$ contains the received reflected signal form a target with noise and $T_{s_2}$ contains only noise and *(iii)* $H_2$ time slot $T_{s_1}$ contains only noise and $T_{s_2}$ contains the received reflected signal with noise. The receiver decision rule for $H_1$ and $H_2$ will be to compare the values of $z_j$ to the orthonormal coefficient values and select the minimum distance to the orthonormal coefficients as illustrated in Table V. However, for $H_0$ all time slots (i.e. $z_j$ values) have comparable energy.

TABLE V SINGLE TARGET DETECTION IN TIME SLOTS

| Observation | Decision |
|---|---|
| $f_z(z_1, z_2 \mid H_1) > f_z(z_1, z_2 \mid H_2) \Rightarrow z_1 > z_2$ | $H_1$ |
| $f_z(z_1, z_2 \mid H_2) > f_z(z_1, z_2 \mid H_1) \Rightarrow z_2 > z_1$ | $H_2$ |
| $f_z(z_1, .. z_N \mid H_j) > f_z(z_1, z_N \mid H_m)\ \forall m \in \{1, .. N\}, m \neq j$ | $H_j$ |

Fig.9a shows the optimum LiDAL receiver structure to be used to detect a single target (see Case I) based on the analysis of Table V and equation (33). Each branch uses one of the orthonormal functions (see shift register) and an integrator to determine the *N* dimensional expansion point collectively between the branches. Therefore, after observing the received signal in $N$ time slots during the listening time ($T = NT_s$), the receiver decides the target presence and range (related to $T_{s_j}$) through the decision circuit. Fig.9b presents an example of the orthonormal functions $\phi_j(t)$ for $N$=4 time slots with $T_s$ =2ns for three radar (LiDAL) scans during the $T$ listening time.

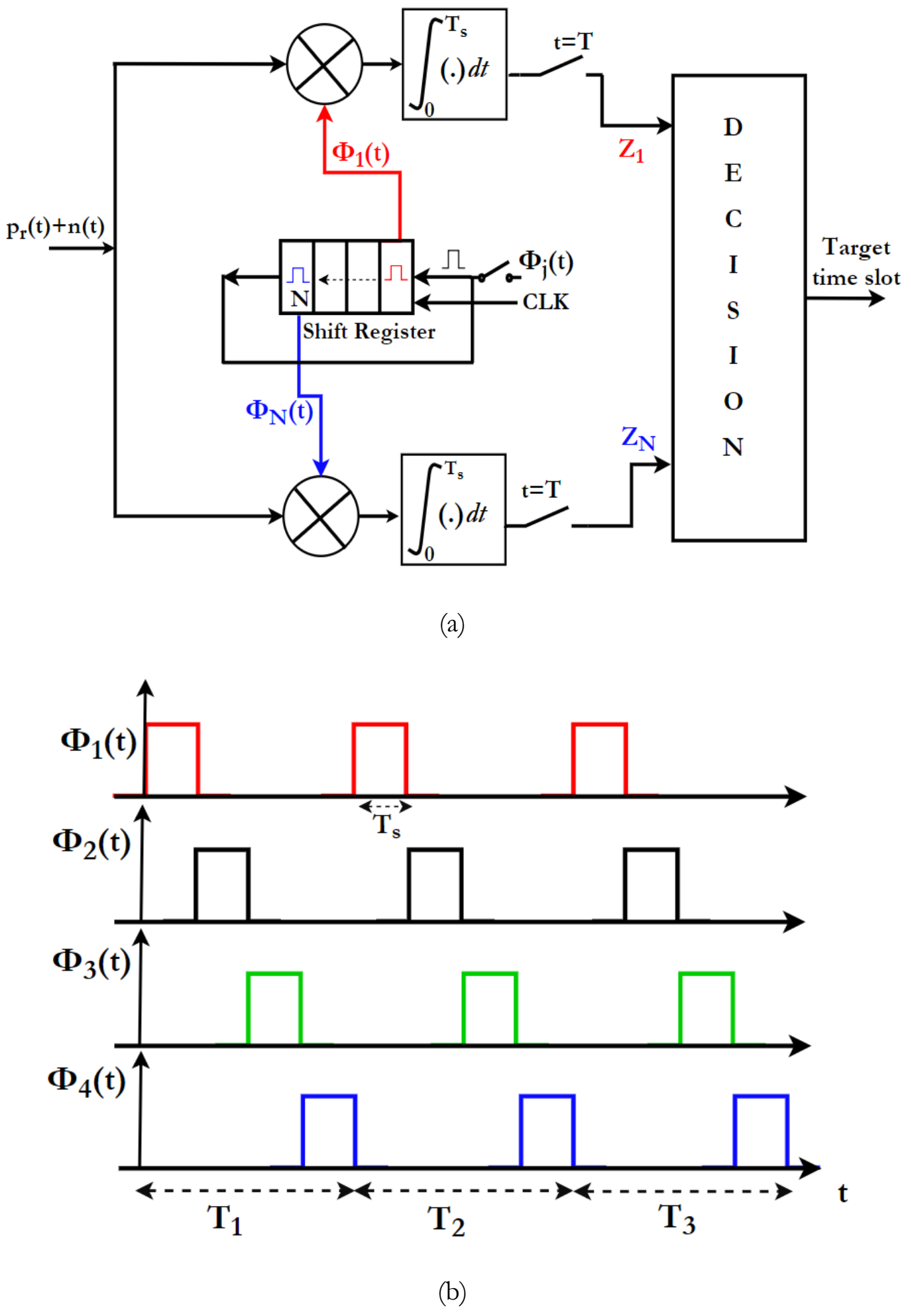

Fig. 9. LiDAL receiver, (a) The LiDAL optimum detector block diagram, single target detection and (b) the orthonormal $\phi_j(t)$ signalling diagram.

We evaluated the performance of the LiDAL receiver through the probability of making a correct decision $P_c$ on $H_i$, where the reflected signal from the target is received as $z_i$; $P_c$ can be derived as:

$$P_c = P(z_i \mid H_j) = P(z_j > z_m) \text{ where } \forall m \in \{1, .. N\}, m \neq j \quad . \quad (43)$$

Substituting equations (33) and (17) in equation (43), we get:

$$P_c = \left( \int_{-\infty}^{z_j} \frac{e^{-\left(\frac{n_j^2}{2\sigma_t^2}\right)}}{\sqrt{2\pi}\,\sigma_t} dn_j \right)^{N-1} \quad . \quad (44)$$

2) Case II assumptions: Multiple targets, target locations are spaced by $\Delta R$ or more, noise is present, no channel dispersion, the receiver $N$ time slots are orthogonal, but the received multiple reflected pulses from $M$ targets $(M \leq N)$ may be shifted depending on the targets locations and hence the received pulses are not orthogonal. We do not consider the case where there are more targets than time slots, which is an extension that warrants further investigation. We consider this situation however in the imaging receiver case in Section V.

*a) Exhaustive Search Receiver (ESR)*

In this section, we propose an optimum receiver for Case II based on an exhaustive search algorithm as follows:

1. The receiver observes the reflected signal $p_r$ and produces the orthonormal expansion $Z$ for the $N$ time slots in the presence of noise.
2. First, the receiver's decision block (as can be seen in Fig.9a) compares these $N$ orthonormal coefficients coordinates to the 'no target' hypothesis as all $N$ time slots contain only noise, where the observed $N$ orthonormal coordinates are $(z_1, z_2 .. z_N)$ and the orthonormal coefficient are $(A_{v_1}, A_{v_2}, .. A_{v_N})$. For the no target case, $A_{i_j} = 0, \forall\, j$ and the error $e_v$ can be defined as:
$$e_v = \sum_{j=1}^{N} ||z_j - A_{v_j}||^2 \quad (45)$$
3. The decision block then compares the observed $N$ orthonormal coefficients coordinates to the coefficients associated with the presence of a single target hypothesis. where there are $N$ possible time slots to receive the reflected signal from the target, yielding $N$ candidate answers; and calculate their errors (see equation (45)).
4. Next, the decision block calculates the errors assuming the presence of two targets, where there are $\left(\frac{N(N-1)}{2}\right)$ candidate answers. Thus, the total candidate answers $(C_A)$ for $N$ time slots and $k$ targets can be defined as:
$$C_A = 1 + \sum_{k=1}^{N} \frac{N!}{(N-k)!\,k!} \qquad N \geq k \quad (46)$$
5. Finally the decision block continues to find the errors for all cases and chooses the $v^{th}$ case (number of targets and their time slots) which has the minimum error:
$$v = \underset{v}{arg\,min} \left( \sum_{v=1}^{C_A} e_v \right) \qquad v \in \{1, .. C_A\} \quad (47)$$

In the exhaustive search receiver, the probability of making a correct decision $P_c^{ESR}$ to detect $k$ targets in $N$ time slots can be written as:

$$P_c^{ESR} = \left( \int_{-\infty}^{z_j} \frac{e^{-\left(\frac{n_j^2}{2\sigma_t^2}\right)}}{\sqrt{2\pi}\,\sigma_t} dn_j \right)^{N-k} \qquad N > k\,. \qquad (48)$$

For example, a LiDAL system with listening time divided into $N = 14$ time slots and maximum counted targets of $k = 10$, the total candidate answers are $C_A = 15914$ . Therefore, the exhaustive search receiver may be very complex to implement for the LIDAL system.

*b) Sub-Optimum Receiver* (*SOR*)

In this section, we introduce a sub-optimum receiver with lower complexity compared to the exhaustive search receiver. Following the analysis of the MAP rules, Fig.10 presents the sub-optimum receiver for Case II. For the sake of simplifying the analysis of Case II, let us assume two target, $k$=2, detection in $N$=2 time slots. Hence, we have four hypotheses: *(i)* $H_0$ noise present only and targets are absent, *(ii)* $H_1$ a single target is present at $T_{s_1}$ with noise, *(iii)* $H_2$ a single target is present at $T_{s_2}$ with noise and *(iv)* $H_3$ two targets present at $T_{s_1}$ and $T_{s_2}$ with noise. Table VI illustrates the four possible hypotheses and receiver observation with the optimum decision. To determine $H_0$ with minimum error, a comparator is connected at the output of each correlator to determine the presence/absence of the received reflected signal at each time slot compared to a lower optimum detection threshold $D_{th_L}$ as can be seen in Fig. 10. In addition, the receiver has to determine whether there is a single reflected pulse located between two neighbouring time slots (i.e. the correct decision is $H_1$ or $H_2$ ) or there are two reflected pulses from two targets received in the two time slots (i.e. the correct decision is $H_3$). Consequently, we set up a second comparator at the output of each correlator with a high detection threshold $D_{th_H} = \frac{\mu}{2}$ as can be seen in Fig. 10. Therefore, the final receiver decision block decides as follows:

1. If the observed received signal $z_j$ is below $D_{th_L}$, then the target is absent in $T_{s_j}$.
2. If the observed received signal $z_j$ is above $D_{th_H}$, then the target is present in $T_{s_j}$.
3. If the observed received signal $z_j$ is above $D_{th_L}$ and below $D_{th_H}$ , then it is a pulse received in two neighbouring time slots $T_{s_j}$, $T_{s_{j+1}}$. Thus the decision circuit compares $z_j$ with $z_{j+1}$ and selects the largest.

Fig.11 shows the LiDAL observation space diagram for multiple (two) target detection in two time slots. Each plane of the observation space is divided into four decision regions. Wherever, (region) the coordinates of the observed received signal fall, the receiver decision is based.

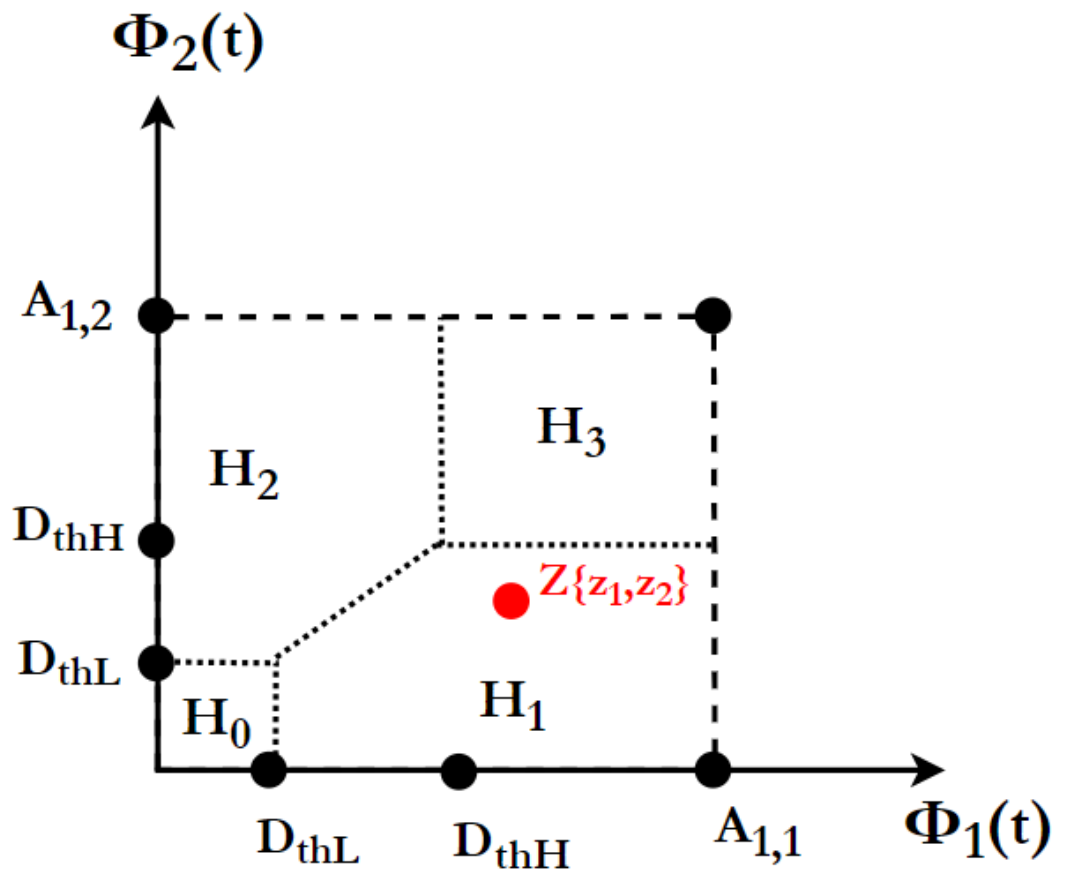

Fig. 11. The LiDAL receiver two-dimensional observation space.

TABLE VI MULTIPLE TARGETS DETECTION HYPOTHESES WITH N TIME SLOTS.

| Hypothesis | Observation | Decision |
|---|---|---|
| No target $(T_{s_1})$ and $(T_{s_2})$ | $z_1, z_2 < D_{th_L}$ | $H_0$ |
| One target $(T_{s_1})$ | $f_Z(z_1,z_2\|H_1) > f_Z(z_1,z_2\|H_i) \quad \forall\, i, i \neq 1$ | $H_1$ |
| One target $(T_{s_2})$ | $f_Z(z_1,z_2\|H_2) > f_Z(z_1,z_2\|H_i) \quad \forall\, i, i \neq 2$ | $H_2$ |
| Two targets at $(T_{s_1})$ and $(T_{s_2})$ | $z_1, z_2 > D_{th_H}$ | $H_3$ |

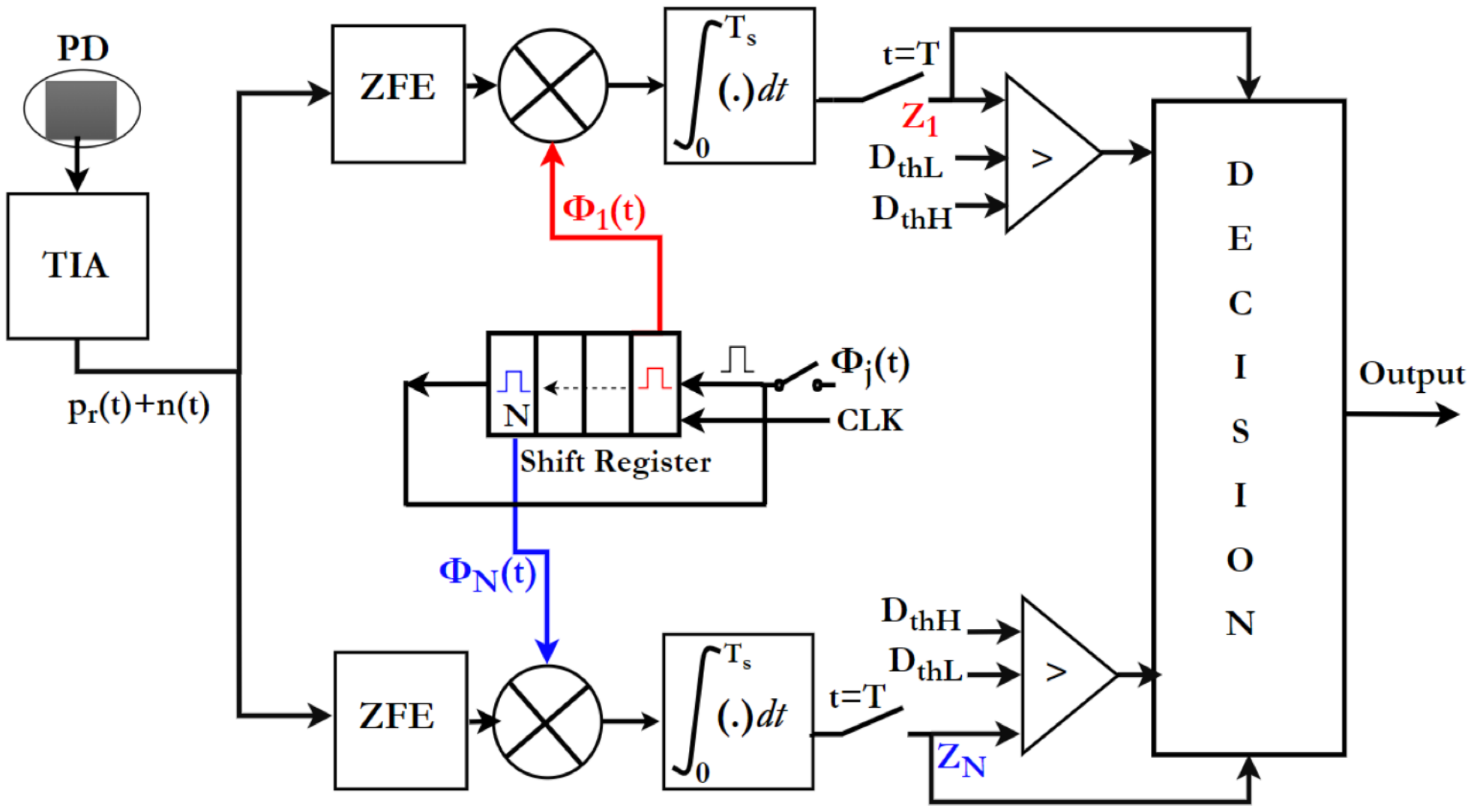

Fig. 10: The LiDAL sub-optimum receiver block diagram.

The probability of a correct decision on target detection in a time slot $P_{c_T}^{SOR}$ for the SOR can be given as:

$$P_{c_T}^{SOR} \geq \int_{D_{th_H}}^{\infty} \frac{e^{-\left(\frac{(z_j-\mu)^2}{2\sigma^2}\right)}}{\sigma\sqrt{2\pi}} dz_j \qquad (49).$$

where the use of the high detection threshold ($D_{th_H}$) establishes an upper bound on the detection errors and hence a lower bound on the probability of correct detection. This is therefore conservative. The total probability of correct decisions, $P_c^{SOR}$, when detecting $k$ targets in $N$ time slots for the SOR can be derived as:

$$P_c^{SOR} = P(H_0)P_{c_Z}^{SOR} + P(H_1)P_{c_T}^{SOR} + \cdots P(H_k)P_{c_T}^{SOR} \qquad (50)$$

where, $P(H_k)$ is the prior probability of having $k$ targets in $N$ time slots, $P(H_k) = \frac{1}{N}$ and $P_{c_Z}$ is the probability of correct decision of detecting zero targets which is written as:

$$P_{c_Z}^{SOR} = \int_{-\infty}^{D_{th_L}} \frac{e^{-\left(\frac{(z_j)^2}{2\sigma_t^2}\right)}}{\sigma_t\sqrt{2\pi}} dz_j. \qquad (51)$$

The probability $P_c^{SOR}$ can therefore be given as:

$$P_c^{SOR} = \frac{1}{N+1}\left(P_{c_Z}^{SOR} + \sum_{k=1}^{N} \frac{N!}{(N-k)!\,k!} \left(P_{c_T}^{SOR}\right)^k \left(P_{c_Z}^{SOR}\right)^{N-k}\right) \qquad (52)$$

3) Case III assumptions: These are the same as Case II but now we consider the effects of the optical channel propagation. The LiDAL channel can be heavily dispersive as shown in Figs 3c, d. The narrow-transmitted pulse and receiver time slot widths cause; (*i*) pulse spreading (over two or more neighbouring time slots) of the received pulse reflected from a single target. This leads to a decrease in the probability of correct decision for the proposed LiDAL optimum receivers (ESR and SOR); (*ii*) ambiguity in target location due to the pulse spread over multiple time slots.

To eliminate the effect of the inter-time slots interference (ITI), the receiver time slot width must be selected according to the minimum LiDAL channel bandwidth where the optimum time slot width $T_{s_{Op}}$ for ITI free operation can be chosen as $T_{s_{Op}} = \frac{1}{BW_{ch_{\min}}}$. The optimum time slot width for ITI free operation is $T_{s_{Op}}$=12ns in the room in Section II.B using the system parameters in that section. However, for $T_s$ =12ns, the radar (LiDAL) detection resolution $\Delta R$ will decrease significantly by a factor of 6 (from $\Delta R$ =0.3m to $\Delta R$ =1.8m). Thus, the time slot width was chosen in Section II.C to maintain the desired radar detection resolution of $\Delta R$ =0.3m with $T_s$=2ns. Therefore, we implemented a zero forcing equalizer (ZFE) in the LiDAL receiver to equalize the channel [48], [49], [50]. In other words, to minimize the inter-time slots interference, while maintaining the selected time slot width ($T_s$ =2ns) for optimum radar detection resolution.

We designed the ZFE to equalize the LiDAL channel at the worst target location. Table VII illustrates the noise enhancement and LiDAL channel delay spread with number of ZFE taps.

The ZFE consists of 7-taps weighted finite impulse response filter (FIR). The weights $c\,[-l, \ldots\, l]$ were optimized according to [50]. The ZFE output signal is written as:

$$y_{ZFE}(t) = \sum_{n=-l}^{l} c_n\, P_r\,(t - nT) \qquad (53)$$

The noise variance after ZFE can be given as [50] :

$$\sigma_{ZF}{}^2 = \sigma_t{}^2 \sum_{n=1}^{l} c_n{}^2 \qquad (54)$$

Note that, for the ZFE design $\sum_{n=1}^{K} c_n{}^2$ is 1.2 and therefore the new variance $\sigma_{ZF}{}^2 = 1.2\,\sigma_t^2$.

Fig.12 depicts the probability of error ($P_E = 1 - P_c^{SOR}$, or $P_E = 1 - P_c^{ESR}$) of detecting single and multiple targets for ESR and SOR after employing the ZFE. The receiver listening time is divided into $N = 4$ time slots (which is the number of time slots needed to cover one optical footprint whose radius is 1.2m, Fig 2, and with $\Delta R$ =0.3m. As can be seen in Fig. 12, the ESR has better performance compared to SOR. For $k = 3$ with 15dB SNR, the $P_E$ was 0.1 and 0.21 for ESR and SOR respectively.

TABLE VII
ZFE DELAY SPREAD AND NOISE ENHANCEMENT

| Number of ZFE taps | Delay Spread (ns) | Noise Variance |
|---|---|---|
| 0 | 4.5 | $\sigma_{ZF}{}^2 = 1\,\sigma_t^2$ |
| 1 | 4.41 | $\sigma_{ZF}{}^2 \approx \sigma_t^2$ |
| 3 | 3.13 | $\sigma_{ZF}{}^2 = 1.15\,\sigma_t^2$ |
| 5 | 1.43 | $\sigma_{ZF}{}^2 = 1.17\,\sigma_t^2$ |
| 7 | 1.02 | $\sigma_{ZF}{}^2 = 1.2\,\sigma_t^2$ |
| 9 | 1.01 | $\sigma_{ZF}{}^2 = 1.22\,\sigma_t^2$ |

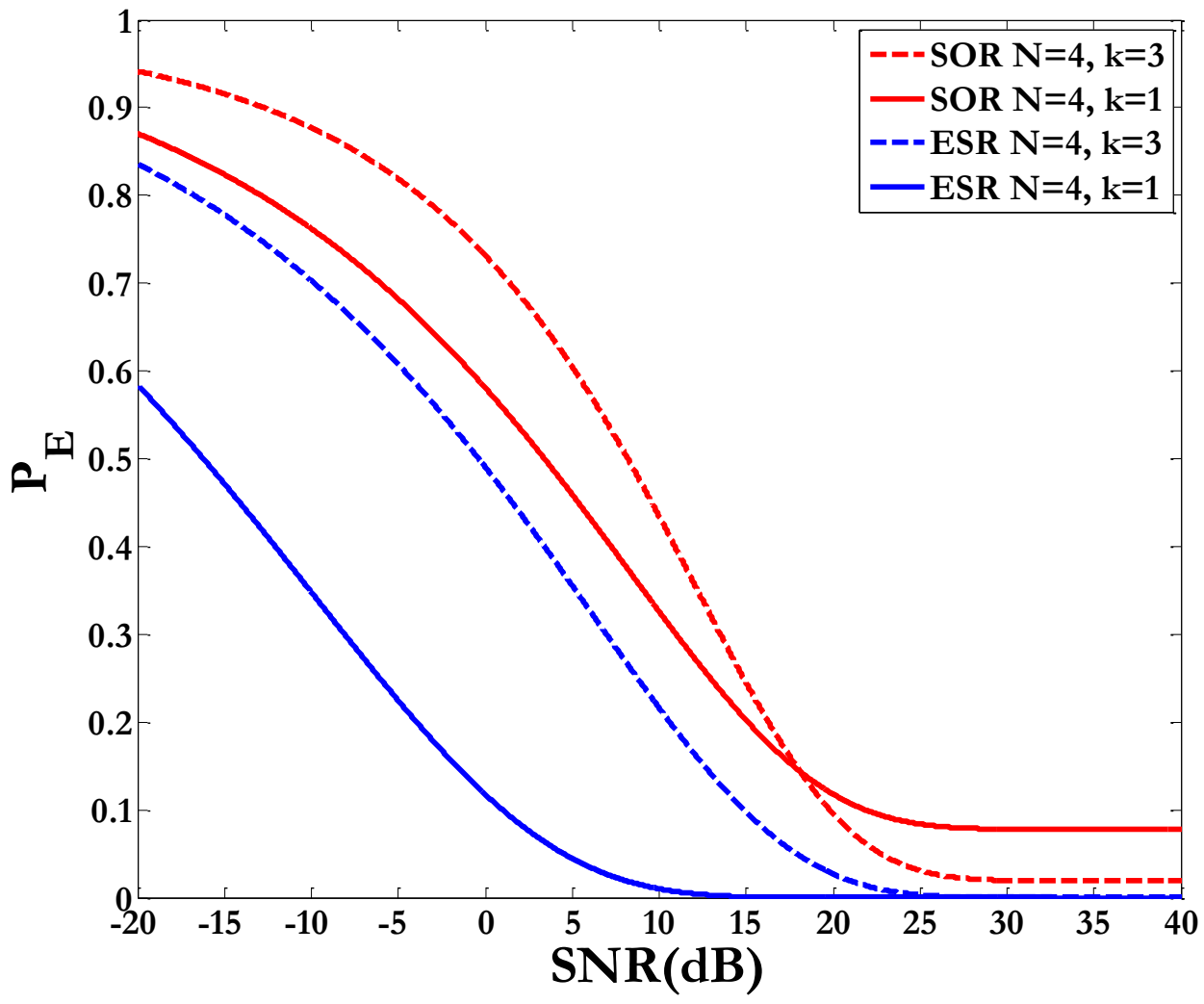

Fig.12: Probability of error IN detecting targets for ESR and SOR.

## IV. Target Distinguishing Approaches and Mobility Modelling in Realistic Environment

To detect the desired targets (humans in our case) using LiDAL, first the unwanted reflected signals from the environment obstacles must be eliminated through signal processing then detection and localization of the target follows using an optimum receiver in conjunction with an operating algorithm. Hence, the most important task in LiDAL is to distinguish the target reflected signal from the background obstacles reflections in a realistic indoor environment. We considered an active target located in a realistic environment (office room in Fig 1a). We define an 'active target' as a target that has the ability to be mobile, standing and sitting which are considered a unique signature that can be used to identify the target from the static obstacles in the realistic environment. In other words, the received reflected signal from the target is time-variant due to target activity while the background obstacles reflections are time-invariant (here we ignore for example the potential slow OW channel variations due to oscillations of indoor fans and the fast variations due to fan blades rotation for example). Thus, by monitoring multiple received signals for a duration of time, it is possible to eliminate the time-invariant signals and detect the changes in the signals reflected from the target movement.

In this paper, we considered and analysed two main approaches for target detection in a realistic environment. Firstly, a background subtraction method was developed to distinguish the target from background obstacles under the assumption that the realistic environment obstacles are static. Here, the target is detected by distinguishing the background reflections in multiple LiDAL measurements / scans. Secondly, a cross-correlation method is used to identify the changes in the LiDAL received signal scans in order to establish the target mobility. Furthermore, we have considered two types of target movement which describe pedestrian and nomadic targets. The target behaviour is modelled as; (*i*) a random walk using a model that avoids obstacles employing Markov chains. This may suit a small environment where a target may move randomly if the environment is mostly empty; (*ii*) a pathway model where the target chooses to walk on certain fixed paths due to the layout of the indoor environment.

### *A. Background Subtraction Method (BSM)*

The background subtraction method was investigated and implemented practically in [51], [52], [53] for UWB radar and camera surveillance systems. This method has poor performance only in cases where a target is moving (i.e. horizontal movement) and its signal reflections arrive at the same time during radar scans leading to ambiguity in single mobile target detection [52], [53], [54]. In LiDAL systems we introduce and make use of collaboration between monostatic and biostatic LiDAL configurations to eliminate the ambiguity in mobile target detection.

To develop the BSM concept in LiDAL we first considered a BSM example under two assumptions (which we remove later) (a) single mobile target with a single stationary background obstacle and zero reflections from the room's floor and walls; (b) there is no ambiguity between the target and the background obstacle (i.e. the target and the obstacle are separated by a minimum distance of $\Delta R$ or more). The received signal is $p_{r_i}(t)$ representing the $i$ th snapshot measurement taken during a time frame of duration $T$ in the presence of noise. The received signal is a superposition of the signals reflected from the target, background object and noise, thus $p_{r_i}(t)$ can be expressed as:

$$p_{r_i}(t) = \alpha_i m\left(t - t_{m_i}\right) + \beta_i b\left(t - t_{b_i}\right) + n_i(t) \quad (55)$$

where $m(t)$ is the reflected signal from the target, $b(t)$ is the reflected signal from the background obstacle, $n_i(t)$ is the noise during the $i^{th}$ snapshot, $\alpha$ and $\beta$ are the attenuation factors due to signal propagation and $t_{m_i}$, $t_{b_i}$ are the time delays for target and background signals respectively. It should be noted that $\left(t_{m_i} - t_{b_i}\right| \geq \tau\,)$ according to assumption (b). The BSM requires at least two snapshots to distinguish a pedestrian target and eliminate the background reflections. Thus, the received signal for the next snapshot $(i+1)$ is given as:

$$p_{r_{i+1}}(t) = \alpha_{i+1} m\left(t - t_{m_{i+1}}\right) + \beta_{i+1} b\left(t - t_{b_{i+1}}\right) + n_{i+1}(t). \quad (56)$$

The subtraction of equations (55) and (56) yields:

$$y_s(t) = \alpha_{i+1} m\left(t - t_{m_{i+1}}\right) - \alpha_i m\left(t - t_{m_i}\right) + \left(n_{i+1}(t) - n_i(t)\right) \quad (57)$$

where $t_{m_{i+1}} \neq t_{m_i}$ as the target is assumed to move while $t_{b_{i+1}} = t_{b_i}$ due to the stationary obstacle. Equation (57) results in perfect elimination of the reflected signal from the background obstacle only if $(\beta_{i+1} = \beta_i)$. However, part of the signal reflected from the target (due to multiple reflections) may contribute to the reflected signal from the obstacle. This is attributed to the presence of the target and its movement which may also block partially the signal reflected by the obstacle. This leads to $\beta_{i+1} \neq \beta_i \rightarrow \beta_{i+1} = \omega_i \beta_i$, where $\omega_i$ is the target impact factor on background reflections due to target presence and/or movement. Thus $y_s(t)$ is written as:

$$y_s(t) = \alpha_{i+1} m\left(t - t_{m_{i+1}}\right) + \alpha_i m\left(t - t_{m_i}\right) + \beta_i(\omega_i - 1)\, b\left(t - t_{b_i}\right) + \left(n_{i+1}(t) - n_i(t)\right). \quad (58)$$

The subtracted signal term $\beta_i(\omega_i - 1)\, b\left(t - \lambda_{b_i}\right)$ of equation (58) may be interpreted as a reflected signal from a target if $\beta_i(\omega_i - 1)\, b\left(t - t_{b_i}\right) \geq D_{th_L}$ and this can lead to false target distinguishing. Furthermore, the subtracted noise term $\left(n_{i+1}(t) - n_i(t)\right)$ has a variance $\sigma_{t_s}^2$ equal to $2\sigma_t^2$. Note that, the lower optimum detection threshold $D_{th_L}$ introduced in this work is based on two hypotheses $H_0$ only noise is present and $H_1$ noise and target are present. Thus, this leads to a new hypothesis which we have not included and will be considered in future work. It is however typically not an issue for the imaging receivers in Section VI due to their narrow FOV.

Fig. 13 shows an example of two snapshot measurements for a mobile target and a stationary obstacle. As can be seen in Fig. 13 the BSM of the snapshots may lead to false target distinguishing due to target movement which affects the signal reflected by the stationary obstacle. The simulation in Fig. 13 was carried out in a room (4m×8m×3m) in the presence of a single target and background obstacle located at ranges of 2m and 3m receptivity. A monastic LiDAL setup was used where the transmitter and receiver are located at the centre of the room's ceiling. Fig. 14 illustrates the proposed LiDAL receiver for target detection and distinguishing using BSM with the sub-optimum receiver.

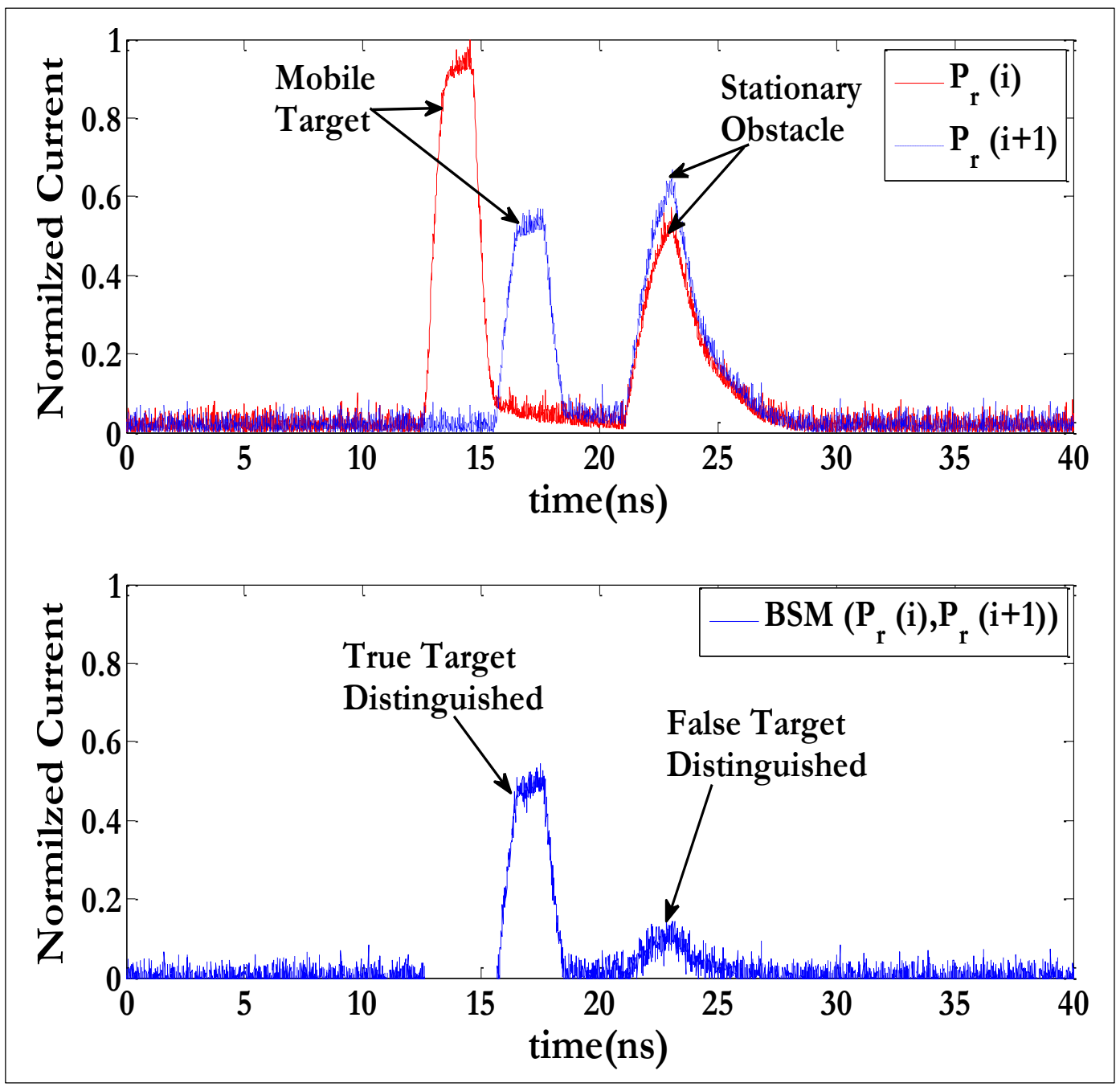

Fig.13: Results of BSM of the received snapshots measurements.

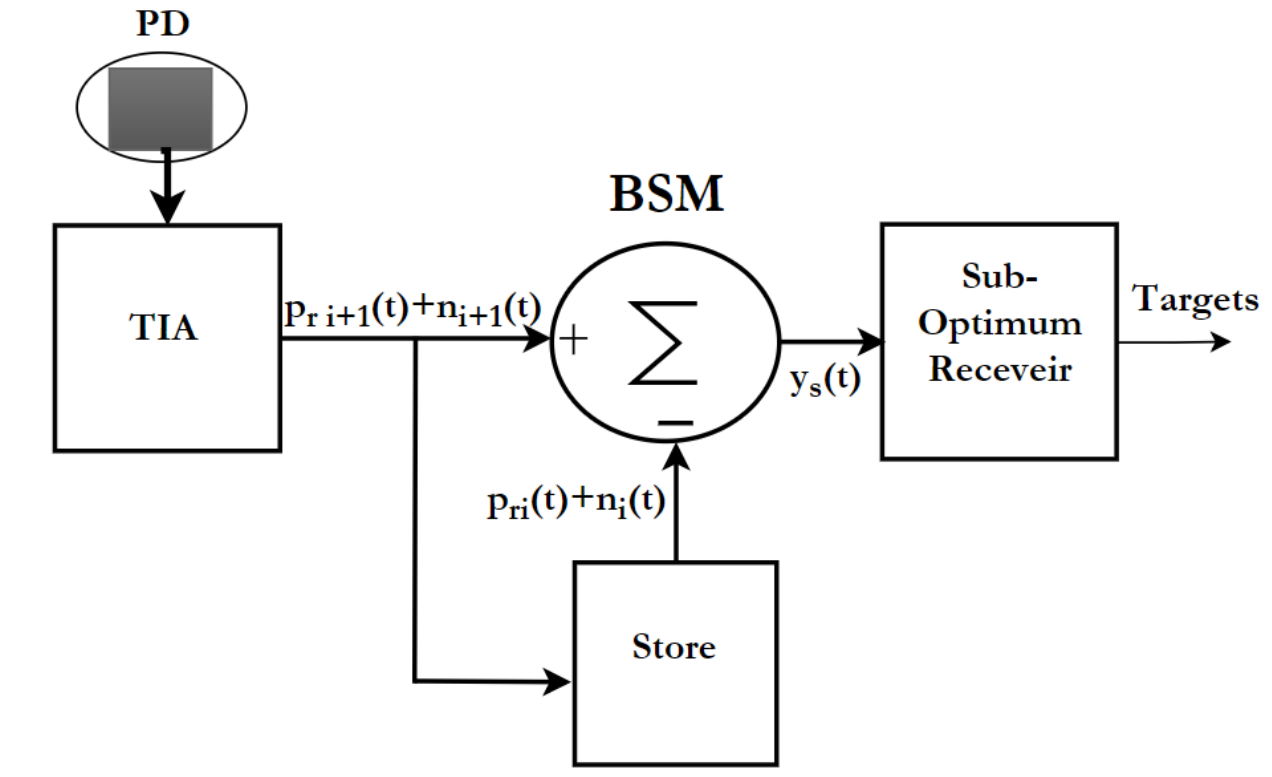

Fig.14: Receiver block diagram of LiDAL with BSM.

*B. Cross-Correlation Method (CCM)*

If there is target motion with continuous velocity (mobile) or discrete velocity (nomadic) in an indoor environment, the target can then be distinguished relative to the stationary background furniture by monitoring the changes in the received reflected signals through the use of multiple snapshots. We employed cross-correlation to identify the correlation between the snapshot measurements of the received reflected signals. Although there is relative motion between the target and $T_{Rx}$ unit, Doppler techniques cannot be used in LiDAL systems due to the limited target speed. Furthermore, cross-correlation is better than Doppler methods at low speeds, for example to estimate low velocity dispersion using ultrasound signals [55]. Also, cross-correlation has the advantage of detecting weak signals [56]. The peak displacement resulting from the cross-correlation between the two snapshots indicates target movement as the background obstacles are stationary and can also be used to determine target range.

In using cross-correlation we firstly look at coarse time scales to determine if there is a mobile target. We refer to this as fast cross-correlation. Here two snapshots are correlated over the full observation time window *T*. If target movement is detected, then a finer time scale cross-correlation is carried out at the slot level comparing two or more time slots, and carrying out each time a cross-correlation of up to *S* snap shots. We refer to this finer cross-cross-correlation as slow cross-correlation. We furthermore define a binary Target Movement Indicator (TMI) whose value is equal to one if the fast or the slow cross-correlations show a change, TMI is equal to zero otherwise. Fig.15 presents the proposed LiDAL snapshot measurements cube for target movement and shows the values of TMI. In Fig. 15, the *y* axis represents time and shows one time frame of duration *T* subdivided into *N* time slots. The *z* axis of Fig. 15 represents the TMI values associated with fast cross-correlation when two snapshots are cross-correlated. Finally, the *x* axis represents TMI values for each time slot when the slow cross-correlation is evaluated. Note that the values of S indicate the number of snapshots cross-correlated. As can be seen in Fig. 15, the first snapshot measurement ($i$=1) is stored until the next measurement ($i$=2) is collected. Then a cross-correlation between the two snapshots for the whole time duration $T$ is carried out to determine the TMI ('0' and '1') ie to determine the 'fast cross-correlation'. In this case, cross-correlating the ($i$=1) and ($i$=2) snapshots yields TMI=0. If TMI is equal to zero, the fast cross-correlation is continued, to carry out cross-correlation between the current snapshot ($i$=2) and the next snapshot ($i$=3). However, if TMI is equal to one, multiple cross-correlations are implemented between the identical time slots of the consecutive snapshots yielding the slow cross-correlation. The slow cross-correlation determines the TMI values associated with each time slots. The value of the TMI associated with slot *j* is referred to as a weight ($w_j$) which represents change / no change in each time slot. For example, *S*=4 represents cross-correlation between snapshots ($i$=1), ($i$=2), ($i$=3) and ($i$=2) and yields a TMI value for each time slot where the TMI values ($w_j$) are ($w_1$, $w_2$ and $w_3$=1; $w_4$, $w_5$ and $w_6$ =0). The values of the TMI weights are integrated in the proposed LiDAL sub-optimum receiver to detect and localize the targets as will be discussed in conjunction with Fig. 17.

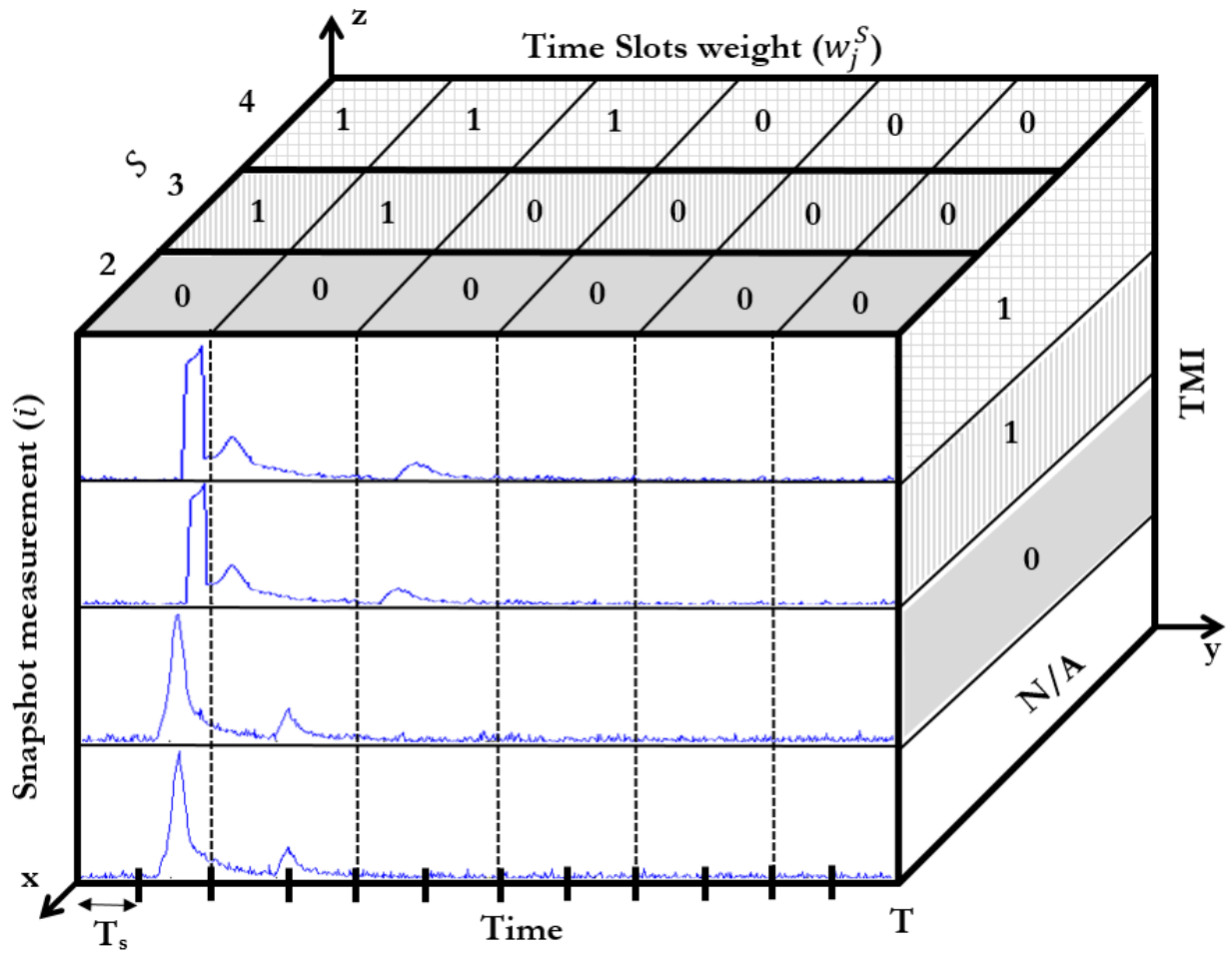

Fig.15: LiDAL snapshots measurement cube.

*i. Fast Cross-correlation*

To investigate the performance of the proposed cross-correlation method let us consider (*i*) a single mobile target with a stationary background obstacle, (*ii*) no ambiguity (i.e. the minimum distance between the mobile target and the background obstacle is $\Delta R$ or more) and (*iii*) white Gaussian noise due to the receiver and ambient noise as discussed in Section III. Here, we analyse the key scenarios of interest and in particular we consider five propositions / scenarios to test the fast cross-correlation method to decide the TMI.

**Proposition I**: we assume that there is no target in the environment, only (background) an obstacle in the two snapshot measurements $(i, i+1)$ as can be seen in Fig. 16a. The received signal reflected from the obstacle in the presence of noise in the $i^{th}$ snapshot, $p_{r_i}(t)$, can be expressed as:

$$p_{r_i}(t) = \beta_i b\left(t - t_{b_i}\right) + n_i(t) \tag{59}$$

and the received signal $p_{r_{i+1}}(t)$ is given as:

$$p_{r_{i+1}}(t) = \beta_{i+1} b\left(t - t_{b_{i+1}}\right) + n_{i+1}(t). \tag{60}$$

The fast cross-correlation function ($\mathcal{R}_{p_{r_i}, p_{r_{i+1}}}$) of equations (59) and (60) over the listening time $T$ is:

$$\mathcal{R}_{p_{r_i}, p_{r_{i+1}}}(\tau) = \mathcal{R}_{bb}(\tau) + \mathcal{R}_{bn}(\tau) + \mathcal{R}_{nn}(\tau) \tag{61}$$

where the term $\mathcal{R}_{bb}$ is an auto-correlation function of the received signal from the obstacle which is defined as:

$$\mathcal{R}_{bb}(\tau) \triangleq \int_{-T}^{T} \beta_i b\left(t - t_{b_i}\right)\, \beta_{i+1} b\left(t - t_{b_{i+1}} + \tau\right) dt \tag{62}$$

and $\mathcal{R}_{sn}$ is the cross-correlation of the received signal (from the obstacle) with noise; and $\mathcal{R}_{nn}$ is the noise auto-correlation. These two correlations are given as:

$$\mathcal{R}_{bn}(\tau) \triangleq \int_{-T}^{T} \beta_i b\left(t - t_{b_i}\right) n_{i+1}(t+\tau)\, dt + \int_{-T}^{T} \beta_{i+1} b\left(t - t_{b_{i+1}}\right) n_i(t+\tau)\, dt \tag{63}$$

and

$$\mathcal{R}_{nn}(\tau) \triangleq \int_{-T}^{T} n_i(t)\, n_{i+1}(t+\tau)\, dt. \quad (64)$$

The correlation factor $\hat{t}$ (i.e. displacement factor which represents the time delay) can be calculated by determining $\tau = \hat{t}$ for which $\mathcal{R}_{bb}$ is maximized. Therefore, $\hat{t}_{bb}$ is defined as:

$$\hat{t}_{bb} = \arg\max_{\tau}\big(\mathcal{R}_{bb}(\tau)\big) \quad (65)$$

It should be noted that the noises in the snapshot measurements are assumed uncorrelated and orthogonal, thus $\mathcal{R}_{nn} \approx 0$ [57] [58]. Also, the value of $\mathcal{R}_{bn}$ can be assumed very small and can thus be neglected [57], [58]. Hence, $\mathcal{R}_{bb}(\hat{t}_{bb})$ identifies whether there is a change or not between the snapshot measurements. For proposition I, the obstacle is stationary ( $t_{b_i} = t_{b_{i+1}}\ \forall i$ ). Therefore $\hat{t}_{bb} = 0$ , see Fig. 16b, indicates that no change took place in the "target" location (TMI=0). Note that the received signal is sampled with $T_s$=0.01ns as discussed in Section III-C, which yields the *x* axis scale of Fig. 16b.

**Proposition II**: We assumed that the target appears in the environment in the $i^{th}+1$ snapshot measurement while the $i^{th}$ snapshot includes only the stationary obstacle as depicted in Fig.16c. The received signal reflected from the target and the obstacle in noise, $p_{r_{i+1}}(t)$, is given as:

$$p_{r_{i+1}}(t) = \alpha_{i+1} m\big(t - t_{m_{i+1}}\big) + \beta_{i+1} b\left(t - t_{b_{i+1}}\right) + n_{i+1}(t) \quad (66)$$

while $p_{r_i}(t)$ is as given in (59). Thus using (59) and (66) $\mathcal{R}_{p_{r_i},p_{r_i+1}}(\tau)$ is given by:

$$\mathcal{R}_{p_{r_i},p_{r_i+1}}(\tau) = \mathcal{R}_{bm}(\tau) + \mathcal{R}_{bb}(\tau) + \mathcal{R}_{bn}(\tau) + \mathcal{R}_{mn}(\tau) + \mathcal{R}_{nn}(\tau) \quad (67)$$

where $\mathcal{R}_{bm}(\tau)$ is the cross-correlation function between the signal received from the target and that received from the obstacle, while $\mathcal{R}_{mn}(\tau)$ is the cross-correlation between the signal reflected from the target and noise. Thus $\mathcal{R}_{bm}(\tau)$ is written as:

$$\mathcal{R}_{bm}(\tau) \triangleq \int_{-T}^{T} \beta_i b\left(t - t_{b_i}\right)\ \alpha_{i+1} m\big(t - t_{m_{i+1}} + \tau\big)\, dt \quad (68)$$

and the $\mathcal{R}_{mn}(\tau)$ is:

$$\mathcal{R}_{mn}(\tau) \triangleq \int_{-T}^{T} \alpha_{i+1} m\big(t - t_{m_{i+1}}\big)\, n_i(t+\tau)\, dt. \quad (69)$$

It should be noted that, $\mathcal{R}_{mn}(\tau)$ can be neglected in a similar fashion to the decision to neglect $\mathcal{R}_{bn}$. The peak in the target-obstacle cross-correlation occurs at $\hat{t}_{bm}$ which can be calculated as $\hat{t}_{bm} = \arg\max_{\tau}\big(\mathcal{R}_{bm}(\tau)\big)$ . For proposition II, $\hat{t}_{bm} \neq 0$ and $\hat{t}_{bb} = 0$. Thus, $\hat{t}_{bm}$ indicates the change that occurred due to the target presence (TMI=1) as can be seen in Fig. 16d.

**Proposition III**: Here we assumed the presence of a mobile target in two successive snapshot measurements with a stationary obstacle as shown in Fig16e. The received reflected signals are:

$$p_{r_i}(t) = \alpha_i m\big(t - t_{m_i}\big) + \beta_i b\left(t - t_{b_i}\right) + n_i(t) \quad (70)$$

and:

$$p_{r_{i+1}}(t) = \alpha_{i+1} m\big(t - t_{m_{i+1}}\big) + \beta_{i+1} b\left(t - t_{b_{i+1}}\right) + n_{i+1}(t) \quad (71)$$

The cross-correlation ($\mathcal{R}_{p_{r_i},p_{r_{i+1}}}$) of equations (70) and (71), gives:

$$\mathcal{R}_{p_{r_i},p_{r_{i+1}}}(\tau) = \mathcal{R}_{mm}(\tau) + \mathcal{R}_{mb}(\tau) + \mathcal{R}_{bm}(\tau) + \mathcal{R}_{bb}(\tau) + \mathcal{R}_{bn}(\tau) + \acute{\mathcal{R}}_{mn}(\tau) + \mathcal{R}_{nn}(\tau) \quad (72)$$

where $\mathcal{R}_{mm}$ is the auto-correlation function of the received reflected signal from the target given as:

$$\mathcal{R}_{mm}(\tau) \triangleq \int_{-T}^{T} \alpha_i m\big(t - t_{m_i}\big)\ \alpha_{i+1} m\big(t - t_{m_{i+1}} + \tau\big)\, dt \quad (73)$$

The cross-correlation $\mathcal{R}_{mb}(\tau)$ is given as:

$$\mathcal{R}_{mb}(\tau) \triangleq \int_{-T}^{T} \alpha_i m\big(t - t_{m_i}\big)\ \beta_{i+1} b\left(t - t_{b_{i+1}} + \tau\right) dt. \quad (74)$$

The cross-correlation $\acute{\mathcal{R}}_{mn}(\tau)$ is given by:

$$\acute{\mathcal{R}}_{mn}(\tau) \triangleq \int_{-T}^{T} \alpha_{i+1} m\big(t - t_{m_{i+1}}\big)\, n_i(t+\tau)\, dt + \int_{-T}^{T} \alpha_i m\big(t - t_{m_i}\big)\ n_{i+1}(t)\ (t+\tau). \quad (75)$$

The time $\hat{t}_{mm}$ is defined as $\hat{t}_{mm} = \arg\max_{\tau}\big(\mathcal{R}_{mm}(\tau)\big)$ while $\hat{t}_{mb} = \arg\max_{\tau}\big(\mathcal{R}_{mb}(\tau)\big)$. In proposition III, we are interested in observing the values of $\hat{t}_{mm}$, $\hat{t}_{mb}$, $\hat{t}_{bm}$ and $\hat{t}_{bb}$, as seen in Fig. 16f, to determine whether a change has occurred or not between the snapshot measurements.

**Proposition IV**: In this proposition, we assume that the target and the obstacle are stationary during the snapshot measurements as presented in Fig. 16g. Here the cross-correlations will have the same definitions as in proposition III, however, $t_{m_i} = t_{m_{i+1}}$ (ie a stationary target). Therefore, as can be seen in Fig. 16h $\hat{t}_{mb}$=$\hat{t}_{bm}$ and the corresponding (side) peaks have the same magnitude.

**Proposition V**: We assumed in this case that in the $i^{th}$ snapshot the target and the obstacle are present, while in the $i^{th}+1$ snapshot, only the obstacle is present (i.e. the target left the environment). This is similar to Proposition II, the case shown in Fig. 16 c, but with the $p_{r_i}(t)$ and $p_{r_{i+1}}(t)$ exchanging their roles. Here $\hat{t}_{mb} \neq 0$ and $\hat{t}_{bb} = 0$. The cross-correlation will be similar to that shown in Fig. 16 d.

Table VIII summarizes the fast correlation outcomes and the value of TMI associated with two consecutive snapshot measurements in LiDAL.

TABLE VIII: TARGET MOVEMENT INDICTOR DECISION

| Proposition | Arrival times $t_i, t_{i+1}$ | Correlation factor $\hat{t}$ | Decision TMI |
|---|---|---|---|
| I | $t_{b_i}$=$t_{b_{i+1}}$ | $\hat{t}_{bb}$ =0 | 0 |
| II | $t_{b_i}$=$t_{b_{i+1}}$, $t_{m_{i+1}} \neq t_{b_i}$ | $\hat{t}_{bb}$ =0, $\hat{t}_{bm} \neq$0 | 1 |
| III | $t_{b_i}$=$t_{b_{i+1}}$, $t_{m_i} \neq t_{m_{i+1}}$ | $\hat{t}_{bb}$ =0, $\hat{t}_{mm} \neq$0, $\lvert\hat{t}_{mb}\rvert \neq \lvert\hat{t}_{bm}\rvert$ | 1 |
| IV | $t_{b_i}$=$t_{b_{i+1}}$,$t_{m_i} = t_{m_{i+1}}$ | $\hat{t}_{bb}$ =0, $\hat{t}_{mm}$ =0, $\lvert\hat{t}_{mb}\rvert = \lvert\hat{t}_{bm}\rvert$ | 0 |
| V | $t_{b_i}$=$t_{b_{i+1}}$, $\lvert t_{m_i}\rvert \neq \lvert t_{b_{i+1}}\rvert$ | $\hat{t}_{bb}$ =0, $\hat{t}_{mb} \neq$0 | 1 |

*ii. Slow Cross-correlation*

The slow cross-correlation is employed over the duration of the time slot $T_{s_j}$ where the same time slot in the frame is considered over several (S) consecutive snapshots measurements. The cross-correlation $\mathcal{R}_{xy}\left(\tau,\ T_{s_j}\right)$ can be given as:

$$\mathcal{R}_{p_{r_i},p_{r_{i+1}}}\left(\tau, T_{s_j}\right) \triangleq \int_{-T_s}^{T_s} p_{r_{i+1}}\left(t - T_{s_j} + \tau\right) \sum_{i=1}^{S} p_{r_i}\left(t - T_{s_j}\right) dt \quad (76)$$

where, $i \in [1, .. S]$ and $S$ is the total number of snapshots.

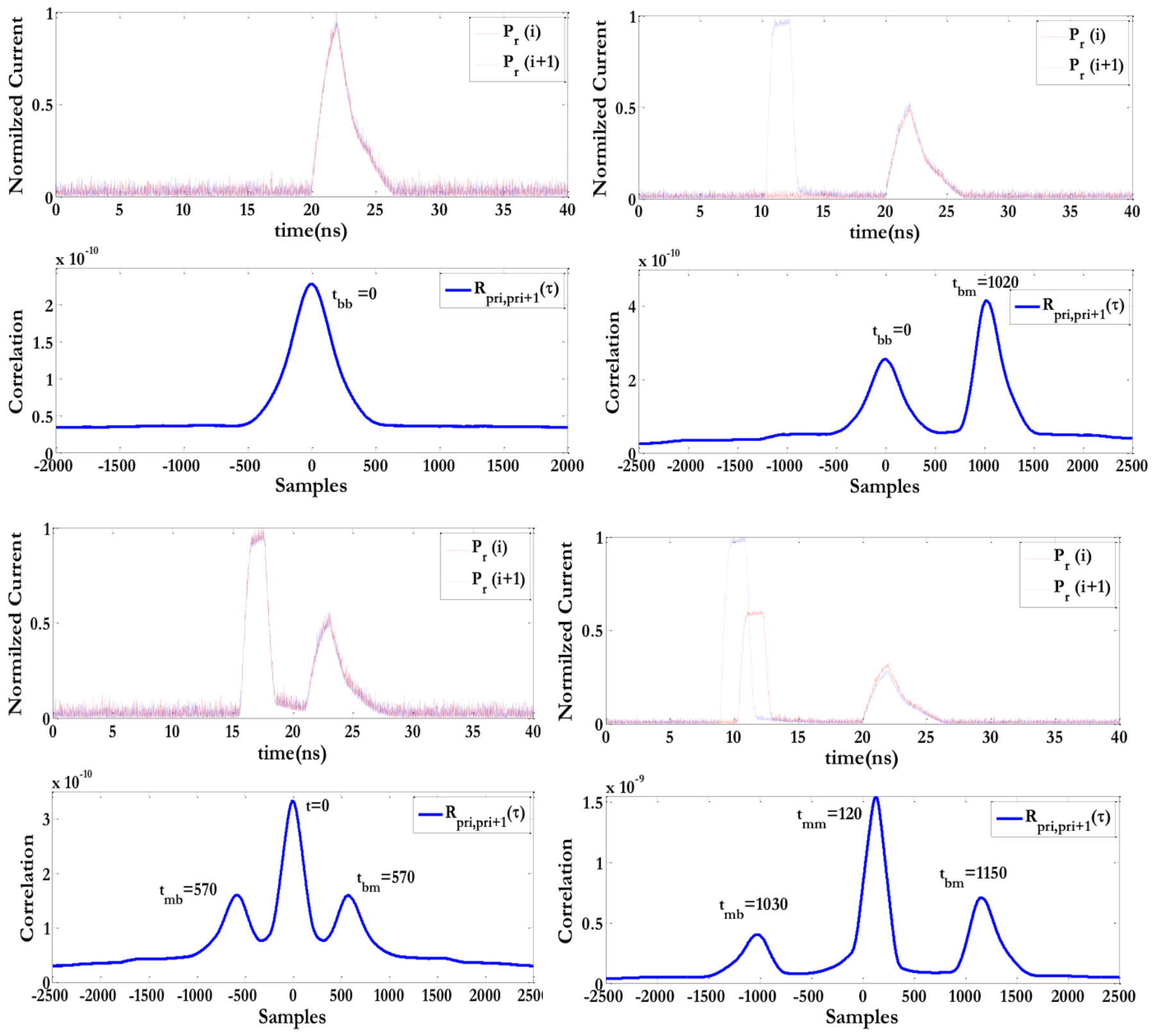

Fig.16: Cross-correlation method (a) CCM of received snapshots measurement of Proposition I, (b) CCM of received snapshots measurement of Proposition II, (c) CCM received snapshots measurements of Proposition III and (d) CCM received snapshots measurement of Proposition IV.

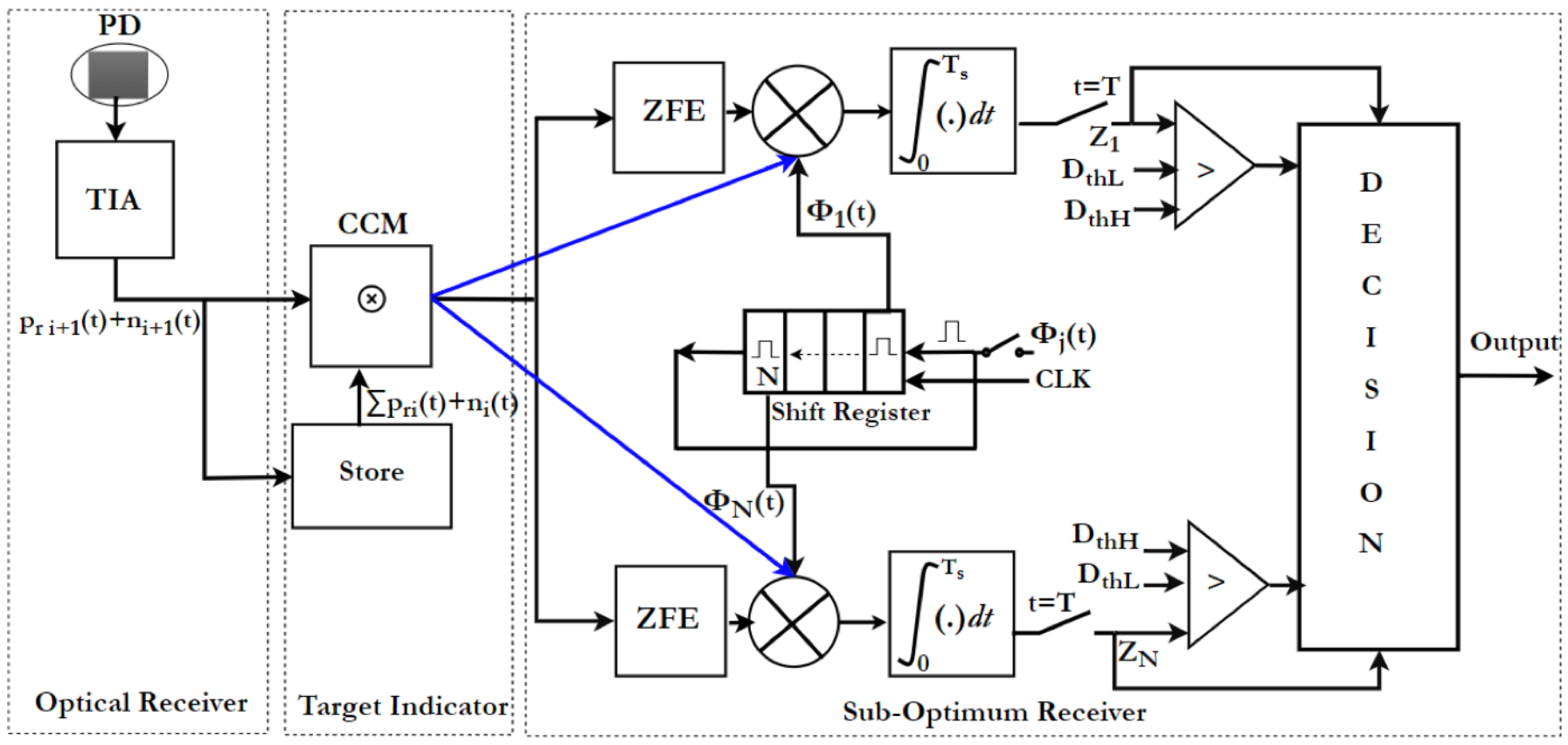

Fig.17. LiDAL receiver block diagram with CCM.

The time slot correlation factor $\hat{t}_{T_{s_j}}$ is calculated as:

$$\hat{t}^{\,S}_{T_{s_j}} = \arg\max_{\tau}\left(\mathcal{R}_{p_{r_i},p_{r_i+1}}\left(T_{s_j},\tau\right)\right) \tag{77}$$

It should be noted that, if the value of $\hat{t}^{\,S}_{T_{s_j}}$ changes for different values of $S$ (i.e. when, more snapshots measurements are considered), then this indicates the presence of the target in a time slot $T_{s_j}$. Thus, when $\hat{t}^{\,S}_{T_{s_j}}$ equals to zero, this indicates no change between the received reflected pulses in $T_{s_j}$ in $S$ consecutive snapshots.

We define a time slot weight $w_j^S$ in CCM to be used in the operation of the LiDAL sub-optimum receiver. The weight $w_j^S$ is defined as:

$$w_j^S = \begin{cases} 0 \ if \ \hat{t}^{\,S}_{T_{s_j}} = 0 \\ 1 \ \ otherwise \end{cases} \tag{78}$$

Equation (78) can be understood by observing that $\hat{t}^{\,S}_{T_{s_j}}$ is the time slot at which the peak of the correlation occurs. If there is no target and hence no motion, then the correlation (76) is an auto-correlation whose peak occurs at $\hat{t}^{\,S}_{T_{s_j}}$ =0 and therefore, the $w_j^S$ is equal to zero in this case indicating the absence of the target.

The time of arrival (TOA) of the received reflected pulse from the target in $T_{s_j}$ can be determined as:

$$TOA_j = \arg\max_{\tau}\left(\int_{-T_s}^{T_s} w_j^S \, p_{ri}\left(t - T_{s_j} + \tau\right) x(t) \ dt \right) \tag{79}$$

Equation (79) can only have a meaningful use if the receiver time slot of interest is large and the received pulse is much narrower than the time slot. In which case equations (76) and (78) identify the time slot in which the reflected pulse from the target occurs (i.e. time slot number); while equation (79) can identity the target pulse location within a time slot.

Fig. 17 presents the LiDAL optimum receiver for target distinguishing and detection using CCM. As can be noted in Fig.17, The output of CCM is represent by time slot weights $w_j^S$ which are multiplied by the orthonormal expansion coefficient, $\Phi_j(t)$, of each time slot. The target indicator block has to be allowed to operate and accumulate $S$ snapshots (see second term of equation (76)) and hence produce $w_j^S$ values for the $j^{th}$ slot and for $N$ time slots before the sub-optimum receiver starts operating. This is only an initialization phase. Furthermore, the $w_j^S$ weights cause the $j^{th}$ slot to produce zero energy in the SOR if there is no target motion, hence stopping the SOR from reporting the reflected pulse from an obstacle as a target.

## C. *Target Mobility Modelling*

Target distinguishing relies on the target movement in the indoor environment in conjunction with the use of the BSM and CCM distinguishing approaches in our study. Target movement leads to a change in the observed signals received by LiDAL. Therefore, modelling the target mobility behaviour is essential to examine the performance of the proposed LiDAL systems. A random walk approach that avoids obstacles is considered for pedestrian and nomadic targets in the realistic indoor environment. For pedestrian targets, we assumed continuous movement at a speed of 1m/s, while for the nomadic targets, discrete movement is assumed.

Three distinct additional studies can be conducted in this area. We address two of these and leave the third for future work. Firstly, mobility helps distinguish targets, however not all locations may be allowed in the room or indoor environment, due to obstacles and furniture. To account for this, we define a space utilization factor (SUF) that effectively reflects the reduction in the allowed target mobility. Secondly, some spaces may be more popular than others, for example a working desk surface in a room. We account for this in simulations by using different transition probabilities from location to another. This is also used to reflect possible target nomadic behaviour. Thirdly, the probability of correct decisions at the output of the receiver, such as that given in (52) can be combined with the probability of detecting target movement as derived in this section to give a combined performance analysis of the receiver and the human mobility pattern and indoor space configuration. In this third study target motion through a number of steps in a given time window (for example a one second time window) provides more repeated opportunities for the receiver to detect the moving target. This can be analysed within the framework of repetition coding. This third study area warrants further research and is not reported here. We consider the first two studies in this section.

### *i. Probability of mobility detection* ($P_{MD}$)

The BSM and CCM employ snapshot measurements to distinguish the target. This relies on the target's motion where a minimum step distance of $\Delta R$ (LiDAL resolution) is assumed. The calculation of the probability of detection is related to: (*i*) the probability that the target moves from location ($L_1$) to location ($L_2$) and (*ii*) the number of target steps required to achieve a $\Delta R$ distance. In order to determine the probability of detection, the following setting was considered:

1) A Markov chain is considered as a representation of the random walk process on a graph. This models the target mobility behaviour in the indoor environment in two dimensions. Markov chain models allow the target walking behaviour to be represented either on directed or on undirected paths [59], [60]. The presence of obstacles was accounted for through the elimination of certain transitions in the Markov chain.
2) The indoor environment floor of the interest $G\ (x,y)$ is divided into a 2D grid with size $j \times i$ and $L$ locations where $L = \frac{x \times y}{\Delta_l{}^2}$, $i = \frac{x}{\Delta_l}$ and $j = \frac{y}{\Delta_l}$; here $\Delta_l$ is the inter-locations distance as shown in Fig. 18.
3) The target can move in space to one of $N_D$ neighbour destinations ($N_D \in [l_1..,l_{N_D}]$) or can stay at the current location ($l_c$) as shown in Fig. 18. The Markov chain considered is a stochastic process on states defined in terms of a transition matrix ($P$) ($N_D+1$ rows and $N_D+1$ columns). The transition matrix of the graph in Fig.18 is given as:

$$P = \begin{bmatrix} p_{s(1)} & p_{m(1,2)} & p_{m(1,N_D)} \\ p_{m(2,1)} & p_{s(2)} & p_{m(2,N_D)} \\ \vdots & \ddots & \vdots \\ p_{m(L,1)} & \cdots & p_{s(L)} \end{bmatrix} \tag{80}$$

where $p_{s(i)}$ is the probability of the target staying in the current state (location *i*) which is related to the target's behaviour, $p_{m(i,j)}$ is the probability of the target moving from current location (*i*) to one of the neighbour locations (*j*).
4) We have considered an undirected target motion pattern. Typically, the target walks to one of $L$ random locations inside an indoor environment where at each location the probability of the target staying at the current location (*i*), $p_{s(i)}$, can be written as:

$$p_{s(i)} = 1 - \sum_{j=1}^{N_D} p_{m(i,j)} \qquad j \neq i \tag{81}$$

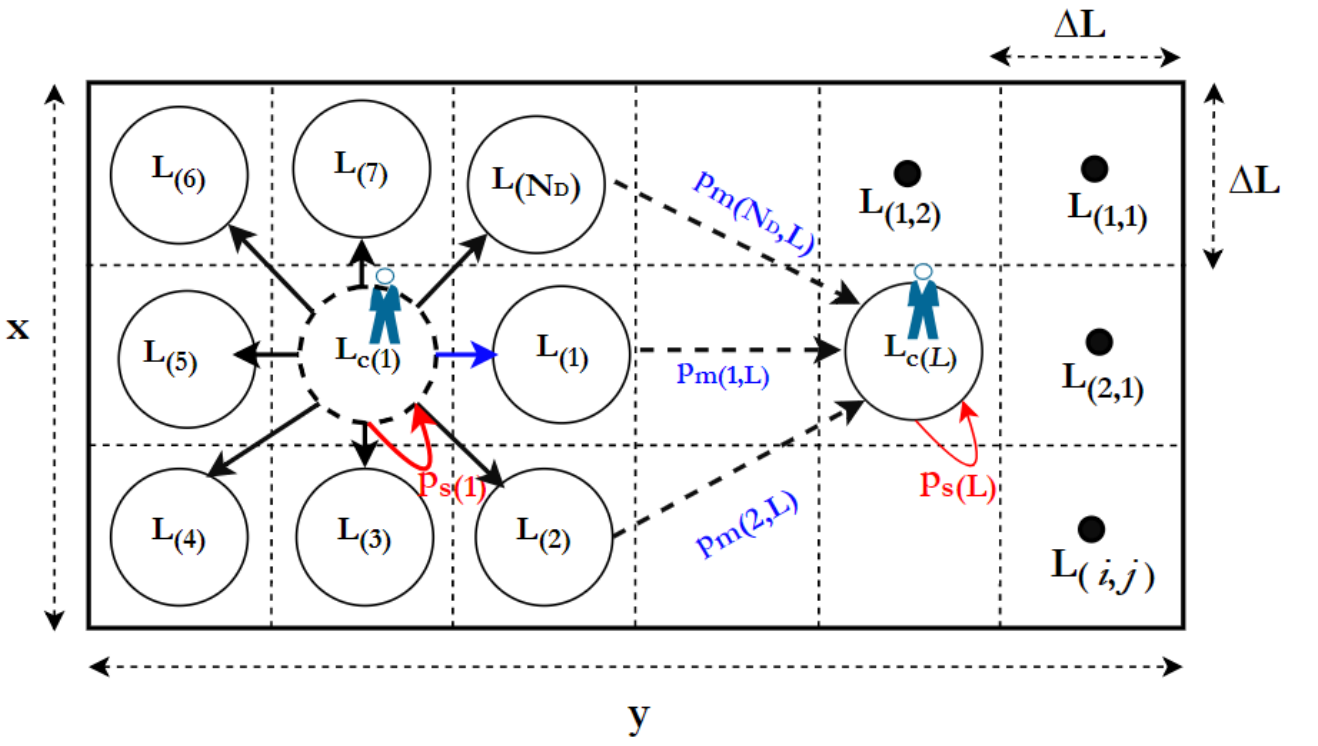

Fig.18. Target random walk model in $G(x,y)$ space.

To simplify the setup of modelling the indoor environment, let the inter-locations distance equal to the LiDAL resolution (i.e. $\Delta_l = \Delta R$). Thus the grid size considered is $\left(\frac{x}{\Delta R} \times \frac{y}{\Delta R}\right)$. This is reasonable as $\Delta R$ is typically about 30 cm where we set this design parameter for LiDAL resolution and it is the minimum typical expected distance between people in an indoor environment. Also, we will assume that pedestrians move at a speed that is an integer multiple of $\Delta R$ m/s to simplify the analysis. Therefore, if the pedestrian speed is $v$ m/s, then in one second the pedestrian visits $\frac{v}{\Delta R}$ locations at most. At each location the target can be distinguished since it has moved at least $\Delta R$ which is a change that can be captured in the snapshot measurements. Therefore, the probability of target movement detection taking into account the target speed in an empty indoor environment, $G_E(x, y)$, $P^E_{MD_T}$ can be expressed as:

$$P^E_{MD} = \left(\frac{\Delta R^2}{xy}\right)\sum_{j=1}^{L}\sum_{i=1}^{N_D} p_{(i,j)}\,. \tag{82}$$

Equation (83) describes the probability of target movement in an empty environment that has $L$ possible locations with $N_D$ neighbour destination to each current location. The probability $p_{(i,j)}$ depends on target activity behaviour (nomadic, continuous motion etc). It is worth mentioning that, we assumed for all $L$ possible locations an equal probability of being in that location, given by $\left(\frac{\Delta R^2}{xy} = \frac{1}{L}\right)$.

For a realistic indoor environment $G_R(x, y)$, free flow in the space is hindered by obstacles (i.e. furniture and walls) where the target movement is restricted and mobility detection can be harder. Therefore, we introduce a 'space utilization factor' in realistic environments to determine the target probability of detection. The space utilization factor $SUF$ can be written as:

$$SUF = 1 - \left(\frac{\Delta R^2}{xy}\right)\left(\frac{1}{N_D}\right)\sum_{j=1}^{L}\left(N_D - \frac{1}{p_{(j)}}\right) \tag{83}$$

where, $p_{(j)}$ is a property of the current location $j$ and is given as $p_{(j)} = \frac{1}{N_A}$. Note that $N_A$ is the number of neighbour locations of location $j$ allowed for the target to move to; with $N_A \leq N_D$. The space utilization factor, $SUF$, has a unity value for a room that has no obstacles ($N_A = N_D$), while for a room with obstacles ($N_A < N_D$), $SUF$ is less than one. The probability of target mobility detection in a realistic environment $P^R_{MD_T}$ can be given as:

$$P^R_{MD_T} = SUF\left[P^E_{MD}\right] \tag{84}$$

Fig.19 presents the probability of target mobility detection in a realistic environment for different values of LiDAL resolutions and space utilized by background obstacles (furniture). The results are obtained for a pedestrian target walking randomly with a speed of $v$=1 m/s in space of $G_E$(4m, 8m). The Markov transition matrix for the pedestrian behaviour selected has $p_s$=0.02, $\sum_i^{N_D} p_m$ =0.98 and $N_D$=8. As can be seen in Fig.19, the space utilization $SUF$ significantly affects the $P^R_{MD_T}$ due to variation in the space allowed for the target to be mobile.

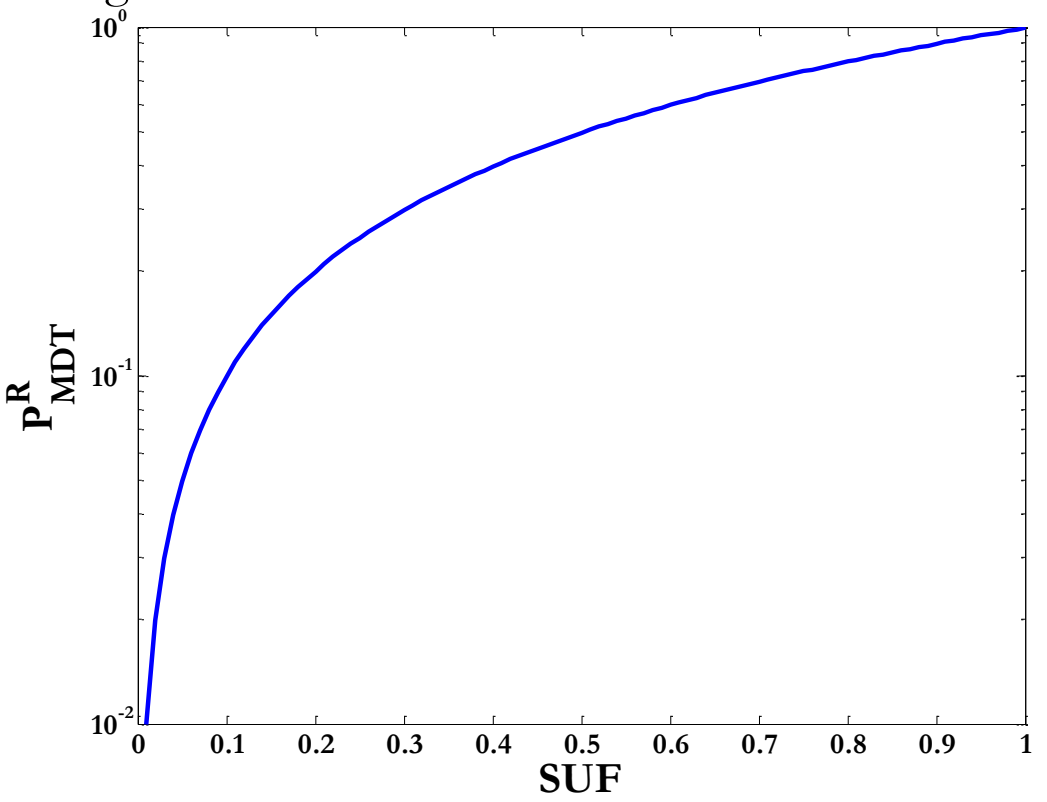

Fig.19. Probability of target mobility detection in a realistic environment.

*ii. Directed random walk with obstacle avoidance*

In this model, we assume that the pedestrian and nomadic targets walk freely inside the room in all the directions except directions that lead to obstacles. In addition, we considered a common scenario where the targets arrival into the room follows a Poisson distribution [61] [62] and the time spent in the room follows a negative exponential distribution [63] [64].

Initially, targets reach the room's entrance at different arrival times $t_a$ with an arrival rate of $\lambda$ and mean time spent in the environment of $1/\gamma$ and therefore in a simulation, the leaving times $t_d$ can be determined. Targets spend times $t_{sp}$ in the environment.

For nomadic behaviour in an indoor environment such as an office room, the nomadic target continuously walks inside the room until it reaches one of the interesting destinations (for instance an office desk). For each nomadic target, $L_D$ interesting destinations are generated randomly where $L_D \in [1, .. L]$. It is assumed that the nomadic target has a speed of (0.5-2 m/s). A Markov transition matrix is then created for the current location to describe the probability of transition to its neighbours. We considered $N_D$=8 neighbours that are equi-probable if no obstacle is present. In the presence of obstacles, some of the $N_D$ directions have zero transition probabilities, while the rest are equi-probable. The decision of staying in the current location or moving to the next destination relies on the allocated probabilities in the transition matrix. Let us assume that the nomadic target has the same behaviour in terms of staying at the interesting destinations (i.e. the staying probability is equal among the locations of interest $L_D$). Thus the probability of staying at a location $l_D$ of interest ($l_D \in [1, .. L_D]$) for a nomadic target is $p^{no}_{s(l_D)} = \frac{1}{L_D}$ and the probability of moving is $p^{no}_{m(l_D,j)} = \frac{1-p^{no}_{s(l_D)}}{N_D}$. For the locations other than the $L_D$ locations of high interest, ie for the $l(i)$ general locations where $(i \neq l_D)$, the nomadic target moves with a speed $v$, thus the probability of staying at $l(i)$ should be very small due to lack of interest. We thus set the $p^{no}_{s(i)} = 0$ and $p^{no}_{m(i)} = \frac{1-p^{no}_{s(i)}}{N_D}$.

During the simulation the nomadic target follows the path with the highest probabilities. Note that, the neighbour destinations to the location of interest have equal probabilities, therefore, the next neighbour destination is decided on equi-probable basis.

The simulation starts with the arrival of targets into the environment following a Poisson distribution and proceeds by determining the time each target spends in the environment where this time follows a negative exponential distribution. The motion of the targets with the environment is then governed by the transition matrix probabilities.

Let the arrival rate be $\lambda$ per hour and let the average dwell time be $1/\gamma$ in hours. Let $T_{ob}$ be the observation window, i.e the simulation time. The probability of having $k$ arrivals in $T_{ob}$ is given as:

$$p_a(k) = \frac{(\lambda T_{ob})^k}{k!} e^{-\lambda T_{ob}}. \tag{85}$$

The probability of a target leaving after $t_d$ is:

$$p_d(t_d) = \gamma e^{-\gamma t_d} \qquad t_d < T_{ob} \tag{86}$$

The room is thus considered a form of $M/M/1$ queue and therefore the maximum number of targets $K$, given $\lambda$ and $\gamma$, can be written as:

$$K = \frac{\lambda/\gamma}{1 - \lambda/\gamma}. \tag{87}$$

Note that, the European standards for the minimum workplace space required per person is 3.7m² for an office environment and 2m² for a meeting room [65]. Thus, in this work we set $K = 6$ for the office room presented in Fig.1a (with an area of 8m × 4m) where the space left unoccupied by obstacles is 24 m². We have used $\lambda = 12$ arrivals per hour and $\gamma = 14$, giving the average time spent in the room as $\frac{K}{\lambda}$ =30 minutes.

*iii. Pathway mobility model*

In this model, the targets move on pre-determined indoor pathways as shown in Fig. 20. Note that, the targets' behaviour in terms of arrival rate, departure rate and number of destination of interest are similar to the setup discussed in the 'random walk with obstacle avoidance' model. However, in this model, there is no random target motion, the targets follow the pre-determined paths.

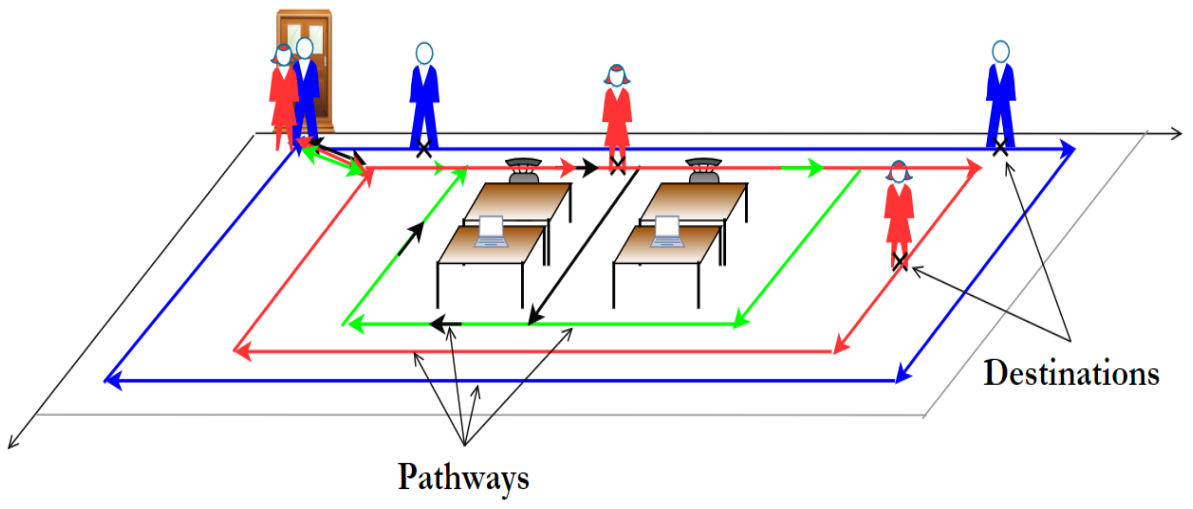

Fig. 20: Pathways Mobility Model.

## V. MULTIPLE-INPUT MULTIPLE-OUTPUT LIDAL (MIMO-LIDAL)

Target ambiguity is the main challenge when using monostatic or bistatic LiDAL systems in an indoor environment. Whenever, the distance between targets or between a target and a background obstacle is less than the LiDAL resolution, target detection ambiguity occurs. Increasing the LiDAL resolution by decreasing the transmitted pulse width improves the target detection resolution, however this requires a higher transmitter and receiver bandwidth and calls for a more complex optical receiver (for example in terms of equalization). Target localization requires determination of the target range and/or the direction (angle) of the received reflected pulse from the target. Unlike the work reported in the literature, our localization approaches in this paper are passive in the sense that the target does not have to carry an optical (VLC) receiver. In the literature [4], [66], [67], many techniques have been proposed for VLC mobile user localization such as triangulation, scene analysis and proximity using angle of arrival, time difference of arrival and received signal strength form multiple transmitters. Our passive approach in LiDAL relies on detecting signals reflected from the target, and therefore received signal strength indicator (RSSI) is not a good detection strategy as discussed in Section III. In this passive localization approach the reflected signals experience heavy fluctuations when reflected signal from the target owing to the environment, target cloth colours and the potential loss of the line of sight component.

In this section, a new multiple-input multiple-output LiDAL system (MIMO-LiDAL) is introduced for target detection, counting and localization. The proposed system is designed to mitigate the ambiguity of multi-target detection to distinguish the targets correctly from the background obstacles in a realistic indoor environment. To tackle the ambiguity of target detection, a collaboration of multiple transmitters and receivers is employed. The detection floor is divided into multiple optical footprints using multiple single-photodetector receivers which provide spatial selection for target detection, see Fig. 21. In addition, we integrated the MIMO-LiDAL system with the proposed target distinguishing approaches of Section IV and the sub-optimum receiver of Section III to optimise the targets detection, counting and localization supported by an algorithm (Section V-C) executed in a connected controller. Furthermore, time-of-arrival (TOA) is employed in the MIMO-LiDAL system for target localization. A simulation is reported in Section VII for the MIMO-LiDAL system in order to identify the accuracy of detecting, counting and localizing multiple targets in a realistic environment.

### *A. System Configurations*

We introduce the MIMO-LiDAL system, to detect, count and localize multiple targets. We implemented multiple narrow-FOV receivers collocated with the light units. The system is designed to tackle the ambiguity of target detection, maximize the number of counted targets and minimize false target distinguishing by employing both monostatic and bistatic LiDAL systems. Fig. 21 presents the setup of the MIMO LiDAL system with the controller. The MIMO-LiDAL system includes eight receivers that are collocated with the eight VLC transmitter units on the room ceiling. The room setup and transmitters' configuration is similar to that in [19] which is a versatile setup used to realise a multi-gigabit/s VLC system. In this work, we assumed that the LiDAL system has access to and can use all the VLC transmitters. The room detection floor is divided into eight optical footprints as shown in Fig. 21. The transmitters and their FOV have to be selected to comply with the illumination levels recommended by the standards [19]. Therefore, we have created the LiDAL optical detection zones through design and selection of the LiDAL receivers FOVs. Each receiver is chosen as a single narrow-FOV photodetector with $\Psi_c$=43.8° which is the acceptance semi-angle of the compound parabolic concentrator (CPC). This FOV is determined based on the required maximum LiDAL range, $R^{FOV}_{Max}$, and is equal to 1.25m in our system, see Section III-A. The collocated transmitter-receiver (i.e. transceiver) unit covers an optical footprint area of 4.91 m². It is worth mentioning that the VLC transmitters designed in [19] are spaced by a distance of 2m. Therefore, the maximum spatial overlap between two neighbouring optical footprints, $\Delta x$, is 0.5m as can be seen in Fig. 21.

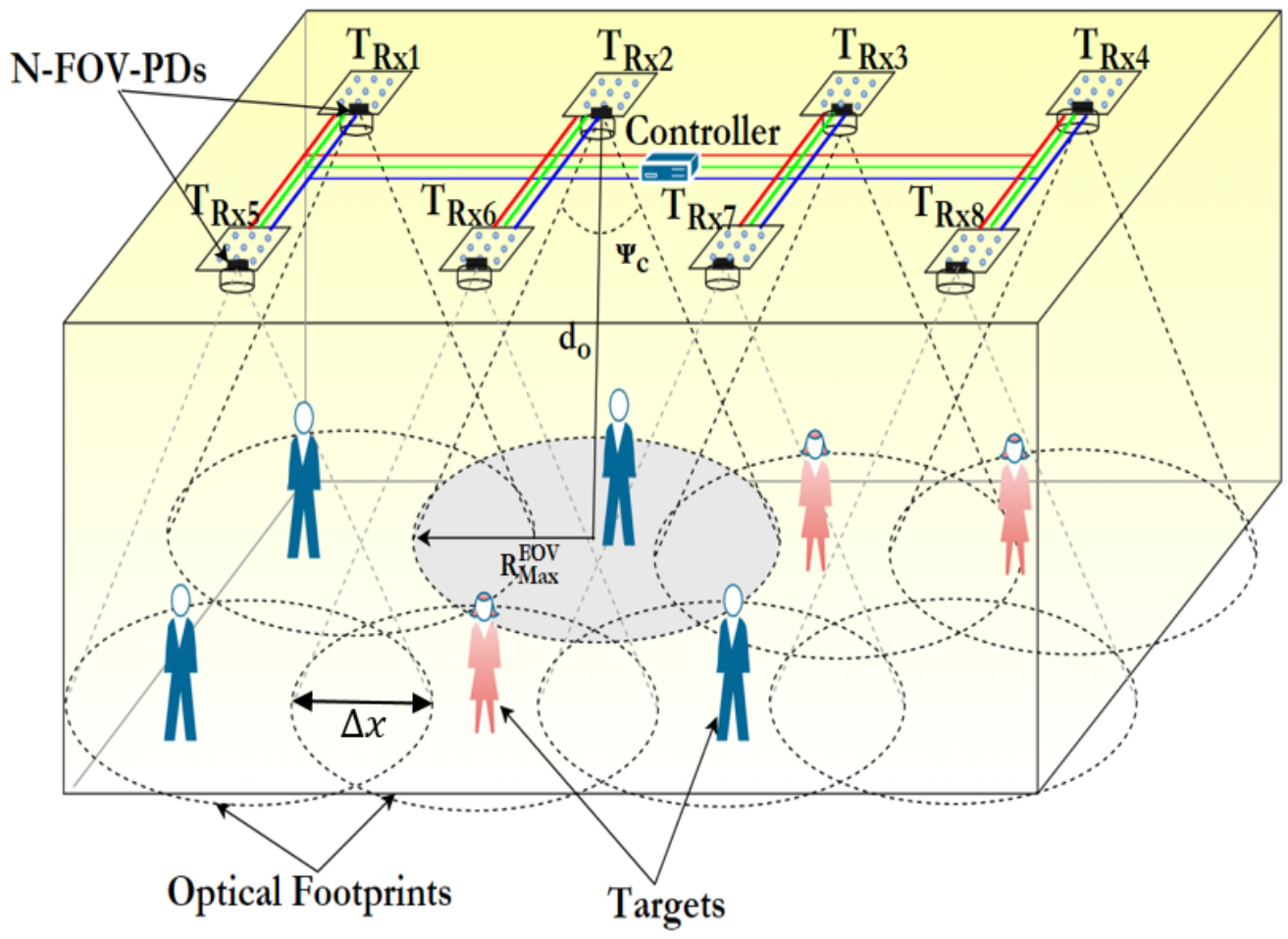

Fig.21: MIMO-LiDAL system.

In MIMO LiDAL, each transceiver unit (collocated $T_X$ and $R_X$) represents a monostatic configuration. Thus, the mean reflected received signal power ($\bar{P}^{M}_{r R^{FOV}_{Max}}$) from a target located at the edge of the optical footprint at a distance of $R^{FOV}_{Max}$ can be derived as:

$$\bar{P}_{r_{R^{\text{FOV}}_{\text{Max}}}}^{M} = \frac{C \ \ \mu_{\rho} \ (d_o - h)^{n+3}}{4\pi^2 \left( R^{\text{FOV}^2}_{\text{Max}} + (d_o - h)^2 \right)^{\frac{n+7}{2}}} \tag{88}$$

where, $C = (n+1)(n_{ele}+1)P_t \, d_A A_R d_A T_f(\Psi_c) G_c(\Psi_c)$ .

The standard deviation of the received signal, $\hat{P}_{r_{R^{\text{FOV}}_{\text{Max}}}}^{M}$, is given as:

$$\hat{P}_{r_{R^{\text{FOV}}_{\text{Max}}}}^{M} = \frac{C \ \ \sigma_{\rho} \ (d_o - h)^{n+3}}{4\pi^2 \left( R^{\text{FOV}^2}_{\text{Max}} + (d_o - h)^2 \right)^{\frac{n+7}{2}}}. \tag{89}$$

In this work, for the MIMO-LiDAL design we set the LD beamwidth for illumination purpose [19] as $n$=0.5; and set the power transmitted by each light unit as $P_t$=18W (as discussed in Section II). The PD area is $A_R$=20mm$^2$, $T_f(\Psi_c)$=1 (a lossless optical filter was assumed), $G_c(\Psi_c)$ =6 obtained using the concentrator gain equation (5) with $N$=1.7 [27] and $\Psi_c$=43.8$^o$. The target effective cross section area was set as $d_A$=0.29m$^2$, which is the minimum target cross section area using Fig. 1b. This minimum area occurs when the target orientation is such that the human (left or right) side faces the transceiver unit (a larger target cross section area results if the person faces or gives their back to the transceiver). The target height selected was $h$=1.7m and , $d_o$=3m (room height). The receiver bandwidth is 315 MHz which corresponds to the maximum channel bandwidth according to the monostatic LiDAL system analysis in Section III-B. Thus the TIA thermal input noise current $\sigma_{thermal}$ is about 2.6 pA/√Hz [41].

Fig. 22, shows the ROC depicting the trade-off between $P_D^M$ and $P_{FD}^M$ of the monostatic LiDAL system for two locations where the targets are located at ranges of $R^{\text{FOV}}_{\text{Max}}$ and $\frac{1}{2} R^{\text{FOV}}_{\text{Max}}$ respectively. The impact of selecting the detection threshold $D_{th}^M$ on the target false detection can be seen in Fig. 23. In this work, we consider MIMO-LiDAL for people detection, counting and localization applications. Therefore, we adjusted the $D_{th}^M$ to maximize the value of $P_D^M$ which implies that high false alarms are accepted to ensure that every target is counted and localized. We chose $P_{FD}^M$=0.1 which leads to $P_D^M$ =0.92 and therefore the optimum detection threshold for the monostatic LiDAL is $D_{th}^M$ = 0.32 $\bar{P}_{r_{R^{\text{FOV}}_{\text{Max}}}}^{M}$ in this case.

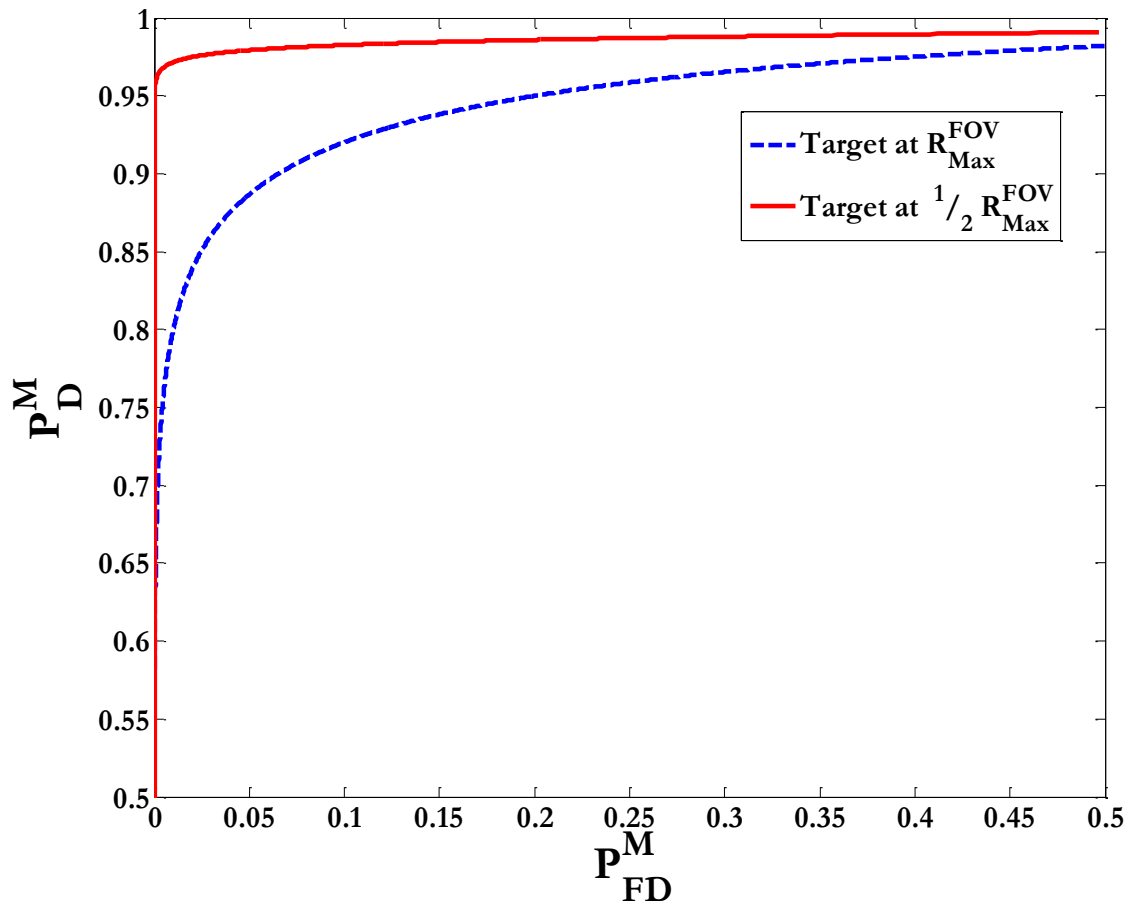

Fig.22: ROC of Monostatic MIMO LiDAL.

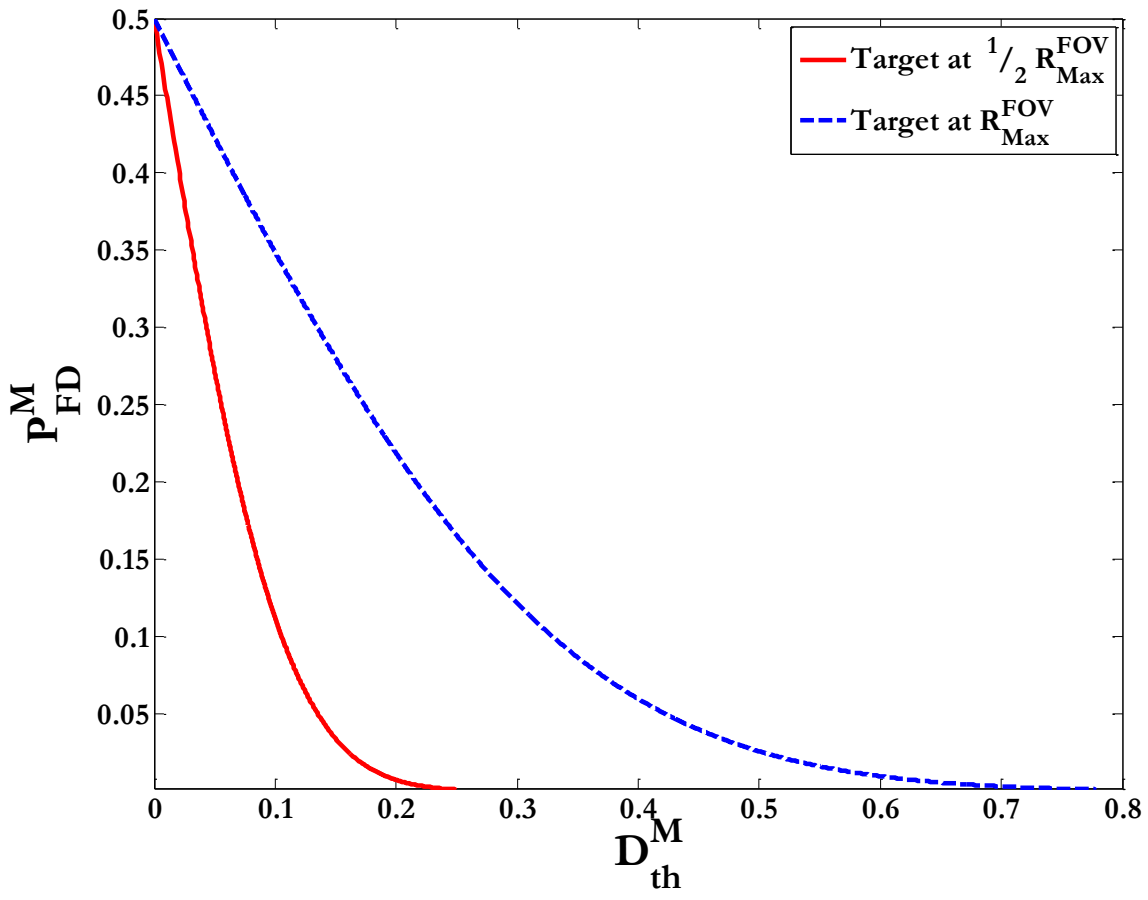

Fig.23: Monostatic MIMO LiDAL false detection with optimum $D_{th}^M$.

### *B. Collaboration of the MIMO-LiDAL Transceivers Units*

In a realistic environment, the ambiguity of target detection can be divided to into two types. Firstly, when a single target moves along a circle centered at the centre of the receiver optical footprint, the reflected pulses from the target arrive at the same time. Therefore, the exact location of a target on this circle (where on the circle) beneath the receiver cannot be established. Secondly, two or more stationary targets cannot be distinguished if they are located at different locations but their distances to a monostatic transceiver ($T_{RX1}$ in Fig. 24a) are the same as can be seen in Fig. 24 a (targets 1 and 2).

These forms of target ambiguity can be resolved if bistatic transceivers are used, see Section III-A and Fig. 24a. In this case, the target position has to be covered by multiple transmitters (at least three transmitters, for spatial localization) that act as anchors, and by at least one receiver.

The footprint coverage radius of each VLC transmitter unit is 4.8m (transmitter beamwidth was set as 75$^o$ for illumination purposes [19]) which results in a minimum coverage overlap of 3.8m between the neighbouring transmitters (ie between the circular optical zones covered by each transmitter). Consider target 2 in Fig. 24a located at the maximum range of $R_{X1}$, ie located at $R^{FOV}_{Max}$ of $R_{X1}$. This target is illuminated by LiDAL $T_{X2}$. Therefore, this collaboration between neighbouring transceivers ($CoN_{TRx}$) has established the second anchor in a bistatic configuration where $T_{X2}$ is now an anchor. The first anchor is $T_{X1}$, where $T_{X1}$ and $R_{X1}$ act as a monstatic LiDAL sub-system. The third anchor is established in the example in Fig. 24a through a bistatic LiDAL subsystem formed by $T_{X3}$ and $R_{X1}$. Therefore, the MIMO LiDAL system in Fig. 24a acts to establish the target location by removing the location ambiguity.

To illustrate the removal of target ambiguity through the joint use of the three anchors, consider Fig.24a which depicts a worst case scenario with two targets located at positions $P_1$ (target 1) and $P_2$ (target 2). Observed through the field of view of the MIMO LiDAL sub-system $T_{X1}$-$R_{X1}$ (i.e. monostatic LiDAL) and $T_{X3}$-$R_{X1}$ MIMO LiDAL sub-system (i.e. bistatic LIDAL), both targets are at same distance to $R_{X1}$ and therefore ambiguity occurs. Considering $T_{X1}$-$R_{X1}$, the round trip time of the reflected pulse from target 1 ($2R_{1(1)}$) is equal to the round trip time associated with the pulse reflected from target 2, $2R_{1(2)}$, resulting in the pulse seen in Fig.24b. Similarly, considering $T_{X3}$-$R_{X1}$ and the trip distances ($R_{3(1)}$ + $R_{1(1)}$) and ($R_{3(1)}$ + $R_{1(2)}$) result in the pulses seen Fig.24d. Thus, ambiguity exists. However, if $T_{X2}$-$R_{X1}$ are used, the distinct trip distances ($R_{2(1)}$ + $R_{1(1)}$) and ($R_{2(2)}$ + $R_{1(2)}$) result in ambiguity resolution as seen in Fig. 24c.

For the bistatic LiDAL, the mean received reflected signal power, $\bar{P}_{r_{R_{\text{Max}}^{\text{FOV}}}}^{B}$, from a target located at the detection edge, ie at a distance of $R_{\text{Max}}^{\text{FOV}}$ (see Fig.24a target located at $P_2$) can be derived as:

$$\bar{P}_{r_{R_{\text{Max}}^{\text{FOV}}}}^{B} = \frac{C\,\mu_\rho\,(d_o - h)^{n+3}}{4\pi^2\left((3\,R_{Max}^{FOV})^2 + (d_o-h)^2\right)^{\frac{n+3}{2}}\left({R_{\text{Max}}^{\text{FOV}}}^2 + (d_o-h)^2\right)^2}. \quad (90)$$

The standard deviation of the received signal, $\hat{P}_{r_{R_{\text{Max}}^{\text{FOV}}}}^{B}$, is given as:

$$\hat{P}_{r_{R_{\text{Max}}^{\text{FOV}}}}^{B} = \frac{C\,\sigma_\rho\,(d_o - h)^{n+3}}{4\pi^2\left((3\,R_{Max}^{FOV})^2 + (d_o-h)^2\right)^{\frac{n+3}{2}}\left({R_{\text{Max}}^{\text{FOV}}}^2 + (d_o-h)^2\right)^2}. \quad (91)$$

Fig. 25 presents the ROC of the bistatic MIMO-LiDAL. Note that unlike the monostatic LiDAL system, the bistatic LiDAL system makes use of distant anchor points, to help resolve the localization ambiguity. Therefore, the mean received signal is low when a distant anchor point is used. To maintain high detection probability in this case, a higher false detection probability, $P_{FD}^{B}$, is used, $P_{FD}^{B}$=0.25. Here higher false alarms are accepted to ensure that the probability of people detection is high. This results in an optimum threshold $D_{th}^{B}$of (0.35 $\hat{P}_{r_{R_{\text{Max}}^{\text{FOV}}}}^{B}$) as can be noted in Fig.26, with $P_{D}^{B}$ =0.7 from Fig. 25. To improve the performance of MIMO-LiDAL system, we (*i*) implemented different optimum detection thresholds $D_{th}^{B}$ which are adjusted adaptively in the sub-optimum detector for the both the monostatic and the bistatic LiDAL systems; (*ii*) optimized the ZFE for the monostatic and the bistatic LiDAL systems.

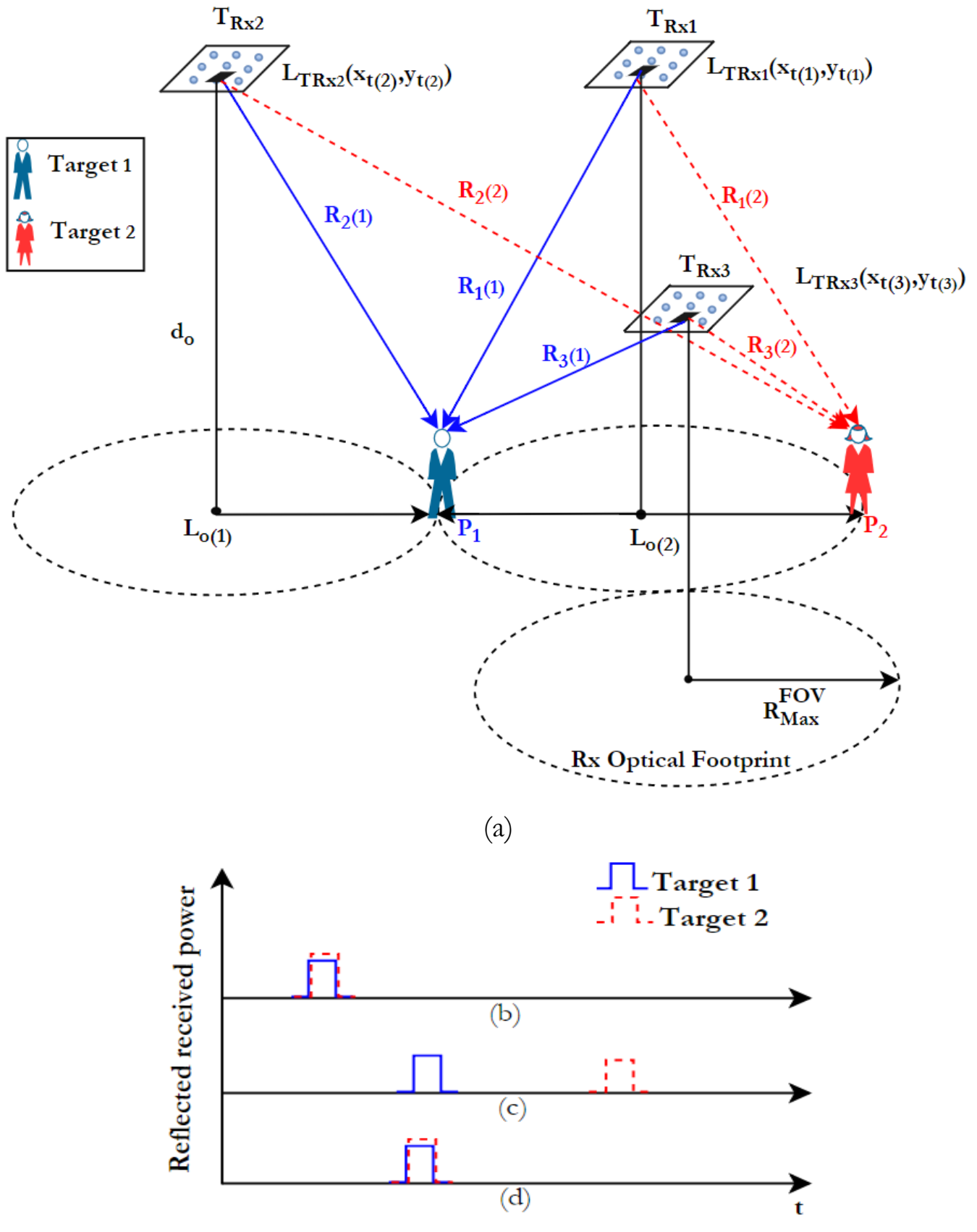

**Fig.24:** Target detection ambiguity: (a) target detection ambiguity in MIMO-LiDAL system with targets ranging, (b) the reflected pulses from targets when Tx1-Rx1 are active, (c) the reflected pulses from targets when Tx2-Rx1 are active and (d) the reflected pulses from targets when Tx3-Rx1 are active.

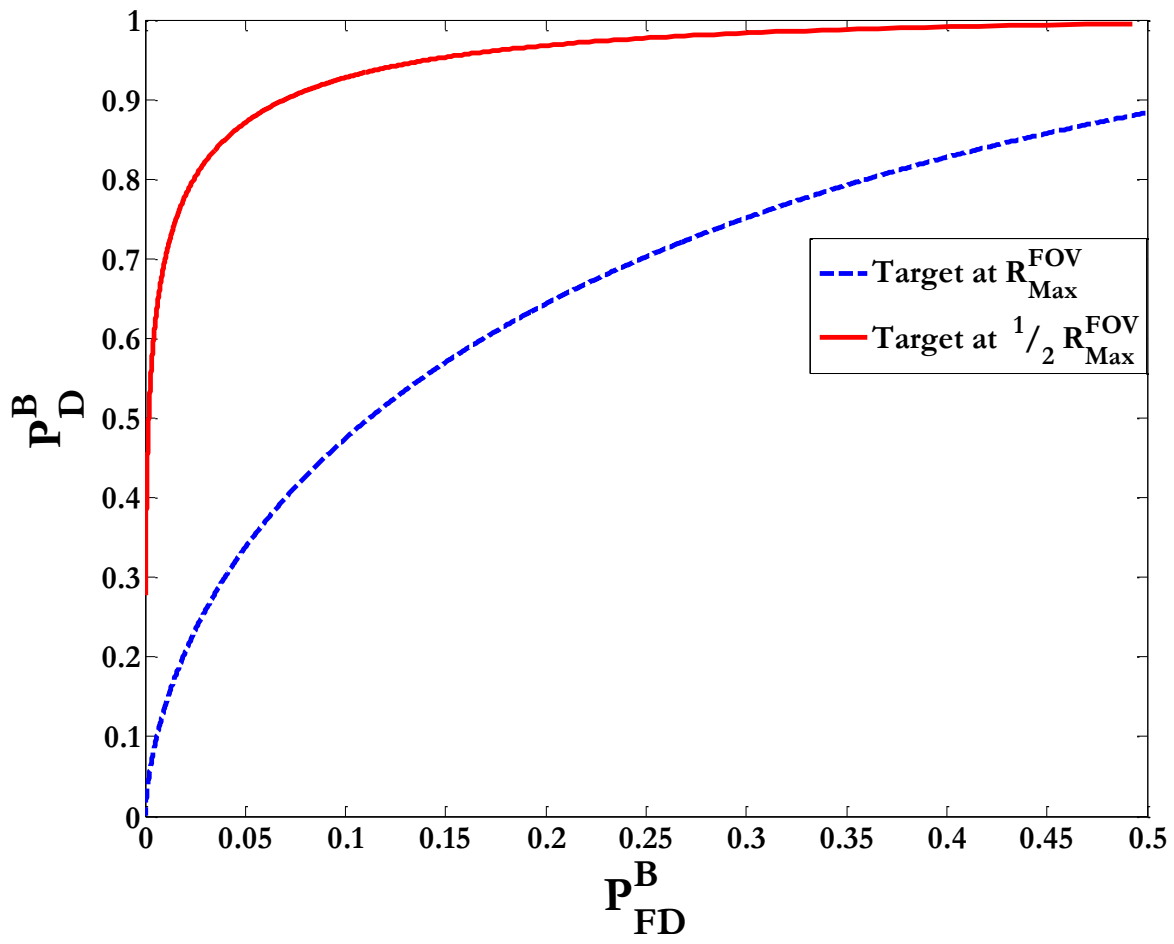

Fig.25: ROC of Bistatic MIMO LiDAL.

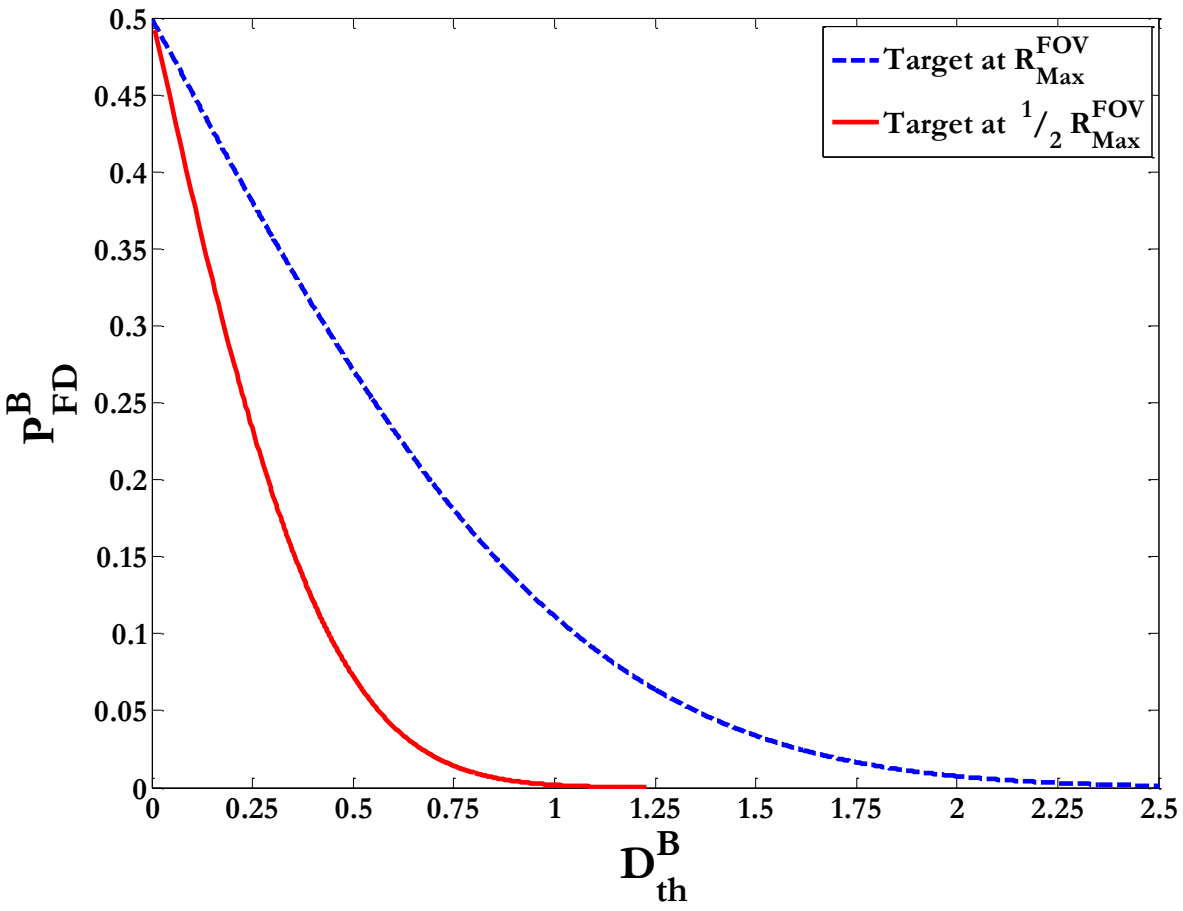

Fig.26: Bistatic MIMO LiDAL probability of false detection with optimum $D_{th}^{B}$.

There is a finite probability that a target is present, but is missed. This probability of miss-detecting a target located at $R_{\text{Max}}^{\text{FOV}}$ (see for example the target at $P_2$ in Fig.24a which is located at the maximum range, $R_{\text{Max}}^{\text{FOV}}$) of MIMO-LiDAL ($P_{MD_{(MIMO)}}^{R_{\text{Max}}^{\text{FOV}}}$) can be derived as:

$$P_{MD_{(MIMO)}}^{R_{\text{Max}}^{\text{FOV}}} = \left(\frac{1}{2}\right)^{K_n+1}\left(1+\text{erf}\left(\frac{D_{th}^{M} - \bar{P}_{r_{R_{\text{Max}}^{\text{FOV}}}}^{M}}{\hat{P}_{r_{R_{\text{Max}}^{\text{FOV}}}}^{M}\sqrt{2}}\right)\right)\prod_{k=1}^{K_n}\left(1+\text{erf}\left(\frac{D_{th}^{B}(k) - \bar{P}_{r_{R_{\text{Max}}^{\text{FOV}}}}^{B}}{\hat{P}_{r_{R_{\text{Max}}^{\text{FOV}}}}^{B}\sqrt{2}}\right)\right) \quad (92)$$

where $\text{erf}(\cdot)$ is the error function and $K_n$ is the number of neighbour transceiver units (bistatic LIDAL). The term $\left(1 - P_{MD_{(MIMO)}}^{R_{\text{Max}}^{\text{FOV}}}\right)$ represents the probability of the target detected by at least one transceiver unit (for example (Tx2-Rx1) as in the case shown in Fig.24).

Let $T_{c_M}$ be the maximum number of targets that can be counted successfully when the targets are located at different distances from the LiDAL transceiver with minimum separation distance of $\Delta R$. This number, $T_{c_M}$, for the MIMO LIDAL system is given as:

$$T_{c_{M(MIMO)}} = \frac{T_{w(MIMO)}}{\tau}\,N_{OF} \quad (93)$$

where $N_{OF}$ is the number of optical footprint zones and $T_w$ is the LiDAL channel time window which corresponds to the difference in the round-trip times of a target placed at $R_{Max}^{FOV}$ and a target placed underneath the transceiver, ie at the centre of the LiDAL transceiver optical detection zone. Thus $T_{w(MIMO)}$ of the MIMO-LIDAL is determined as:

$$T_{w(MIMO)} = \frac{\left(\left(\left(\sqrt{(3\,R_{Max}^{FOV})^2 + (d_o - h)^2}\right) + \left(\frac{R_{Max}^{FOV}}{\sin(\Psi_c)}\right)\right) - 2(d_o - h)\right)}{c} \quad (94)$$

where $d_0$ is the perpendicular distance between of the $i$ *th* transceiver unit coordinates $L_{TRx(i)}$ and the centre of the transceiver illumination $L_{o(i)}$ (see Fig.24a).

### C. Target Localization

To localize a target, the time of arrival of the reflected pulse and its direction are required. However, in the MIMO-LiDAL system considered, the direction of the reflected pulse from the target cannot be determined due to the wide-FOV of the receiver. An angle diversity receiver can be used which can help determine a coarse direction of arrival based on the receiver face that detects the signal. The angular resolution is however typically coarse as the number of faces in the angle diversity receiver are typically limited and small. An even coarser localization can be achieved with a single wide FOV LiDAL receiver which can provide an estimated range, thus placing the target (human) on a circle on the floor in an indoor environment.

For accurate target localization, collaboration of neighbouring MIMO-LiDAL transceiver units can be utilized with a time of arrival (TOA) approach to localize the target. To determine the exact target location, ranges to at least three transmitters (anchors) must be obtained. In Fig. 24a $R_1(j)$, $R_2(j)$ and $R_3(j)$ are the ranges of target $j$ to the three transceivers. The location of the target is calculated as the intersection of the three (circles) ranges. Any target in the indoor environment will lie in the coverage area of at least one monostatic receiver (LiDAL system), see Fig. 21. Therefore, this localization technique relies on the success of target detection by at least $K \geq 2$ neighbouring bistatic LiDAL sub-systems. The monostatic MIMO-LiDAL range can be written as:

$$R_1(j) = \frac{t_{trip}^{j}(T_{x1}, R_{x1})\, c}{2}. \quad (95)$$

The bistatic MIMO-LiDAL range is given as:

$$R_k(j) = \left(t_{trip}^{j}(T_{xk}, R_{x1})\right) c - R_1(j) \quad \forall k \in K, k \neq 1 \quad (96)$$

where $R_1(j)$ is the range in metre of the $j$ *th* target from the monostatic LiDAL subsystem (subsystem number 1), $R_k(j)$ are the ranges in metres of the $j$ *th* target from LiDAL bistatic subsystem $k$, $t_{trip}^{j}(T_{x1}, R_{x1})$ and $t_{trip}^{j}(T_{xk}, R_{x1})$ are the trip times between the identified transmitter and receiver units, which are monostatic and bistatic respectively in this case.

Consider a target $j$ whose position is $P_j(x_j, y_j)$, and consider the $kth$ transmitter anchor located at $(x_{t(k)}, y_{t(k)})$, we have:

$$(x_{t(k)} - x_j)^2 + (y_{t(k)} - y_j)^2 = R_k^2(j). \quad (97)$$

A least squares approach [67], [68] can be used to solve (98) to provide an approximate location of the $j$ *th* target at the intersection of $K + 1$ circles is given as:

$$x_j(x_{t(k)} - x_{t(1)}) - y_j(y_{t(k)} - y_{t(1)}) = \frac{1}{2}\left(R_1^2(j) - R_k^2(j) + x_{t(k)}^2 + y_{t(k)}^2 - x_{t(1)}^2 - y_{t(1)}^2\right). \quad (98)$$

Equation (99) can be written in matrix form where $A$ and $B$ are location matrices [69], [70]:

$$A = \begin{pmatrix} x_{t(k)} - x_{t(1)} & y_{t(k)} - y_{t(1)} \\ x_{t(K)} - x_{t(1)} & y_{t(K)} - y_{t(1)} \end{pmatrix} \quad (99)$$

$$B = \frac{1}{2}\begin{pmatrix} (R_1^2(j) - R_k^2(j)) + (x_{t(k)}^2 + y_{t(k)}^2) - (x_{t(1)}^2 + y_{t(1)}^2) \\ (R_1^2(j) - R_K^2(j)) + (x_{t(K)}^2 + y_{t(K)}^2) - (x_{t(1)}^2 + y_{t(1)}^2) \end{pmatrix} \quad (100)$$

and $X$ is:

$$X = [x_j\ y_j]^T\ . \quad (101)$$

The target position $P_j(x_j, y_j)$ can be determined as [70] :

$$X = P_j(x_j, y_j) = (A^T A)^{-1} A^T B. \quad (102)$$

In MIMO LiDAL, target localization depends on collaboration of neighbouring transceivers. Thus, to localize target 2, located at $P_2$, in Fig24.a , $T_{X1}$, $T_{X2}$ and $T_{X3}$ work *separately* with $R_{X1}$ to localize target 2. This requires three *separate* LiDAL scans. Hence, the probability of detection of target 2 by $T_{X1}$-$R_{X1}$ (Monostatic LiDAL sub-system) is independent of the probabilities of detection of the same target by $T_{X2}$-$R_{X1}$ and $T_{X3}$-$R_{X1}$ (both are Bistatic LiDAL sub-systems). Consequently, the probability of localizing a target located at the maximum range, $R_{\text{Max}}^{\text{FOV}}$, $P_{L(MIMO)}^{R_{\text{Max}}^{\text{FOV}}}$ can written as:

$$P_{L(MIMO)}^{R_{\text{Max}}^{\text{FOV}}} = P_D^M \prod_{k=1}^{K} P_D^B(k) \quad (103)$$

$$P_{L(MIMO)}^{R_{\text{Max}}^{\text{FOV}}} = \left(\frac{1}{2}\right)^{K+1} \text{erfc}\left(\frac{D_{th}^M - \bar{P}_{r_{R_{\text{Max}}^{\text{FOV}}}}^{M}}{\hat{P}_{r_{R_{\text{Max}}^{\text{FOV}}}}^{M}\sqrt{2}}\right) \prod_{k=1}^{K}\left(\text{erfc}\left(\frac{D_{th}^B(k) - \bar{P}_{r_{R_{\text{Max}}^{\text{FOV}}}}^{B}(k)}{\hat{P}_{r_{R_{\text{Max}}^{\text{FOV}}}}^{B}(k)\sqrt{2}}\right)\right) \quad (104)$$

### D. MIMO LiDAL System Operating Algorithm

To distinguish human targets from other objects (obstacles) and to localize human targets in MIMO-LiDAL, pulses are transmitted from the transmitters in a sequence through $M$ frames (single pulse per frame) which are managed by the controller. The receiver collects the reflected signal from the targets and ambient obstacles including walls, floor and furniture during the receiver listening time $T$. Fig. 27 shows the proposed receiver block diagram in each transceiver unit of the MIMO LiDAL system.

In Fig 27, the controller instructs transmitter (anchor) $k$ to emit a pulse while the other anchors are silent. This action as well as the received reflected pulse (from the target) form the input to the receiver in Fig. 27. The received signal is fed in Fig. 27 firstly to a "distinguishing method" block, this having been discussed in Section IV, where humans are distinguished from obstacles using for example human motion. The output of the distinguishing method block forms the input to the optimum detector block of Section III. The optimum detector output identifies the time slots that contain targets. This information is used to determine the TOA. Furthermore, the slots that contain targets are counted to determine the number of human targets in the environment. Given that a number of LiDAL subsystems collaborate (three or more anchors), the target location is estimated. Finally, duplicate targets are eliminated. These are targets that lie in the overlap areas of the optical zones covered by the LiDAL receivers.

The controller conducts the detection, counting and localization process as follows:

1) The first pulse of the control signal $c(t)$ activates the transceiver monostatic LiDAL sub-system to (i) send an optical pulse $x(\tau)$ from the transmitter $Tx\,(k)$, and (ii) initiate the receiver $Rx\,(k)$ to listen to the reflected signal.
2) The receiver $Rx\,(k)$ collects the reflected optical signal in an observation widow of duration $T$. A distinguishing method (in this work we considered BSM and CCM methods) in conjunction with the designed sub-optimum LiDAL receiver are then used to detect the targets' presence and their ranges and update the counter $V_c(i)$ as can be seen in Fig. 27.
3) For target localization, the controller identifies the $K$ neighbouring bistatic LiDAL sub-systems. In this work we considered $K$ =2. The second and third control pulses activate the neighbouring transmitters $Tx\,(k+1)$, and $Tx\,(k+2)$ with the receiver $Rx\,(k)$. Each control pulse generates a LiDAL pulse from *one* of the LiDAL bistatic sub-systems and results in reflections being observed during a time duration $T$. The second pulse is generated at the end of the observation time $T$. The three trip times (one monostatic and two neighbouring bistatic LiDAL sub-systems) are then used to determine the targets' locations using TOA.
4) Target elimination follows where the targets located in the overlap zones are counted only once. Due to position errors, duplicated targets are eliminated if the Euclidean distance between any two such target locations is less than $\Delta R$. The counter $V_c(i)$ is updated accordingly.
5) For the remaining *I*-1 optical zones, steps (1) to (4) are repeated. The *I* optical zones in the room are shown in Fig. 21.
6) The number of targets, $N_E$, is calculated as $N_E = \sum_{i=1}^{I} V_c(i)$.

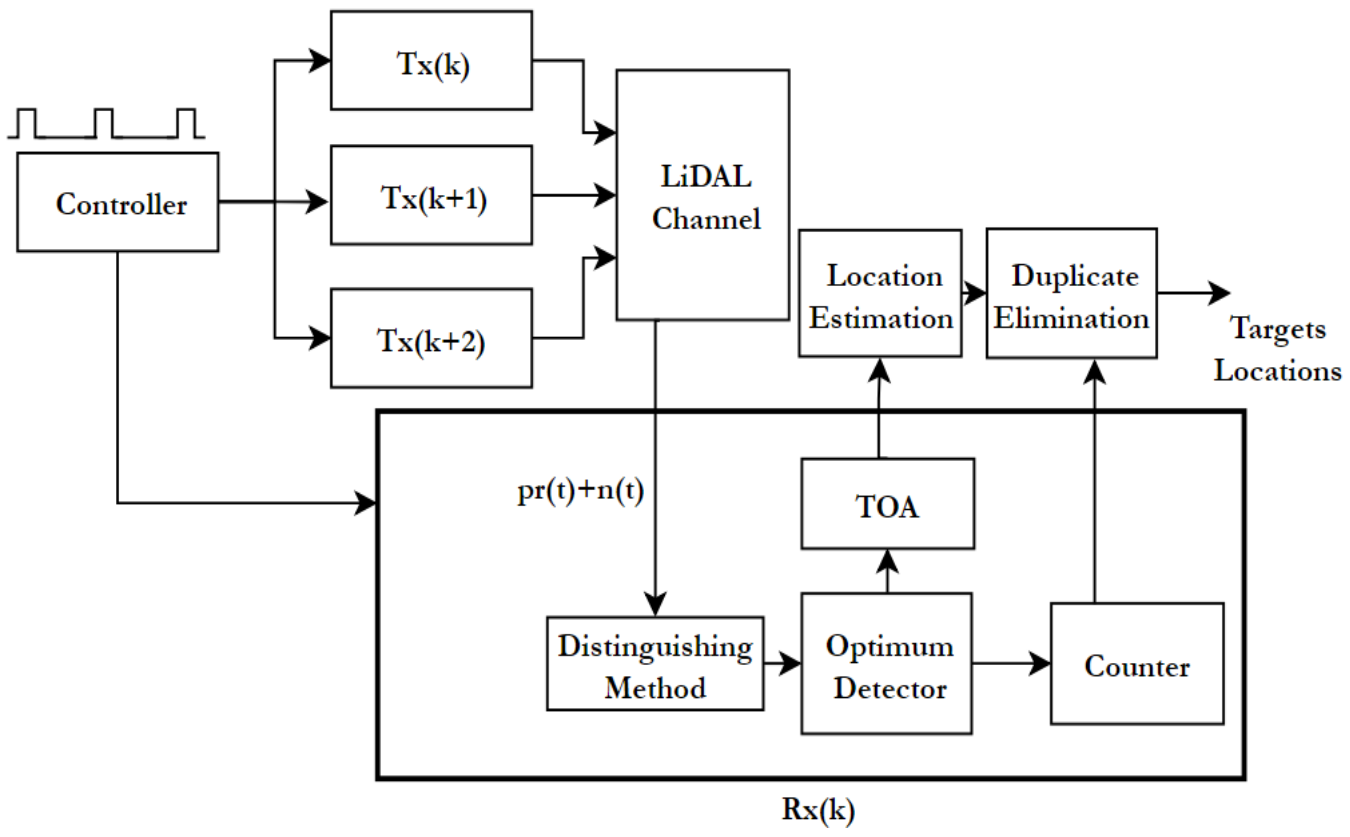

Fig. 27: the receiver block diagram of MIMO-LiDAL system.

In terms of complexity, the number of scans (time frames of duration $T$) needed to cover all the optical zones in the room is key. For the MIMO-LiDAL system, the number of frames, $M$, required to complete one monitoring cycle (i.e. detection, counting and localization of a full room that has $I$ optical zones) is determined as $M = I(k+1)$. Therefore, the VLC MAC overhead, $OH_{ML}$, required to use the same VLC system for communication and MIMO-LiDAL localization is:

$$OH_{ML} = \frac{T\,M}{T_{MAC_{VLC}}} \qquad (105)$$

where $T_{MAC_{VLC}}$ is the VLC MAC frame duration. For instance, if the MIMO-LiDAL system is used for pedestrian (more demanding than nomadic) target monitoring, then target location evaluation every 100 ms is sufficient given a maximum pedestrian speed of 3 m/s, where the 100ms results in motion by $\Delta R = 30cm$, which is the minimum distance of interest in this work. Therefore, a combined MIMO-LiDAL – VLC frame duration of 100ms can be considered. Considering one optical zone in Fig. 21, its dimensions and considering the more demanding (distributed transmitters and receivers) bistatic LiDAL subsystem observation window duration, equation (95) gives this duration, $T$, for our system parameters as 44ns. If there are *I*=8 optical zones as in Fig. 21, then the number of frames needed is M=24 frames leading to an observation time of 1.1 $\mu s$. The other key, non-real time blocks in Fig. 27 are the location estimation which solves the matrix operations in (103) and the duplicate target elimination block which carries out a simple Euclidian distance comparison as in step 4 of the algorithm above. These non-real time operations can be carried out in the remaining part of the 100ms frame duration and may last for few milliseconds depending on the processor used. The key point is that visible light communication can resume after the 1.1 $\mu s$. The communications interruption overhead needed is thus negligible, however a localization result may require 10 snapshots, which are collected in 10 frames and thus a localization result may take one second, which is acceptable for pedestrian movement.

## VI. Multiple-Input Single-Output imaging-LiDAL (MISO-IMG-LiDAL)

In LiDAL, the elimination of target ambiguity is important to detect, count and localize targets correctly. Traditional bistiatic RF radar eliminates target ambiguity by using the estimated target range (round trip time) with the angle of arrival of the received signal reflected from the target, where the angle of arrival is estimated using beam steering based on mechanical rotated RF receivers or phased antenna arrays. In our optical imaging radar, the receiver consists of a photodiode array with an imaging lens that forms an image of the observed region on the receiver detectors. To determine the direction of the received reflected signal from the target, the imaging receiver pixels that observe the target are used together with their FOV. Hence, in this work we introduce an imaging LiDAL system that employs an imaging detection receiver with multiple VLC transmitters (light sources / engines). We refer to this system as multiple-input (multiple LiDAL transmitters) single-output (single LiDAL receiver) imaging LiDAL. The MISO-IMG-LIDAL system can provide; (*i*) target ambiguity elimination where the targets are separated in the optical imaging domain; (*ii*) target localization where the imaging receiver forms an image of the floor and hence each imaging receiver pixel observes a small and finite region on the floor. Most importantly, localization is achieved in this case using one time frame (no need for three anchors); (*iii*) interference minimization, (the interference results from reflections from the background obstacles) which can lead to improvement in the performance of the distinguishing methods such as the BSM method; (*iv*) LiDAL channel bandwidth enhancement due to the narrow FOV of the pixels which reduces the complexity of the optimum receiver without implementing an equalizer to tackle the channel dispersion; (*v*) simplified system design where the localization accuracy / resolution is no longer a function of the pulse width. Instead the localization accuracy can be increased by increasing the number of receiver pixels (and hence also reducing the per pixel FOV). The pulse used for localization can thus have a larger duration compared to the pulse duration used in the LiDAL systems of Section V. This leads to simplified pulsed transmitter design, which is welcome given that commercial high resolution imaging receivers are available with several million pixels (here we use hundreds of pixels); (*vi*) overhead reduction, where the imaging LiDAL overheads are reduced compared to MIMO LiDAL due to the lower number of radar scans required to detect and localize targets.

*A. MISO- IMG - LiDAL System Configurations*

The MISO-IMG-LiDAL system consists of eight LiDAL transmitter units and one imaging receiver installed in the centre of room's ceiling (2m, 4m, 3m) as shown in Fig. 28. The imaging receiver includes a number of pixels, where each pixel is a photodiode (PD) optical receiver. The advantage of the massive number of pixels is in providing spatial selection to separate the targets in the optical domain (i.e. more narrow optical footprints). This results in reduced targets ambiguity and increased resolution in the spatial domain. The imaging receiver lens forms an image of the floor on the receiver pixels thus dividing the floor into an optical grid as can be noted in Fig. 28, where each sub-receiver has a narrow FOV and covers a given optical footprint.

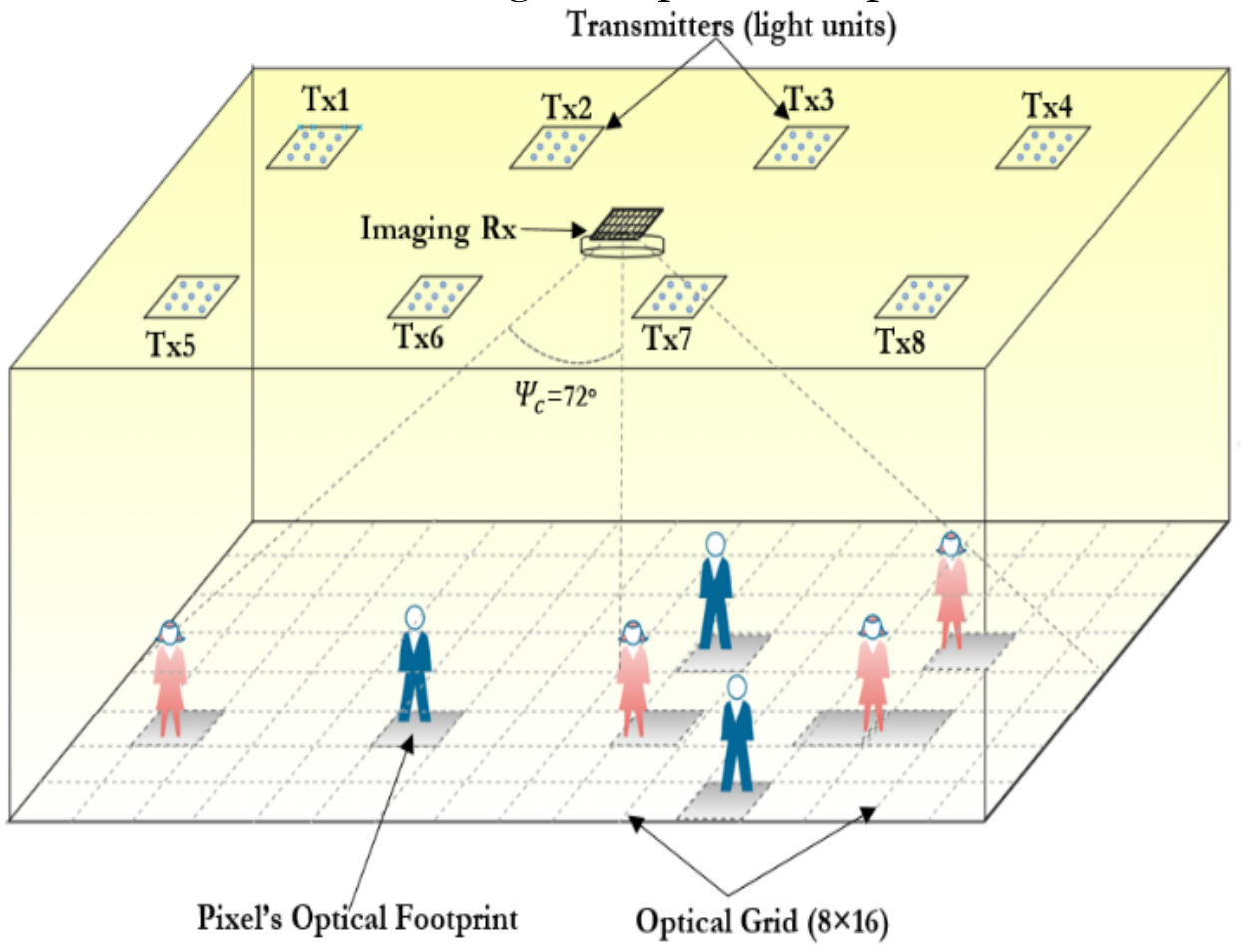

Fig.28: MISO–IMG-LiDAL system.

The configuration of the imaging receiver is defined by (a) the entrance area of imaging receiver lens where $A = \frac{9\pi}{4}$ cm² [29], [71]; (b) the semi-acceptance angle of the imaging lens with semi-angle FOV of $\Psi_c$=72° in our system to enable the imaging receiver to cover the entire floor along room length of 8m; and (c) the lens exit area as defined in [72]:

$$A' = \frac{A\, sin^2(\Psi_c)}{N^2} \tag{106}$$

where $N$ is the lens refractive index. The lens semi-angle FOV can be defined as (see Fig.26a):

$$\Psi_c = \tan^{-1}\left(\frac{D}{2f}\right) \tag{107}$$

where, $f$ is the lens focal length and $D$ is the PD array length as can be seen in Fig. 29a.

The imaging receiver maximum range $R_{Max}^{FOV}$ is related to the target as:

$$R_{Max}^{FOV} = tan(\Psi_c)\ (d_o - h). \tag{108}$$

We define the imaging lens zooming ratio $R_{zoom}$ as:

$$R_{zoom} = \frac{2R_{Max}^{FOV}}{D}. \tag{109}$$

To separate two targets at a distance of $\Delta s$ from each other, as can be seen in Fig.29b, the minimum pixels' distance $\Delta d$ is given as:

$$\Delta d = \frac{\Delta s}{R_{zoom}}. \tag{110}$$

The imaging lens transmission factor $T_f$ is defined as [72] :

$$T_{f(img)}(\delta) = -0.198\delta^2 + 0.0425\delta + 0.8778 \tag{111}$$

where $\delta$ is the incidence angle measured in radians.

We selected an imaging receiver total photodetection area of 2 cm² (2cm length × 1cm width) which approximately fits into the exit area of the lens [29], [72], [73]. The photodetector area is divided into an array of (8 columns × 16 rows) pixels to satisfy the design parameter $\Delta s$ which is chosen as 0.5m. We assumed there is no gap between the pixels. It is worth mentioning that, we change the LiDAL resolution from $\Delta R$ of 0.3m to $\Delta s$ of 0.5m to obtain an integer number of pixels. Each pixel has a square area of 1.56mm² (1.25mm × 1.25mm) and this corresponds to the area of a PD. The pixel's optical detection area can be determined by calculating the viewing angles (azimuth and elevation) corresponding to the receiver location as can be seen in Fig. 29c. The azimuth $(A_z)$ and elevation $(E_L)$ angles of the imaging receiver pixels can be written as [29], [45] :

$$E_{Lj} = \tan^{-1}\left(\frac{\sqrt{d_{xj}^2 + d_{yj}^2}}{d_o - h}\right) \tag{112}$$

$$A_{zj} = \tan^{-1}\left(\frac{d_{yj}}{d_o - h}\right) \tag{113}$$

where $d_x$ and $d_y$ are the horizontal separation distances along the $x$ and $y$ axes as can be seen in Fig. 29c and $j$ is the pixels number.

According to the design parameters of the imaging receiver, each pixel is treated as a PD that covers a typical square optical footprint area of 0.25 m² (pixel's range $P_R$=0.5m i.e. pixel's FOV =11°) on the floor. The optical grid which covers the total detection floor is divided into 128 optical footprints (8×16). We assumed that the imaging lens has no reception distortion with ideally square optical FOV for all pixels. The proposed MISO-IMG-LiDAL can be used for detection, counting, and localization of multiple targets within the optical grid.

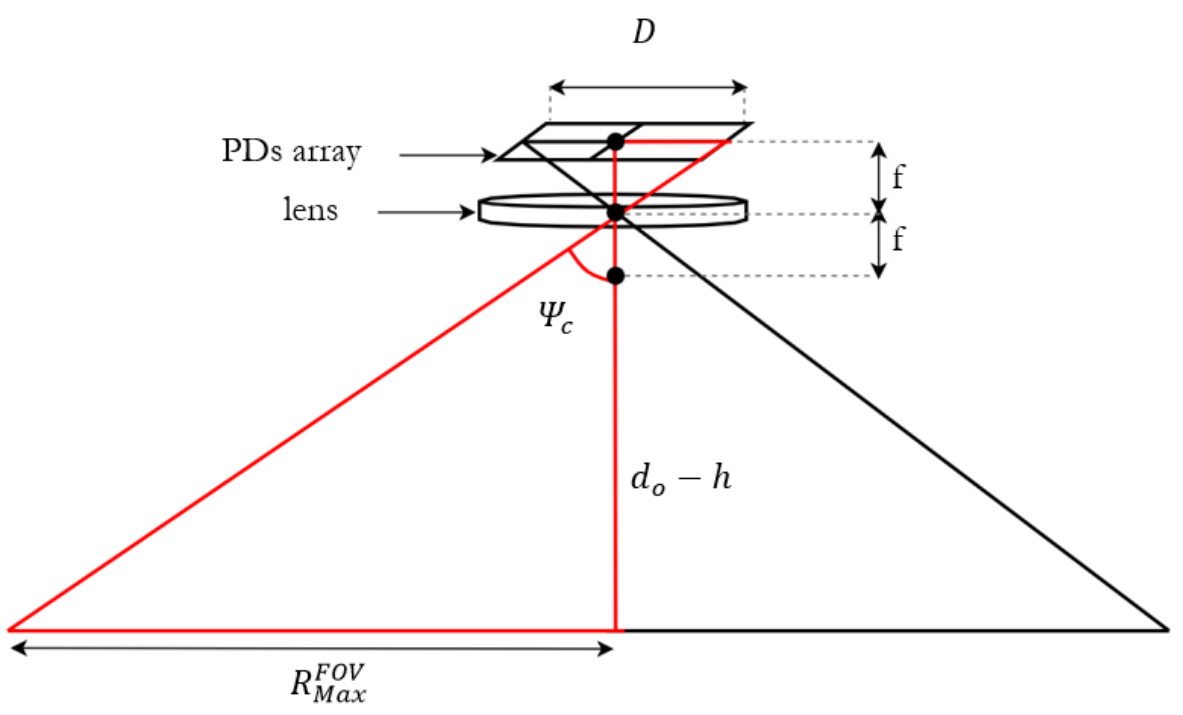

(a)

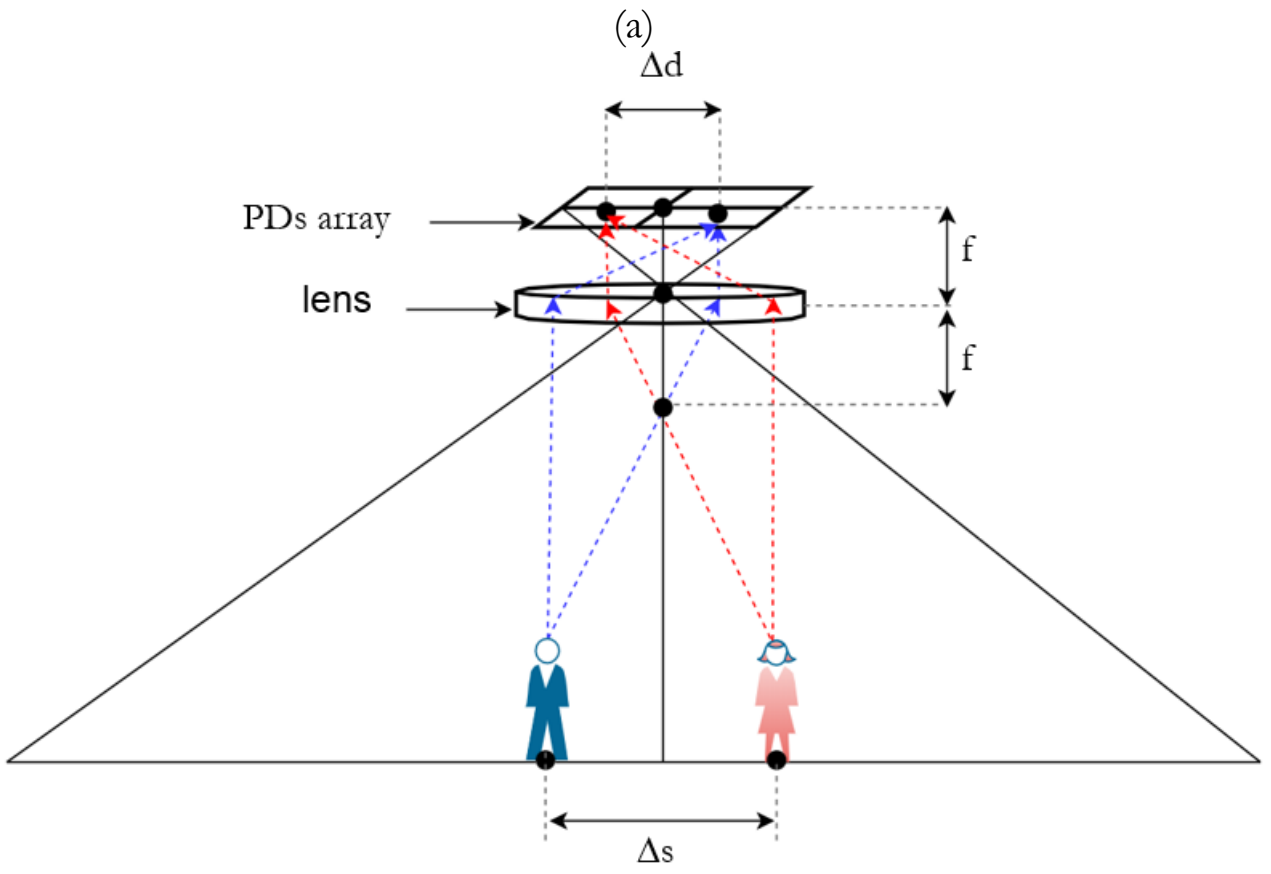

(b)

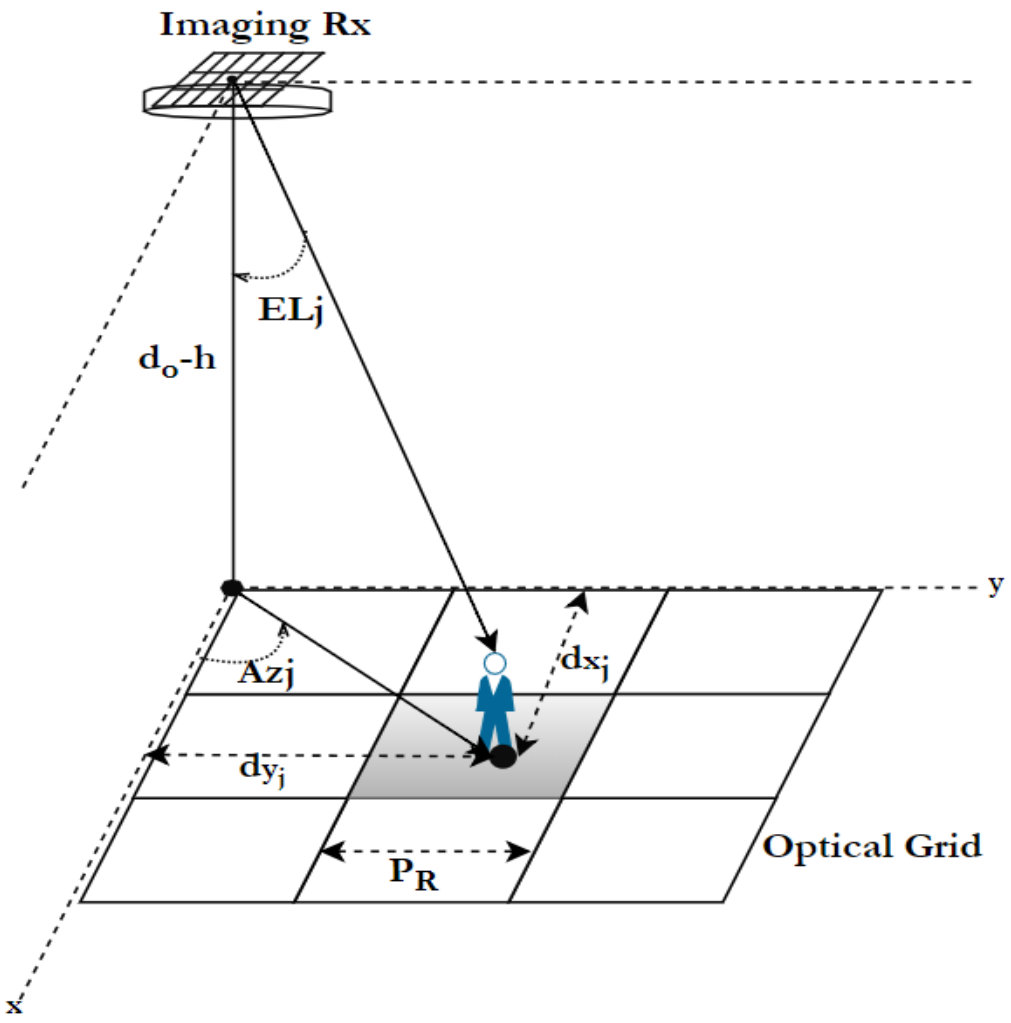

(c)

Fig.29: LiDAL imaging receiver design, (a) lens FOV with $R^{FOV}_{Max}$ , (b) targets optical resolution and (c) pixel's angles and range.

In MISO-IMG-LiDAL, the transmitter unit and the imaging (pixel) receiver are separated and therefore work as bistatic LiDAL. We have calculated the maximum channel bandwidth for MISO IMG-LiDAL using the approach described in Section II-B for the bistatic LiDAL. The maximum channel bandwidth for a single pixel receiver is $Bw_{ch(img)}$=480MHz. We also employed the TIA in [41] for each pixel receiver with input noise current $\sigma_{thermal(img)}$ of 2.6 pA/√Hz.

In a Bistatic imaging MISO LiDAL system, ie $T_{x1}$ working with the imaging $R_x$, the mean received reflected signal ($\overline{P}_{r_{R^{FOV}_{Max}}}^{B_{img}}$) from a target located at the edge of the optical footprint (grid), as can be seen in Fig. 30, can be derived as:

$$\overline{P}_{r_{R^{FOV}_{Max}}}^{B_{img}} = \frac{C_{img}\,\mu_p\,(d_o-h)^{n+3}}{4\pi^2\left(\left(2\sqrt{2}\,P_R\right)^2+(d_o-h)^2\right)^{\frac{n+3}{2}}\left(R^{FOV^2}_{Max}+(d_o-h)^2\right)^2} \qquad (114)$$

where: $C_{img}=(n+1)(n_{ele}+1)P_t\,d_A\,A\,T_{f_{(img)}}(\Psi_c)G_c(\Psi_c)$.

and the standard deviation of the received signal $\hat{P}_{r_{R^{FOV}_{Max}}}^{B(img)}$ is given as:

$$\hat{P}_{r_{R^{FOV}_{Max}}}^{B(img)} = \frac{C_{img}\,\sigma_p\,(d_o-h)^{n+3}}{4\pi^2\left(\left(2\sqrt{2}\,P_R\right)^2+(d_o-h)^2\right)^{\frac{n+3}{2}}\left(R^{FOV^2}_{Max}+(d_o-h)^2\right)^2} \qquad (115)$$

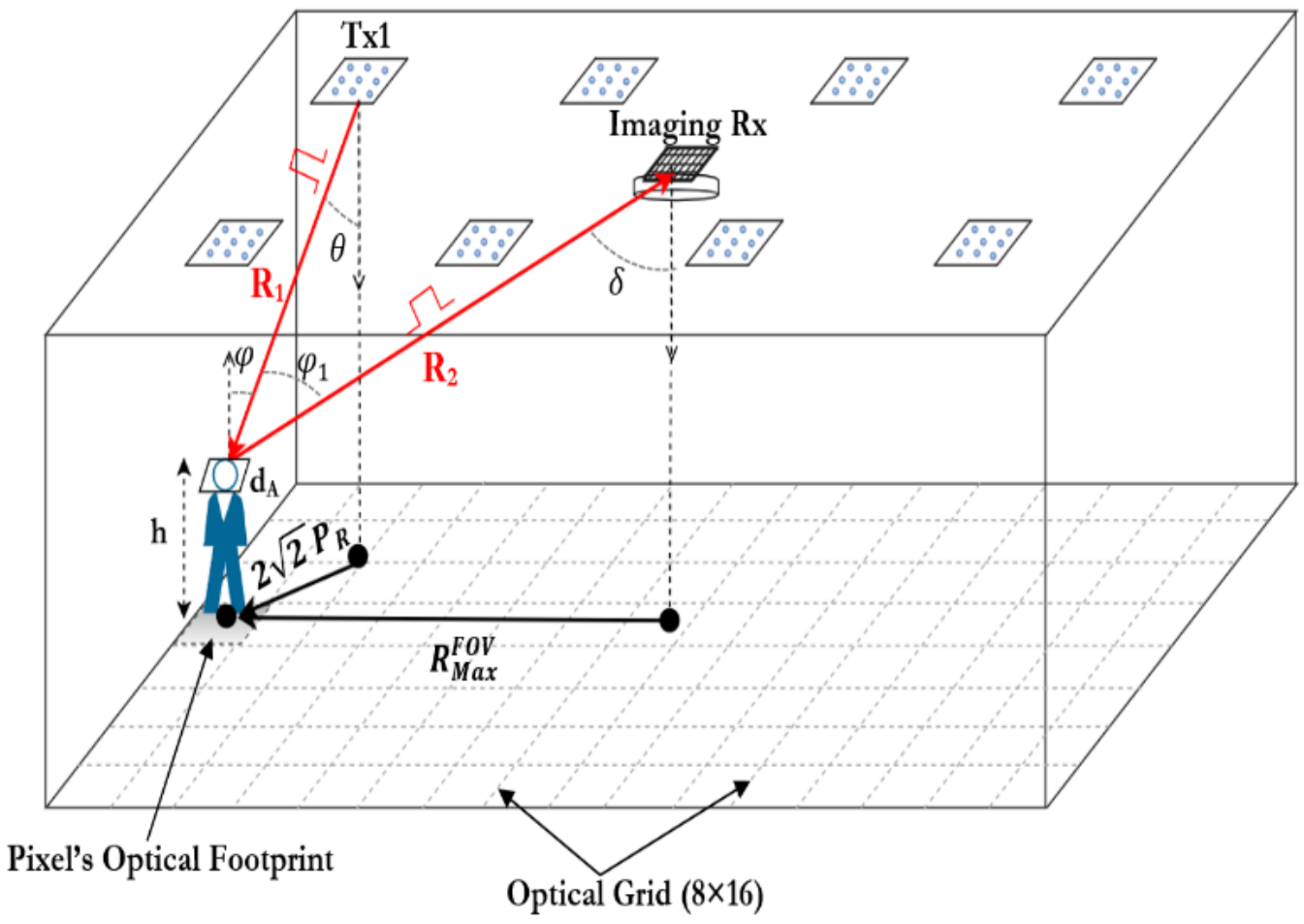

Fig.30: Bistatic MISO-IMG-LiDAL system.

Fig. 31, presents the ROC depicting the trade-off between $P_D^{B(img)}$ and $P_{FD}^{B(img)}$ of the bistatic MISO-IMG-LiDAL system for a target located at a range of $R^{FOV}_{Max}$. The impact of selecting the detection threshold $D_{th}^{B(img)}$ on the target false detection can be seen in Fig. 32. In this work, we consider MISO-IMG-LiDAL for people counting and localization applications. Thus, we selected the detection threshold $D_{th}^{B(img)}$ to maximize the value of $P_D^{B(img)}$. We accept $P_{FD}^{B(img)}$=0.1, thus giving $P_D^{B(img)}$ =0.9 and giving an optimum detection threshold $D_{th}^{B(img)}$= $0.39\overline{P}_{r_{R^{FOV}_{Max}}}^{B_{img}}$ in this case.

The maximum number of targets $C_M$ that can be counted in MISO-IMG-LiDAL system is:

$$C_{M(MISO_{img})} = \frac{T_{w(img)}}{\tau}\,O_{GS} \qquad (116)$$

where $O_{GS}$ is the optical grid size (128 optical footprints) and $T_{w(img)}$ is the channel time window of the imaging receiver's pixel ($j$) which corresponds to the difference in trip times of a target placed at the edge of a pixel's optical footprint $P_R$ (see target location in Fig.30) and a target placed underneath the transceiver. Thus $T_{w(img)}$ is given as:

$$T_{w(img)} = \frac{\left(\sqrt{P_R^2+(d_o-h)^2}\right)-(d_o-h)}{c} \qquad (117)$$

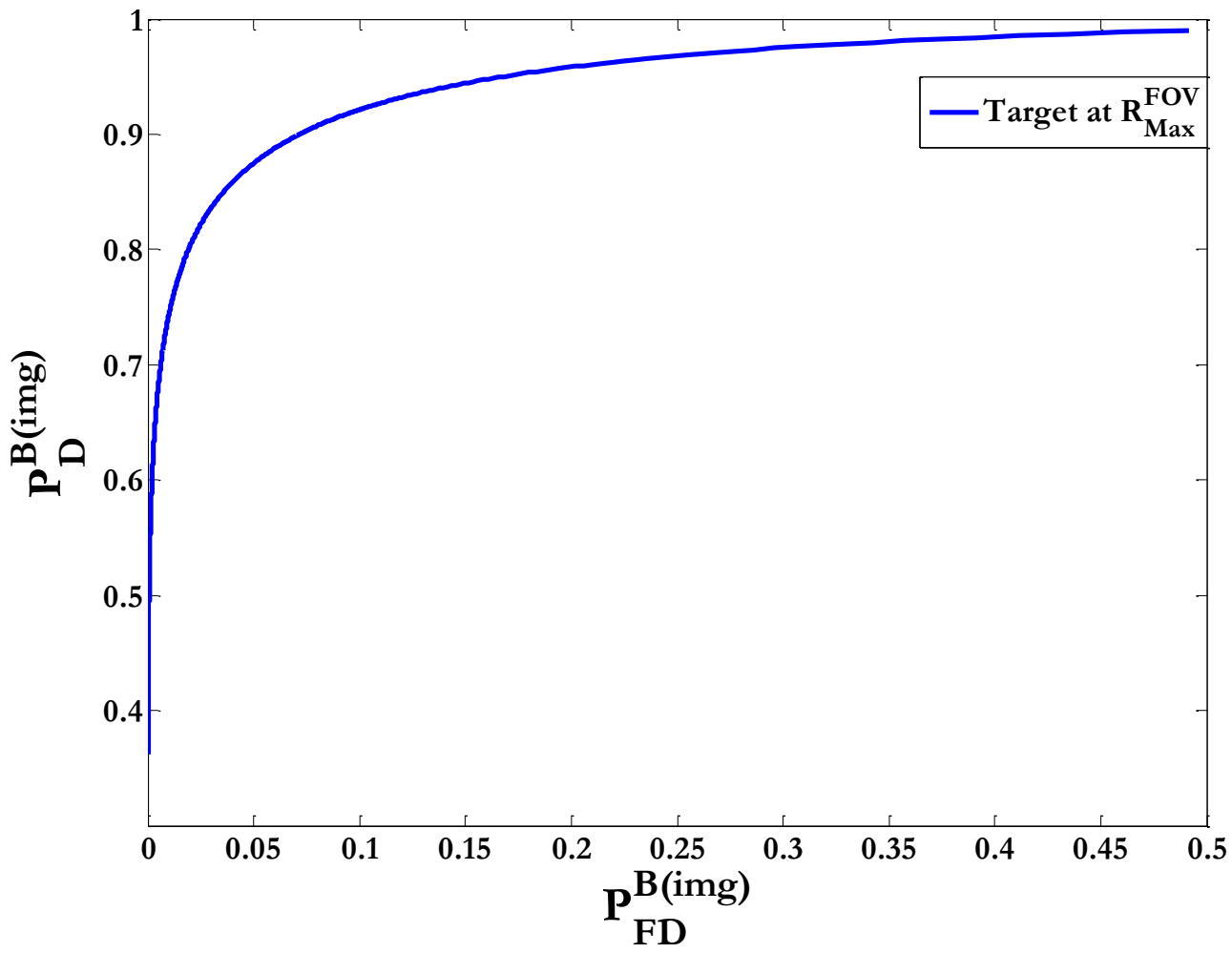

Fig.31: ROC of Bistatic MISO -IMG-LiDAL system.

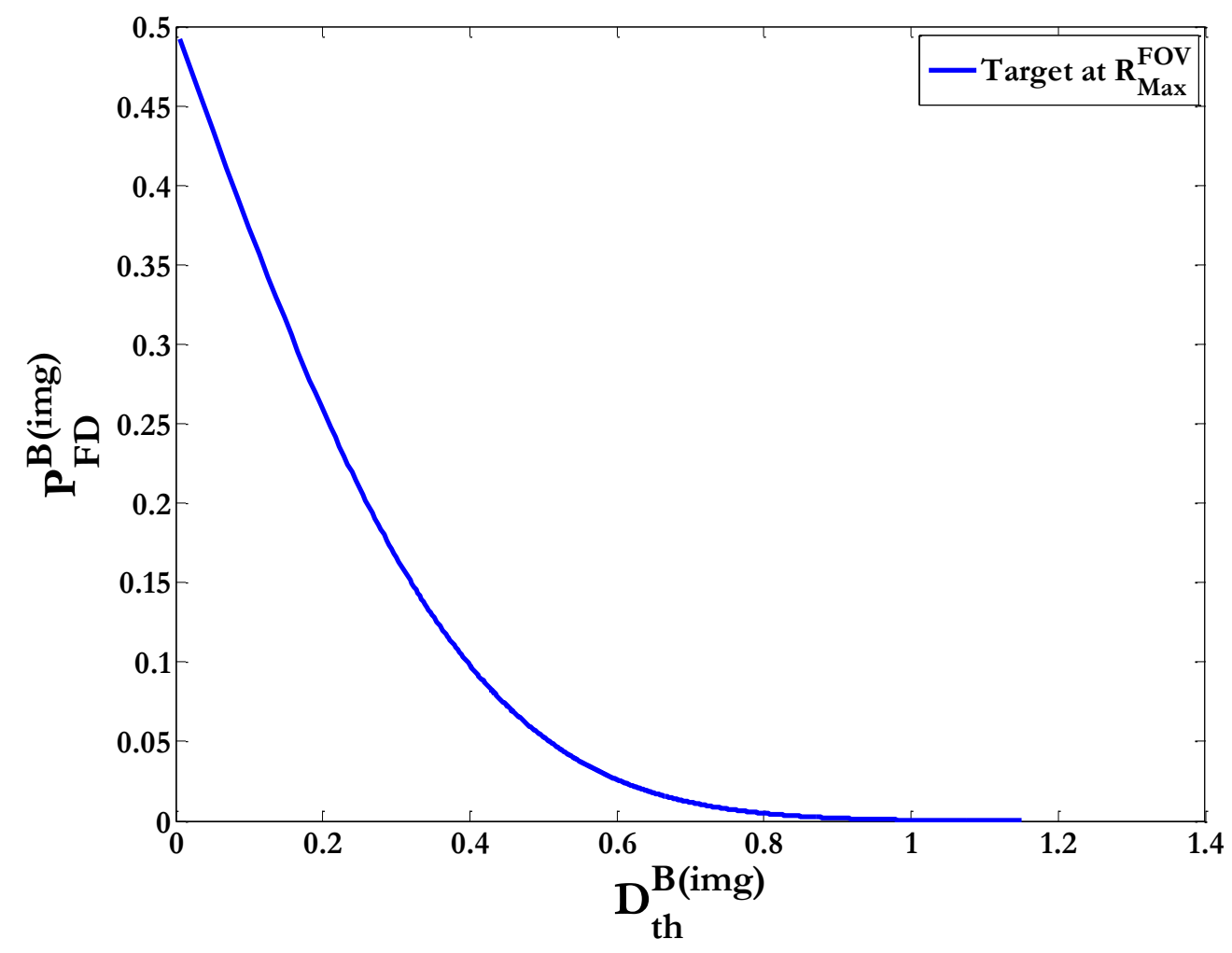

Fig.32: Bistatic MISO-IMG-LiDAL false detection with optimum $D^{B_{img}}_{th}$.

## B. *Target localization*

Target localization in MISO-IMG-LiDAL relies on the calculation of the direction of arrival (DOA) of the reflected signal arriving at the pixel's centre. The elevation and azimuth angles of the pixels are determined based on the design specifications of the imaging receiver with respect to the receiver's coordinates. However, the values of these angles are recalculated whenever the location of the receiver is changed (note that the receiver in our system is fixed in one location for a given room). The target position can be found by calculating the distance between the imaging receiver location $(x_r, y_r, z_r)$ and the centre of the target's pixel as shown in Fig. 33. The (range) distance $R_j$ between the ground reference point and the pixel's centre is given as:

$$R_j = \sqrt{\left(\frac{d_o - h}{\cos\left(E_{Lj}\right)}\right)^2 - (d_o - h)^2} \tag{118}$$

and the pixel $(x_j, y_j)$ coordinates are defined by:

$$x_j = R_j \cos\left(A_{z_j}\right) \tag{119}$$

$$y_j = R_j \sin\left(A_{z_j}\right). \tag{120}$$

The coordinates of target $k$, $P_k(x, y)$, are calculated with respect to the receiver ground reference centre point $L_o(x_o, y_o)$ (see Fig. 33):

$$P_k(x, y) = \left(x_o + x_j\right), \left(y_o + y_j\right) \tag{121}$$

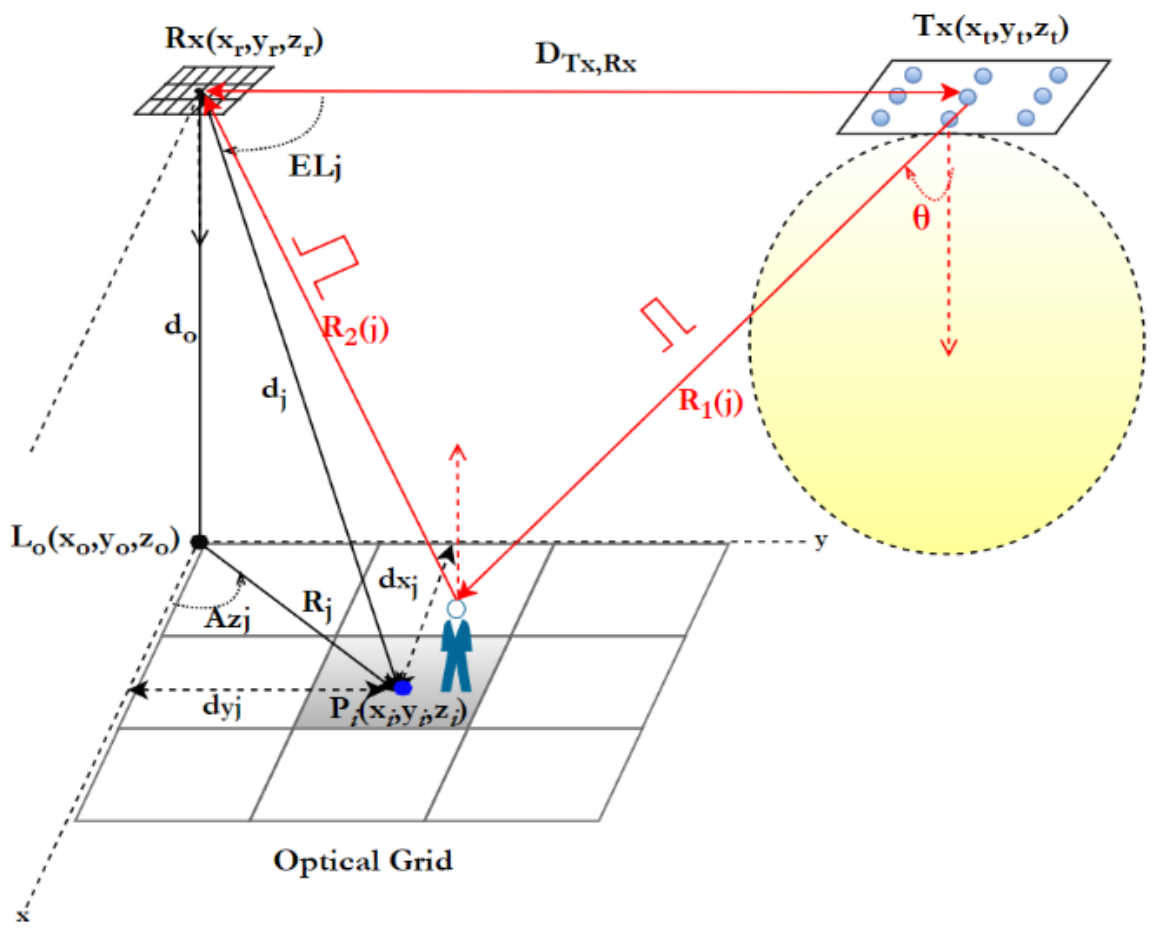

Fig.33: Target localization in MISO imaging-LiDAL.

The probability of localizing a target located at $R^{\text{FOV}}_{\text{Max}}$ ( see target in Fig.30) in imaging LiDAL system, $P_{L(img)}^{R^{\text{FOV}}_{\text{Max}}}$, can written as:

$$P_{L(img)}^{R^{\text{FOV}}_{\text{Max}}} = \text{erfc}\left(\frac{D^{B(img)}_{th} - \bar{P}_{r_{R^{\text{FOV}}_{\text{Max}}}}^{B_{img}}}{\hat{P}_{r_{R^{\text{FOV}}_{\text{Max}}}}^{B(img)} \sqrt{2}}\right). \tag{122}$$

## C. *Targets Detection in MISO-IMG-LiDAL*

The MISO-IMG-LiDAL system (which has many small optical detection footprints) has the ability to detect targets under different mobility schemes by tracking and marking the target in the imaging optical detection grid, see Fig. 34. In MISO-IMG-LiDAL, detection is accomplished using snapshot measurements, considering the change in the received reflected signals observed by the pixels due to target motion. When the target moves a distance more than the spatial distance of the imaging receiver $\Delta s$, the target is distinguished by monitoring the change that occurs in the pixels in at least two Imaging LIDAL scans (snapshots). We identify the change between pixel snapshot measurements using a pixels cross-correlation method (PCCM) and a pixels subtractions method (PSM). PCCM is similar to the slow cross-correlation we discussed in Section IV, however in Imaging LiDAL we employ the correlation between the pixel snapshots instead of the time slots snapshots. Thus, the cross-correlation $\left(\mathcal{R}_{p_{r_k}, p_{r_{k+1}}}\right)$ between the $k^{th}$ received pixels snapshot and $S$ consecutive received pixel snapshots is given as:

$$\mathcal{R}_{p_{r_k}, p_{r_{k+1}}}\left(\tau_p | N_P\right) \triangleq \int_{-N_P}^{N_P} p_{r_{k+1}}\left(x_n - \tau_p\right) \sum_{k=1}^{S} p_{r_k}(x_n)\, dx_n \tag{123}$$

and

$$p_{r_k}(x_n) = \sum_{i=1}^{I_P} \sum_{j=1}^{J_P} \int_0^T p_{r_k}^{(i,j)}(t) + n(t) dt \tag{124}$$

where, $N_P$ is the total number of pixel receivers ($N_P = I_P \times J_P$), $i$ is number of pixels in row $i$, $i \in [1, ..I_P]$, $j$ is number of pixels in column $j$, $j \in [1, ..J_P]$, $x_n$ is the pixel number, $x_n \in [1, ..N_P]$ and $p_{r_k}(t)$ is the received reflected signal power in each pixel receiver. The pixel displacement factor ($\hat{X}_n$ ) can be defined as:

$$\hat{X}^s_n = \arg\max_{\tau_p}\left(\mathcal{R}_{p_{r_k}, p_{r_{k+1}}}\left(\tau_p | N_P\right)\right) \quad \hat{X}^s_n \in [1, .N_P - 1]. \tag{125}$$

When $\hat{X}^s_n$ is zero, this indicates no change between the received reflected pulses in the $x^{th}_n$ pixel. When $\hat{X}^s_n \neq 0$, target motion is observed from the $x^{th}_n$ pixel with a displacement number of $\hat{X}^s_n$ pixels. Similar to the CCM with time slots receiver, we define a weight $w^S_{x_n}$ for each pixel receiver to be employed with the pixel sub-optimum receiver. Thus $w^S_{x_n}$ is defined as:

$$w^S_{x_n} = \begin{cases} 0 \; if \; \hat{X}^s_n = 0 \\ 1 \;\; otherwise \end{cases}. \tag{126}$$

For PSM, the subtraction of the $k^{th}$ received pixel snapshot from $S$ consecutive received pixel snapshots can be written as:

$$p^{x_n}_{r_{S(k,k+1)}} = p_{r_{k+1}}(x_n) - \sum_{k=1}^{S} p_{r_k}(x_n). \tag{127}$$

The computed value $p^{x_n}_{r_{S(k,k+1)}}$ is used in the sub-optimum receiver to decide the presence or absence of the target.

Fig. 34 shows an example of pedestrian targets where the targets move on the detection floor of the MISO-IMG-LiDAL system with different mobility schemes. Targets 1, 2 and 3 are nomadic, pedestrian and 'power walking' (ie fast) targets respectively. As can be seen in Fig. 34, the motion of target 1 is distinguished through snapshots measurements of $k$=1 and $k$=2 where target 1 has moved from pixel (1,1) to pixel (1, 3). While observing snapshots $k$=2 and $k$=3, target 2 is detected and marked in pixel (1, 1) and no change occurs in pixel (1,3), the nomadic target. Thus the total number of marked pixels is two (counter value) indicating the presence of two targets and their locations. In snapshot $k$=4, target 2 moves to the location of target 1 (at the same narrow optical zone). In this case, a counting error occurs as the distance between targets becomes less than the radar resolution, as pixel (1, 3) now contains both targets. In snapshot $k$=4, target 3 enters the environment at (3, 7). In the next snapshots, comparing snapshot $k$=4 and $k$=5, the counter value is updated where the detection error that occurred at $k$=4 is now resolved due to the movement of target 2 away from target 1. Note that the nomadic target 1 has not moved at $k$=4 and at $k$=5, and is still at pixel (1, 3). The pedestrian target 2 has moved from (1, 3) at $k$=4 to (1, 4) at $k$=5. The power walking target 3 has moved from (3, 7) at $k$=4 to (3, 2) at $k$=5. A similar pattern continues, comparing $k$=6 and $k$=7.

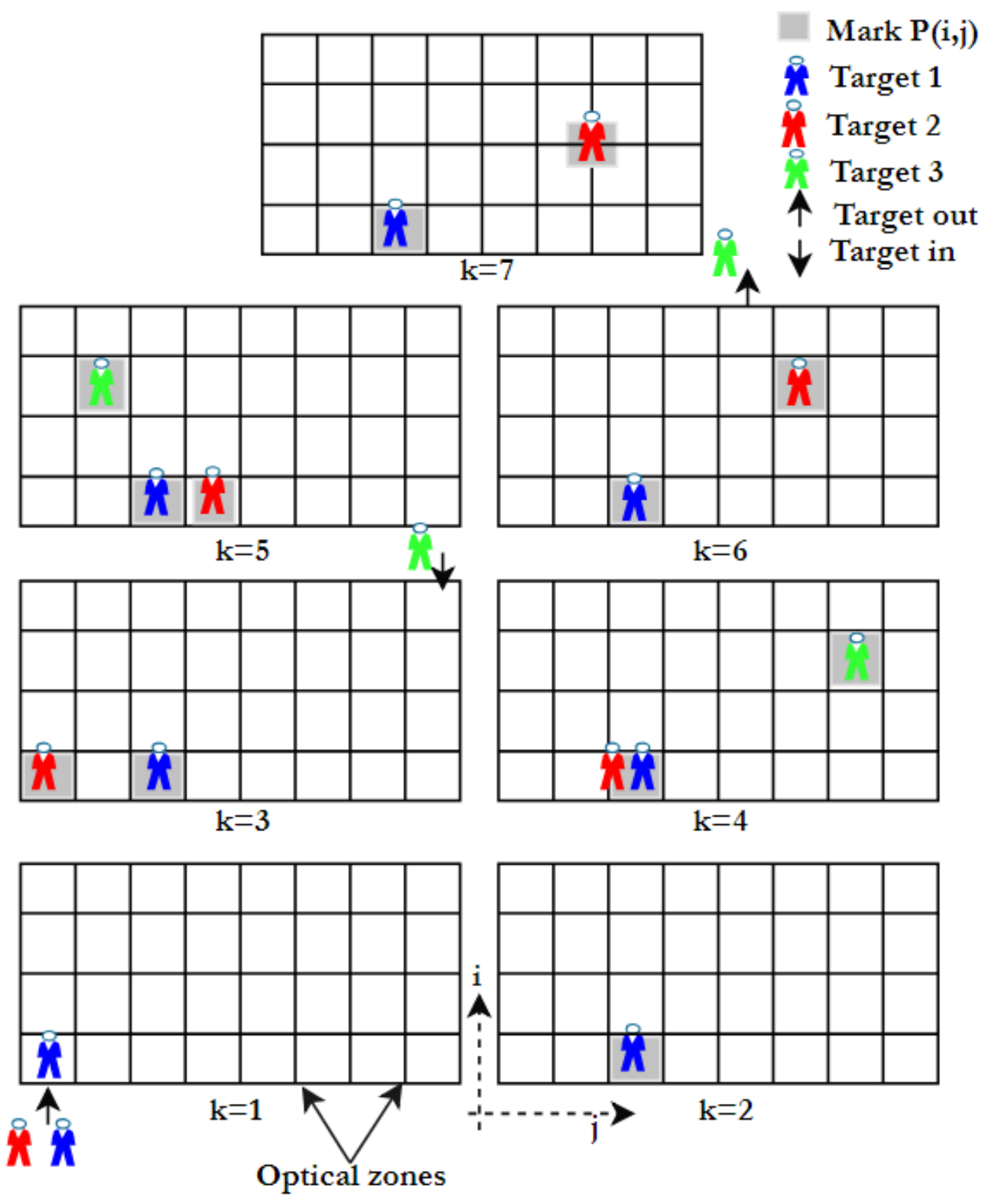

Fig.34: A top view of three targets movement on the detection floor of MISO-IMG-LiDAL radar system during $S$ snapshots measurements.

The main challenges of target detection in MISO-IMG-LiDAL are; (*i*) the transmitters have limited optical footprints and for coverage, these transmitter optical footprints overlap on the floor. Therefore, for target localization, the transmitters have to be turned on in turn to scan the entire room. A target located in the region where the optical footprints overlap, see Section V-B, can be counted more than a single time when it is reported by the scans associated with each transmitter. Such duplicate counting has to be removed; (*ii*) the receiver pixels cover finite regions on the floor, see Fig. 28. A target may be located at the intersection of up to four pixels, see target 2, at $k$=7 in Fig. 34. In this case, the issue is resolved by setting up a low and a high threshold as was done in Section III-B, where we dealt there with target overlap over multiple time slots and here we deal with target overlap in multiple pixels. A pixel reporting an output above the threshold contains a target, a target is absent if the signal is below the threshold and the pixel with the highest output energy is selected when multiple pixels have outputs between the thresholds.

Thus, in relation to challenge (*i*), and in order to eliminate multiple counting of a single target due to $L_{tx}$ active transmitters, we activate each transmitter individually and listen to reflections from the targets using the imaging receiver. To simplify the process, we note that each transmitter covers a finite optical footprint on the floor. Therefore, the only pixels that can possibly report a reflection are a group of pixels that cover the transmitter optical footprint on the floor. As such we divided our imaging receiver 128 pixels into 8 groups with 16 pixels per group. Here each group of receiver pixels (GRP), as can be seen in Fig.35a, covers one transmitter optical footprint, with 8 transmitters in our setup, see Fig. 28.

In relation to challenge (*ii*), the solution was described at top level above. Note that the signal at the output of each pixel is processed using an orthonormal expansion shown in Fig. 35 (b) which is an approach that follows our work in Section III-B translated from a time domain approach in III-B to a spatial approach at the pixel level in this section. Note that, the sub-optimum imaging receiver in Fig. 35 (b) collects signals from $N_P$ pixels. In terms of listening time, we considered one time slot ($T_s = T$) for each pixel receiver. Fig.35 (b) shows the sub-optimum imaging receiver (SOIMR) for the MISO-IMG-LiDAL system. The SOIMR has $N_P$ orthonormal functions $\phi_p(x_n)$ with integrators and comparators. The decision circuit decides as follows:

1. If the observed received signal $z_{x_n}$ is below the lower threshold, $D_{th_L}$, then the target is absent in pixel $(i, j)$, denoted here as pixel $x_n$.
2. If the observed received signal $z_{x_n}$ is above the higher threshold, $D_{th_H}$, then the target is present in pixel $(i, j)$, denoted here also as pixel $x_n$. Note that, both detection thresholds $D_{th_L}$ and $D_{th_H}$ have been optimized for the MISO-IMG LiDAL system in this section following an approach similar to that discussed in Section III-B.
3. If the observed received signal $z_{x_n}$ is above $D_{th_L}$ and below $D_{th_H}$, then it is a received reflected pulse from a target located within the FOVs of multiple neighbouring pixels. Thus the decision circuit compares $z_{x_n}$ with all possible neighbouring pixels and selects the pixel that has the largest $z_{x_n}$ as the pixel that contains the target. We considered a worst case scenario of three neighbour pixels as shown in Fig. 35 (b), where the decision circuit compares $z_1$ with its three neighbouring pixels $z_2, z_3$ and $z_4$ and choses the largest.

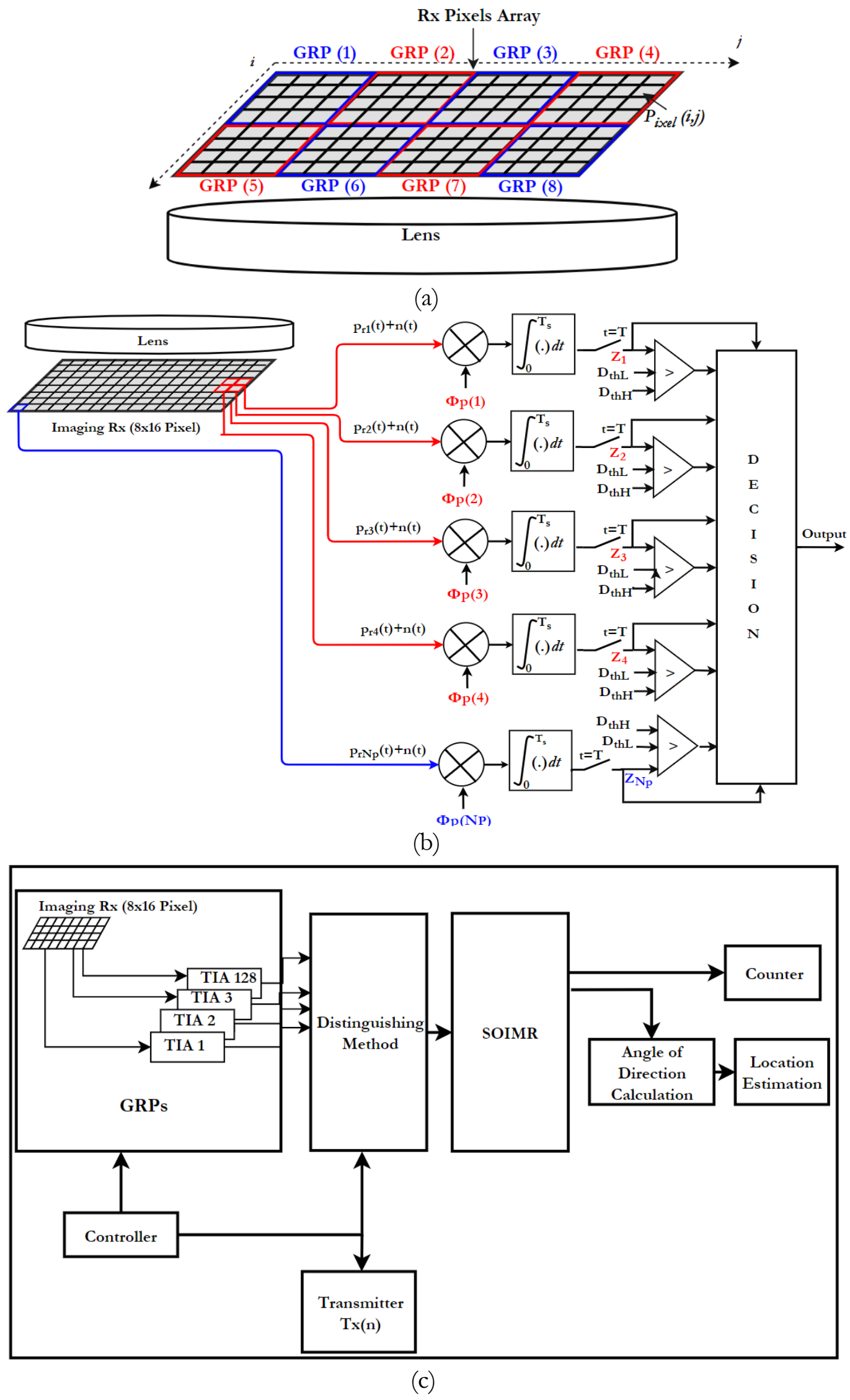

Fig.35: Imaging receiver design: (a) Eight GRPs of the imaging receiver, (b) the proposed sub-optimum imaging receiver (SOIMR) and (c) the receiver block diagram of MISO-IMG-LiDAL.

### *D. MISO-IMG-LiDAL System Operating Algorithm*

Fig. 35 (c) shows the schematic receiver diagram of the MISO –IMG-LiDAL system. The controller coordinates the detection, counting and localization processes as detailed below:

1) The controller activates transmitter $T_x(n)$ which sends an optical pulse, and also initializes the group receiver's pixels $GRP(n)$ to collected the reflected signals. We divided the imaging receiver pixels into $n = 8\ GRPs$ (see Fig. 35 (a)).
2) The controller then updates the value of $n$, and if $L_{tx} > n$ step (1) is repeated, where $L_{tx}$ is the number of active transmitter units ( $L_{tx}$ =8) of the MISO-IMG-LIDAL system.
3) A distinguishing method (PSM or PCCM) is applied with the SOIMR to process the received reflected signals from all pixel receivers to detect and count the targets.
4) Finally, pixel identification is carried out to estimate the target location.

For the MISO-IMG-LiDAL, the number of frames $M$ required to detect and localize targets is equal to 8. Therefore, the overhead occupied in the VLC system MAC frame can be calculated using (106) as discussed in Section V-D. Note that, the MISO-IMG-LiDAL system requires 8 observation frames to complete one monitoring cycle of the room, compared to 24 observation frames for the MIMO-LiDAL system. The reduction in complexity is always a factor of 3 regardless of the number of transmitters (optical footprints) used and regardless of the number of receiver pixels. This factor relates to the need for 3 anchors in the MIMO-LiDAL system.

Note that parallels can be drawn between our MISO-IMG-LiDAL system and camera imaging sensors in the sense that an image sensor mounted on the ceiling can also localize a target. The main differences between our MISO-IMG-LiDAL system and traditional image sensors are: (*i*) simplicity, an imaging sensor based on commercial camera components may typically have 16 million pixels to resolve the minute image details. In our application, localization to within a distance of about 30 cm is sufficient. Therefore the number of pixels is reduced to about 128 pixels, a factor of 125k reduction in complexity, (ii) with reduced number of pixels, high speed photodetectors and wideband optical receivers can be used leading to a localization system that can detect fast moving targets, which become increasingly important in applications such as robotics, (iii) If our space-based MISO-IMG-LiDAL system is combined with our time domain approach of Section V, then the pixels determine the target location in two dimensions (ie on the floor) while the time delay between the transmitted pulse and the pulse received by the pixel determines the distance of the target. Thus this combined system can localize the target in three dimensions while image sensors localize targets in two dimensions.

## VII. SIMULATION SETUP, RESULTS AND DISCUSSION

In this section, we describe the simulation settings, describe three scenarios and a case study which are used in this paper to evaluate the proposed LiDAL systems in terms of targets detection and localization. The LiDAL systems are evaluated through computer simulation using MATLAB®. The three scenarios and case study are as follows: (i) *the first scenario establishes the baseline*, ie the best performance expected in our LiDAL systems. It evaluates the performance of our LiDAL systems in an ideal environment where obstacles (furniture) are absent hence reducing interference from the environment, and reducing the likelihood of confusing a target (human) with furniture (obstacles). It also considers continuous motion, where pedestrians move continuously hence helping the target distinguishing methods; (ii) *the second scenario represents a challenging localization setting,* which is a realistic but also favourable localization environment. It introduces the first major impairment to localization in LiDAL, ie the presence of obstacles. Therefore, this scenario considers a realistic room with furniture, partitions, bookshelves, doors and windows, unlike the empty room of scenario (i). Scenario (ii) however continues to consider continuous pedestrian motion to support the target distinguishing methods, thus allowing the impact of obstacles to be studied in isolation, and in this sense it is a favourable environment; (iii) *the third scenario represents a harsh localization environment.* It adds nomadic motion to the second scenario and therefore considers the two main impairments in LiDAL localization jointly; namely the presence of obstacles and lack of motion (sometimes) which makes target distinguishing harder. In all three scenarios we evaluate the results while using BSM and CCM for target distinguishing where mobility is the input to these methods. We also evaluate results in the three scenarios for the two systems of interest: MIMO-LiDAL and MISO-IMG-LiDAL; (iv) Finally we consider a case study where a realistic office environment is considered with pedestrian arrivals, departures, nomadic behaviour, pathway mobility and a finite evaluation window of one hour when the office is evaluated. In this case we use the better target distinguishing method, namely CCM and evaluate both systems: MIMO-LiDAL and MISO-IMG-LiDAL.

### *A. Systems setup*

In this section we introduce the systems setup and the parameters used. The LiDAL systems were evaluated in two types of environments. Room A is an environment of the same size as the room in Fig. 1(a) but is an empty room (i.e. free from furniture). Room B is a realistic environment, a furnished office, as shown in Fig. 1(a). Table IX illustrates the simulation parameters used in LiDAL systems.

TABLE IX
SIMULATION PARAMETERS

| Parameters | Configurations |
|---|---|
| Room A and B | |
| Length | 8m |
| Width | 4m |
| Height | 3m |
| $\rho$ - ceiling | 0.8 |
| $\rho$ - floor | 0.3 |
| $\rho$ - walls | 0.8 |
| LiDAL Transmitter Units | |
| locations (*x*, *y*, *z*) | (1,1,3), (1,3,3), (1,5,3), (1,7,3) (3,1,3), (3,3,3), (3,5,3), (3,7,3)m |
| Elevation | $90^0$ |
| Azimuth | $0^o$ |
| RGB-LDs in each unit | 9 (3×3) |
| Transmitted optical power per unit | 18 W |
| Transmitted Pulse width $\tau$ | 2ns |
| RGB-LD semi-angle at half power beam width ($\Phi$) | $75^o$ |
| MIMO LiDAL Receiver | |
| Photodetector Area | 20 $mm^2$ |
| Receivers locations | Attached with Tx units |
| Photodetector Responsivity | 0.4 A/W |
| Receiver Acceptance Semi-angle | $43.8^o$ |
| CPC Reflective Index (*N*) | 1.7 |
| TIA Noise Current | 2.5 pA/√Hz |
| MISO IMG LiDAL Receiver | |
| Photodetector Area | 2$cm^2$ |
| Receiver location (*x*, *y*, *z*) | (2,4,3)m |
| Number of pixels | 128 |
| Pixel's area | 1.56 $mm^2$ |
| TIA Pixel Receiver Noise Current | 2.6 pA/√Hz |
| Lens FOV | $72^o$ |
| Time Bin Duration | 0.01 ns |
| Sampling Time $T_{sa}$ | 0.1ns |
| Time Slot Width $T_s$ | 2ns |
| Listening Time $T$ | 1ms |

To evaluate the counting and localization performance of the different LiDAL systems two key metrics are defined: (i) The mean absolute percentage error (MAPE) which is used to quantify the counting accuracy, and (ii) the distance root means square error (DRMSE) which is used to quantify the localization accuracy. The counting performance of the LiDAL systems is measured in terms of MAPE which is given as [74] [75]:

$$MAPE = \frac{1}{J}\sum_{j=1}^{J}\left|\frac{A\,(j)-N_E(j)}{A(j)}\right| \times 100\% \qquad (128)$$

where $J$ is the number of times the experiment is repeated (iterations), $A\,(j)$ and $N_E(j)$ are the targets' actual and estimated (by LiDAL systems) numbers, respectively. In order to evaluate the

**TABLE X: SIMULATION FLOW**

**Inputs:** $i_{max}$ =*K*; (Maximum number of targets)
$j_{max} = Itr$ ; (Number of iterations )
$N_E(j)$ is the estimated number of targets at iteration $j$.
$E_l(k,j)$ is the estimated location of target $k$ at iteration $j$.
$\rho\ (k,i)$ is target $k$ reflection factor when an environment with $i$ targets is considered

**for** $i$ = 1: $i_{max}$;
  **for** $j$= 1: $j_{max}$;
    Generate a random location(s) $l(k,j)$ and $\rho\ (k,i)$ for target(s) $k \in [1,..i]$
    Generate additive white Gaussian noise $n_j(t)$
    Apply LiDAL system detection algorithm
    Hence determine $N_E(j)$ and $E_l(k,j)$
    $j == j_{max}$
  **end for**
  Calculate MAPE
  Calculate DRMSE
  save MAPE and DRMSE at given value of $i$
  $i == i_{max}$
**end for**

localization performance of the LiDAL systems, DRMSE is used to measure the location accuracy, where DRMSE is given as [75] [76]:

$$\text{DRMSE} = \sqrt{\sigma_x^2 + \sigma_y^2}. \qquad (129)$$

Here $\sigma_x$ and $\sigma_y$ are the error standard deviations associated with the estimated $(x_e, y_e)$ and the actual $(x_a, y_a)$ coordinates of the target, respectively.

The three scenarios and the case study were evaluated using the simulation flow shown in Table X. The simulation starts by considering an indoor environment that has $i$ targets where the maximum number of targets is $i_{max}$ =*K*. It then considers a number of iterations where each iteration contains the same number of targets, however the targets are located at different random locations in each iteration. The iterations continue to $j_{max}$ = $Itr$. For a given number of targets, each iteration then generates random target locations, noise and reflection coefficients for each target (cloth colour and texture). The reflection coefficient associated with each target remains fixed for the number of iterations considered. The LiDAL system detection algorithm is then invoked resulting in estimated number of targets, $N_E(j)$, and estimated target locations, $E_l(k,j)$. This is finally used at the end of the $j_{max}$ iterations to calculate MAPE and DRMSE. The simulation then continues by considering more targets in the environment (with new reflection coefficients (clothing) for the targets) and full number of $j_{max}$ iterations.

The human target dimensions in Fig. 1 (b) are 48cm × 15cm. If a 50cm spacing is considered between targets, then the area needed per human target is 98cm × 65cm = 0.63$m^2$. This leads to a maximum number of targets in an 8m×4m room of 51 targets. This represents a very dense reception type event. As discussed in Section IV-C, the European standards for the minimum workplace space required per person is 3.7$m^2$ for an office environment and 2$m^2$ for a meeting room [65]. Therefore, we considered a 2$m^2$ space requirement per person, leading to a maximum of 16 targets in an 8m×4m room. Therefore, different number of targets, up to 15 targets, were considered in our simulations.

We next consider the three scenarios and case study.

### *B. Scenario 1: The baseline*

In this scenario we considered a room that has no obstacles, ie the room is empty and no furniture is considered. We also assumed perfect mobility conditions for the mobile targets (i.e. pedestrian targets with a speed of 1m/s). These targets were randomly and uniformly distributed on the detection floor with minimum inter-target-distance of 0.5 m. We considered a normal random distribution for the target reflection factor based on the proposed model in Fig. 6.

Fig. 36. depicts the counting error, MAPE, of the LiDAL systems tested in scenario 1. As can be seen in Fig. 36 the MAPE of the MIMO-LiDAL system with BSM for a single target is about 0.5% which is comparable with the probability of miss-detection of a single target in equation (105) with $P_{M(MIMO)}$ =0.016. This agreement is a useful verification of our analytic results and simulations, where the single target case can experience errors due to the randomness associated with the target reflection coefficient and the noise in the receiver and environment. The MIMO-LiDAL system MAPE reaches 7% in the presence of 15 targets with BSM. Fig. 36 shows an increase in MAPE with increase in the number of targets. This increase in MAPE can be attributed to a number of factors: (i) with increase in the number of targets, the room clutter increases with more objects (targets) acting as reflectors. Signals from LiDAL are reflected by the desired target and by other targets as well as secondary subsequent reflections from the walls. This increases the probability of error in counting the targets; (ii) with a larger number of targets, there is a higher potential for targets to occur either at the optical footprint overlap zones of MIMO LiDAL (see Fig. 21) or between up to four pixels in the MISO-IMG-LiDAL (see Fig. 34). These locations are the most challenging for the LiDAL localization systems.

In addition, the MIMO LiDAL system performance when the BSM is used for target distinguishing (ie using mobility to distinguish human targets from obstacles) is worse than the performance when the CCM is used. This is due to the increase in the inter-targets-interference (due to increased reflections) in the presence of more targets. Note that BSM and CCM perform comparably at lower number of targets, with the performance gap increasing with increase in the number of targets. As can be noted in Fig.36, the MAPE range for the MIMO-LiDAL system with CCM was from 0.3% to 5%. It is clear that the CCM has better performance than BSM as the inter-targets-interference does not affect the performance of CCM to the same extent.

The best localization results are due to our MISO-IMG-LiDAL configuration as can be seen in Fig. 36. The MAPE associated with MISO-IMG-LiDAL with BSM for single target detection is about 0.8% which is comparable to $P_M^{B(img)} = 0.1$ (see Fig. 31 with $P_M^{B(img)} = 1 - P_D^{B(img)}$). The MAPE range of MISO-IMG-LiDAL is from 0.8% to 3.5% with BSM, and 0.6% to 3% with CCM as seen in Fig. 36. Compared to the MIMO-LiDAL system, the MISO-IMG-LiDAL has better performance due to the ability of the latter to use the spatial dimension to resolve the ambiguity of targets (i.e. separate the targets using multiple pixels that have distinct narrow optical footprints). Due to the spatial resolution of targets, the MISO-IMG-LiDAL system has comparable performance under the BSM and the CCM, with a slight difference of 0.5% in MAPE where the CCM performs better.

### *C. Scenario 2: Challenging localization environment*

This scenario represents a challenging environment where obstacles (furniture and other objects) are present as seen in Fig. 1 (a), where the obstacles can reflect the LiDAL signals in a fashion similar to human targets. Continuous pedestrian motion is however considered, and therefore the environment is favourable from the point of view of being able to distinguish human targets from stationary obstacles.

Fig. 37 presents the MAPE associated with the LiDAL systems for targets in scenario 2. One can observe that the MAPE increased significantly for MIMO-LiDAL with BSM from its range of 0.5% to 7% in scenario 1 to a new range of 6% to 35% in scenario 2. Similarly, under MISO-IMG-LiDAL with BSM, the MAPE increased from its previous range of 0.3% to 5% in scenario 1, to a new range of 5.5% to 22%. This is due to the presence of obstacles (furniture) in scenario 2 and due to the poor performance of BSM in a furnished environment due to the interference from the reflections attributed to background obstacles and furniture. Furthermore, in the presence of furniture, the residual space available for human motion is reduced, even when targets move continuously. This leads to impaired performance of BSM and CCM. In the MIMO-LiDAL system with the CCM, the MAPE was 1% to 5% in scenario 1, and increased to 4% to 16% in scenario 2. The best system in both scenarios is the MISO-IMG-LiDAL with CCM. This system saw its MAPE increase from a "0.5% to 3.5%" in scenario 1 to "2% to 12%" in scenario 2 due to the presence of obstacles and their associated reflections and due to the reduced residual space available for human motion. It is worth noting that the other general trends are comparable in the two scenarios, with the MAPE performance deteriorating with increase in the number of targets, and improving with the use of the imaging system and the CCM.

Fig. 38 shows the cumulative distribution function of the DRMSE positioning error for the MIMO-LiDAL system. Fig. 38 presents the CDF of targets successfully detected in scenario 1 and scenario 2. As can be noted, the 95% CDF confidence interval is at 0.45m and 0.5m positioning error for scenarios 1 and 2 respectively, while the average DRMSE is 0.28m and 0.38m respectively. The results in Fig. 38 clearly show that the DRMSE is larger in scenario 2 due to the presence of obstacles and hence the potential for such obstacles to be confused with human targets. The positioning error in MIMO-LiDAL occurs due to wrong decisions in the sub-optimum detector when it identifies the time slot that contains the signal reflected from the target. One wrong time slot leads to a 0.3m $(\Delta R = 0.3m)$ change in the error associated with the range to the anchor.

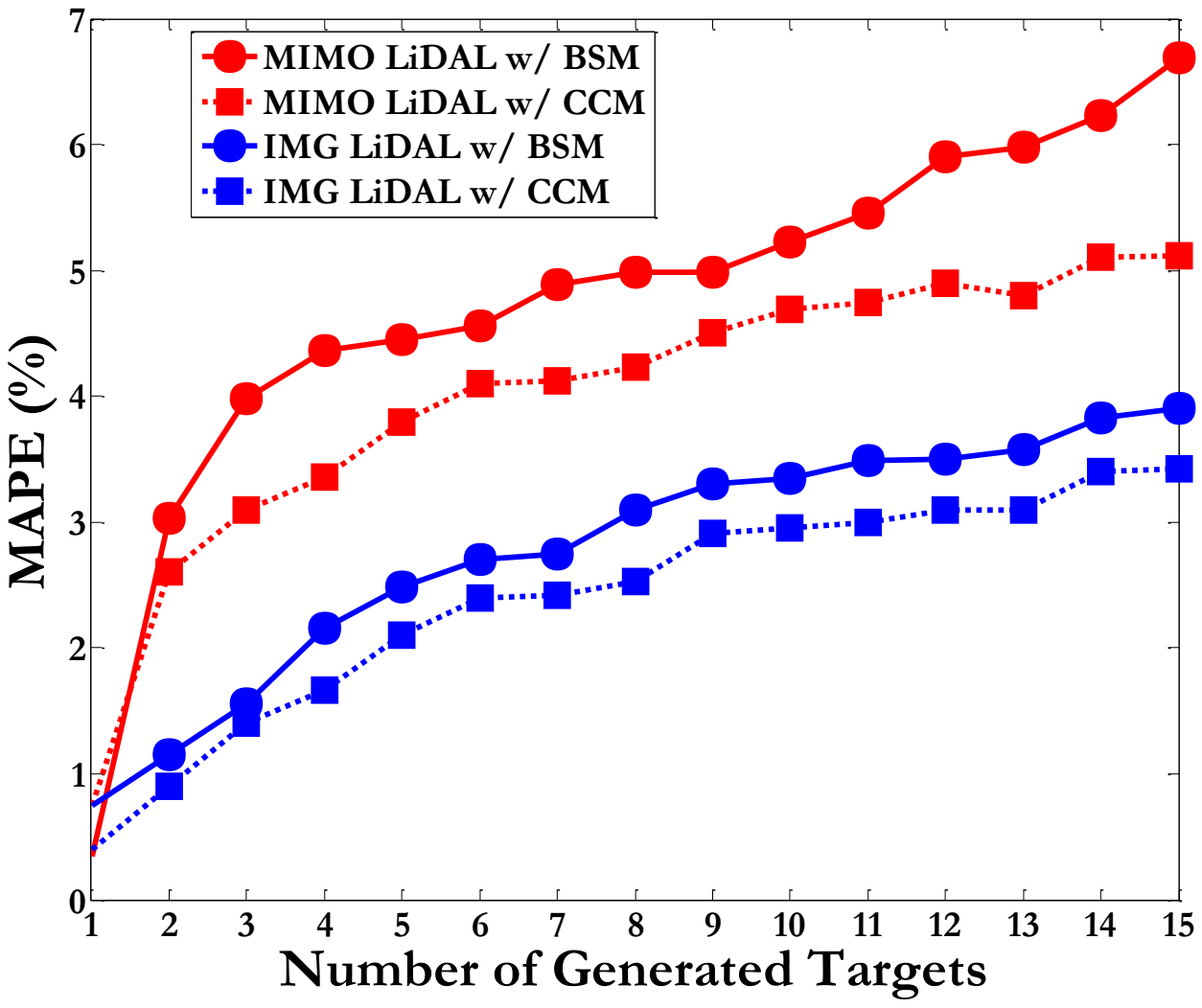

Fig.36: MAPE of LiDAL systems with BSM and CCM in the empty environment of scenario 1.

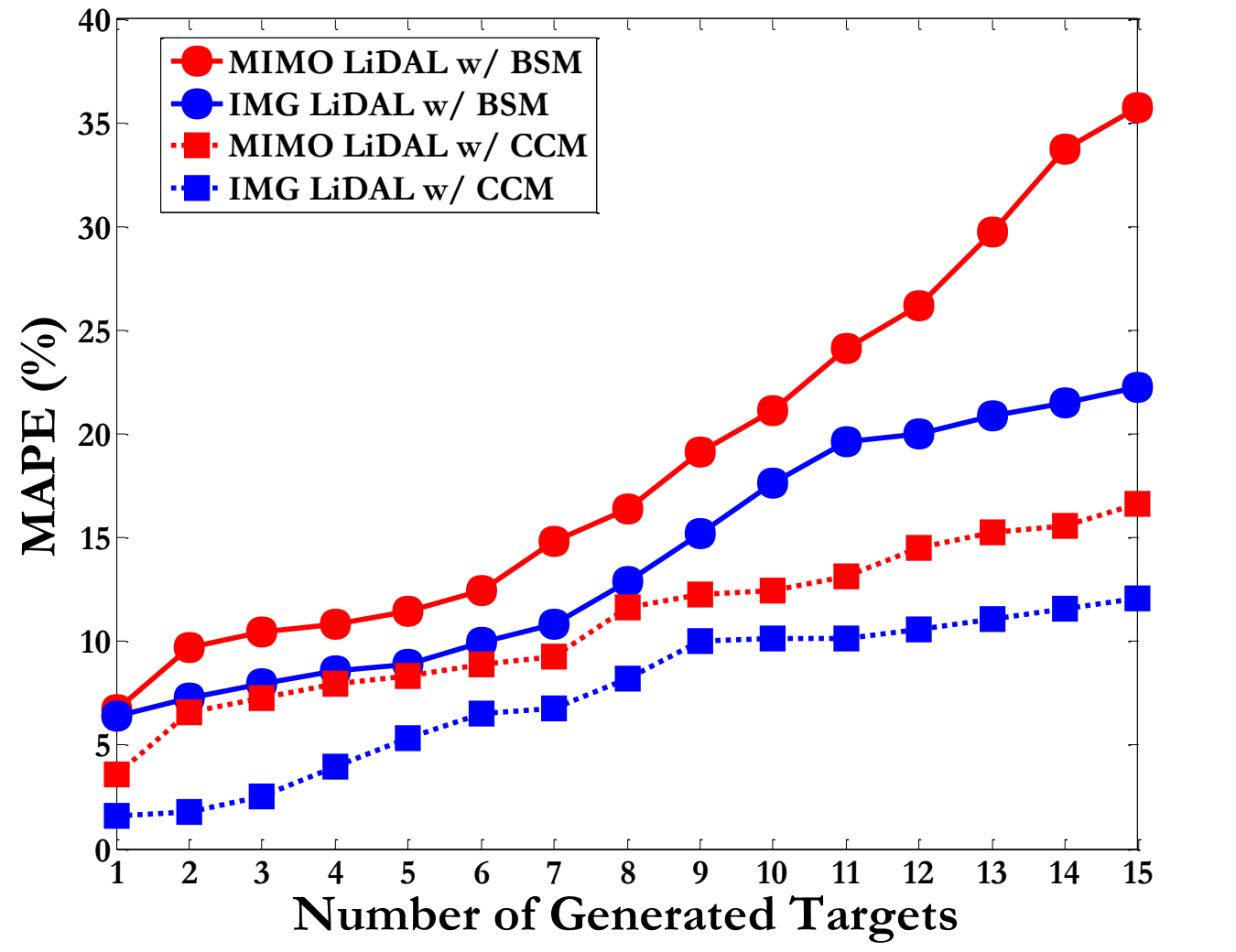

Fig.37: MAPE of LiDAL systems with BSM and CCM in the realistic environment of scenario 2.

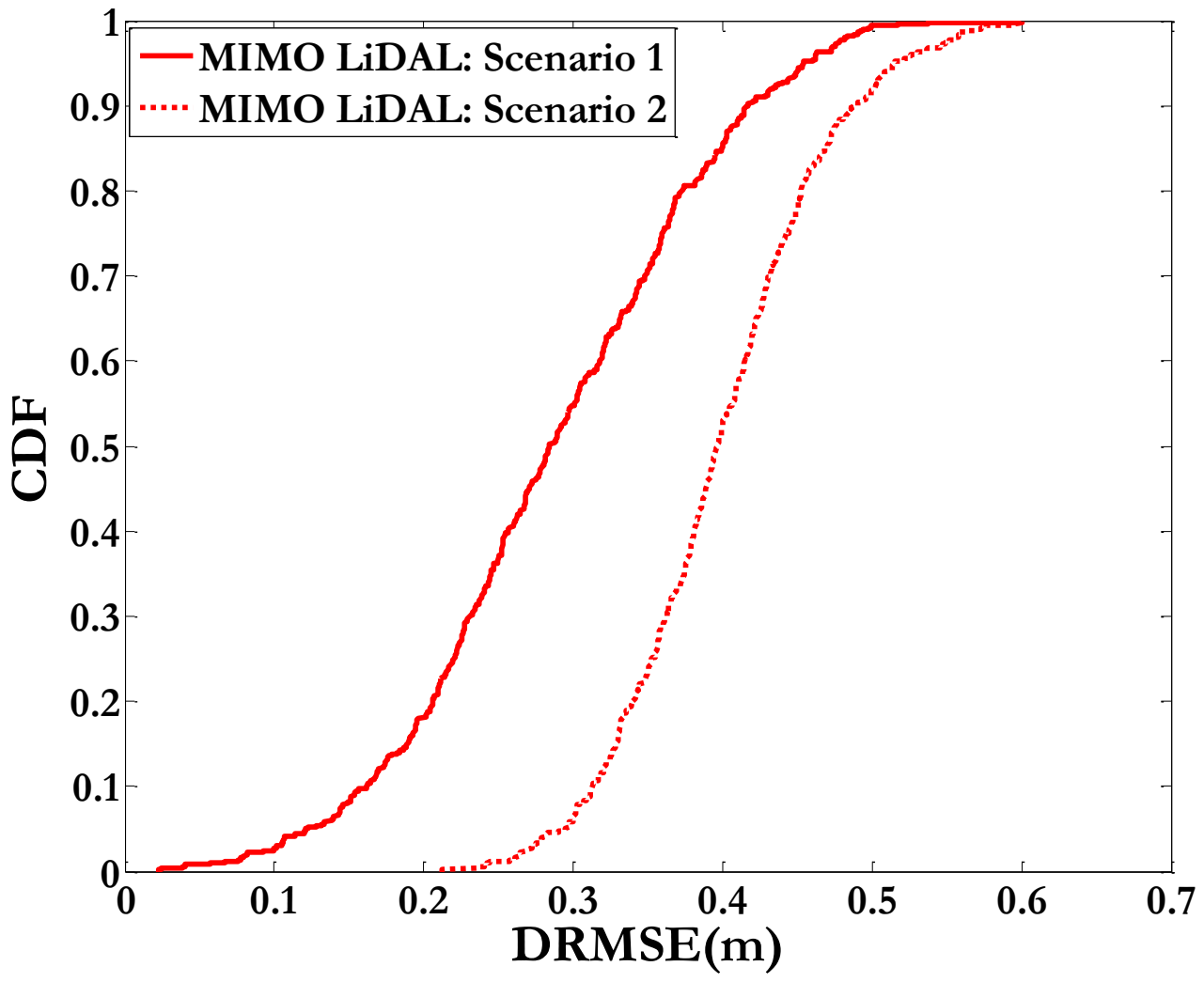

Fig.38: CDF of DRMSE of the proposed MIMO LiDAL system.

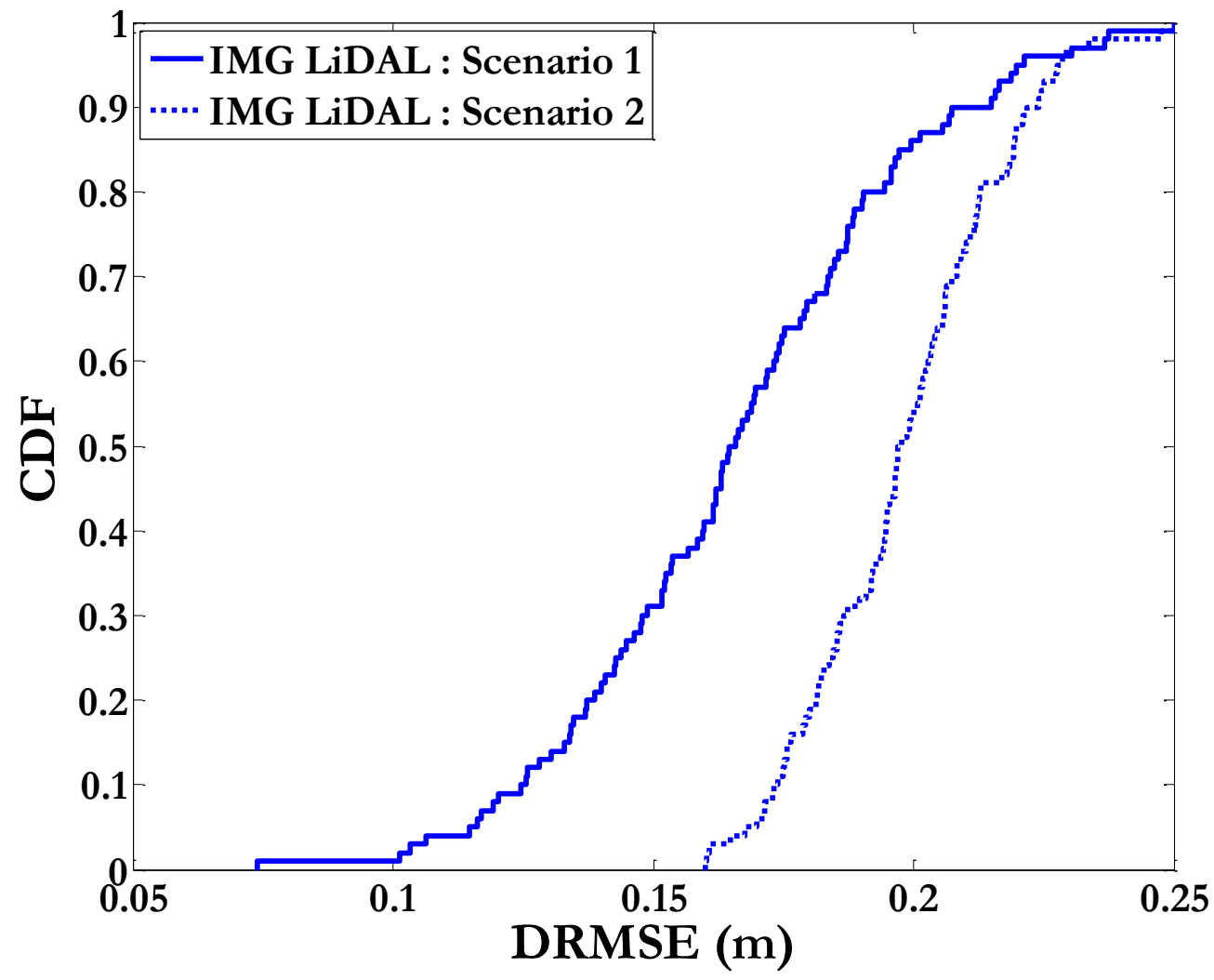

Fig.39: CDF of DRMSE of the proposed IMG LiDAL system.

The DRMSE CDF results associated with MISO-IMG-LIDAL in scenarios 1 and 2 are shown in Fig. 39. It should be observed that overall, the DRMSE values in MISO-IMG-LIDAL are smaller than the corresponding values in MIMO-LiDAL due to the enhanced resolution of the imaging receiver which resolves the target locations spatially into pixels, whereas the MIMO-LiDAL system relies on three ranges that have to be determined accurately, with the potential for wrong slot errors in the three ranges. In the MISO-IMG-LiDAL system, at the 95% confidence interval, Fig. 39, the DRMSE are 0.21m and 0.23m for scenarios 1 and 2 respectively, whereas the average values of DRMSE are 0.16m and 0.19m for scenarios 1 and 2 respectively. The sources of error in MISO-IMG-LiDAL are attributed to noise, reflections, and targets random reflection coefficients. These sources of error can translate in the worst case into targets appearing at the intersection of up to four pixels, or targets assumed to be located at the centre of the coverage area of each pixel on the floor when the target may be at the edge of the pixel coverage area.

*D. Scenario 3: harsh localization environment*

In this scenario, the LiDAL systems experience both impairments, namely the presence of obstacles (as in scenario 2) and nomadic mobility. Therefore, unlike scenario 2, the users can be stationary for periods of time and therefore the LiDAL systems are not able to distinguish such stationary targets from obstacles, i.e. furniture (in most cases, except when tracked using imaging receivers). To quantify the extent of nomadic behaviour, we define a mobility factor (MF) given by

$$MF = \frac{T_{ob} - \sum_{d=1}^{L_D} t_d}{T_{ob}} \quad (130)$$

where $t_d$ is the time spent by the nomadic target in location $d$, which is a location of interest among the $L_D$ locations of interest. Therefore, a MF=1 indicates a pedestrian target, ie a target that is in continuous motion as in scenarios 1 and 2. A MF that approaches zero, indicates a target that is fully nomadic, ie a target that spends most of the time stationary in a number of locations.
Fig. 40 shows the MAPE for a MIMO LiDAL system where obstacles (furniture) are present as well as nomadic target behaviour. The MAPE decreases with increase in the MF as it becomes easier for the target distinguishing methods to distinguish targets from stationary obstacles. The results in Fig. 40 used the CCM for target distinguishing.

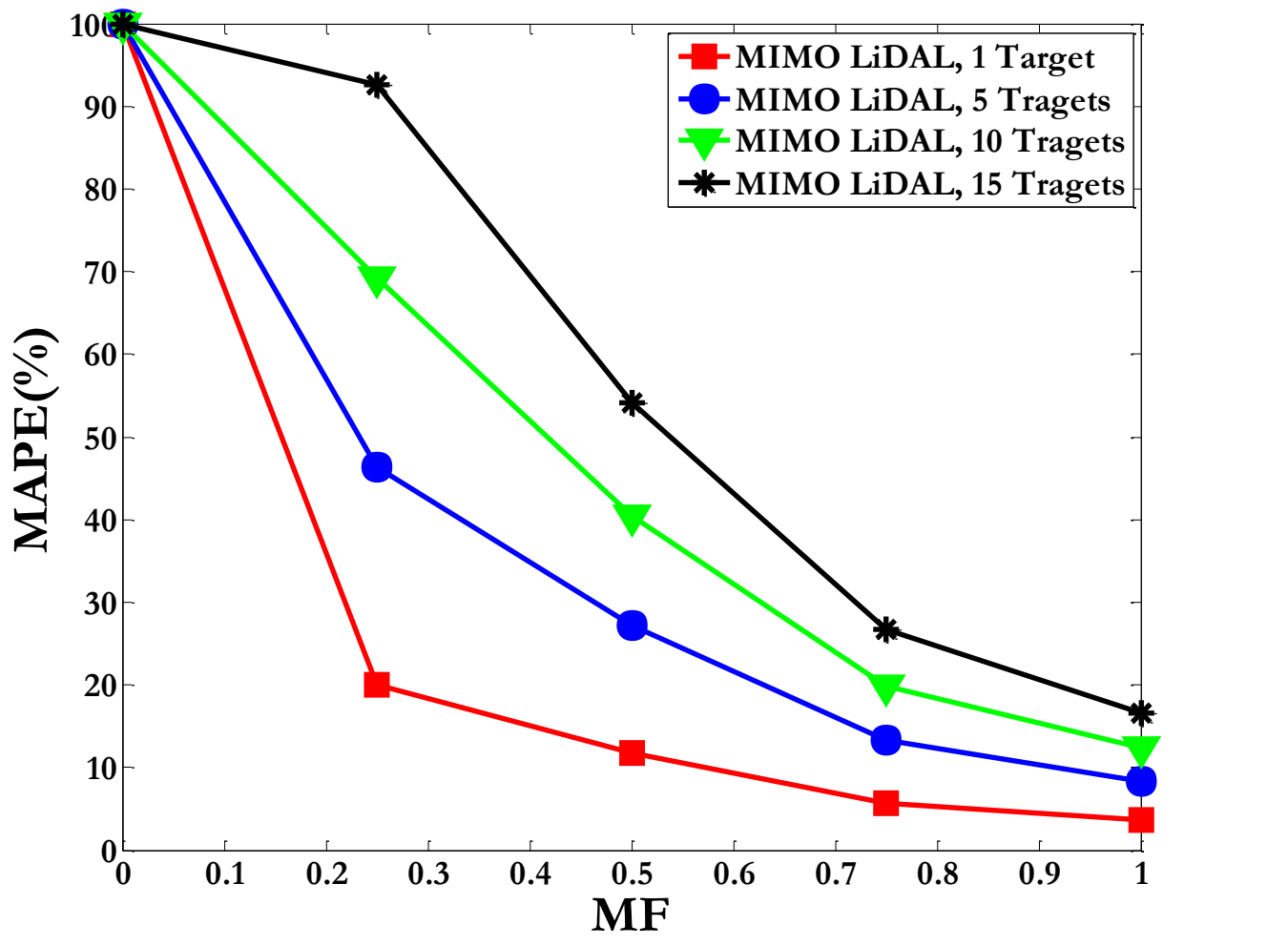

Fig.40: CDF of counting MAPE in the MIMO LiDAL system for nomadic targets with different MF.

For a given value of MF, ie for a given level of nomadic behaviour, the MAPE decreases with decrease in the number of targets as was observed in scenario 2. It is worth observing that a MPAE of 20% or less is only achieved in the MIMO LiDAL system for mobility levels that correspond to MF approaching one.

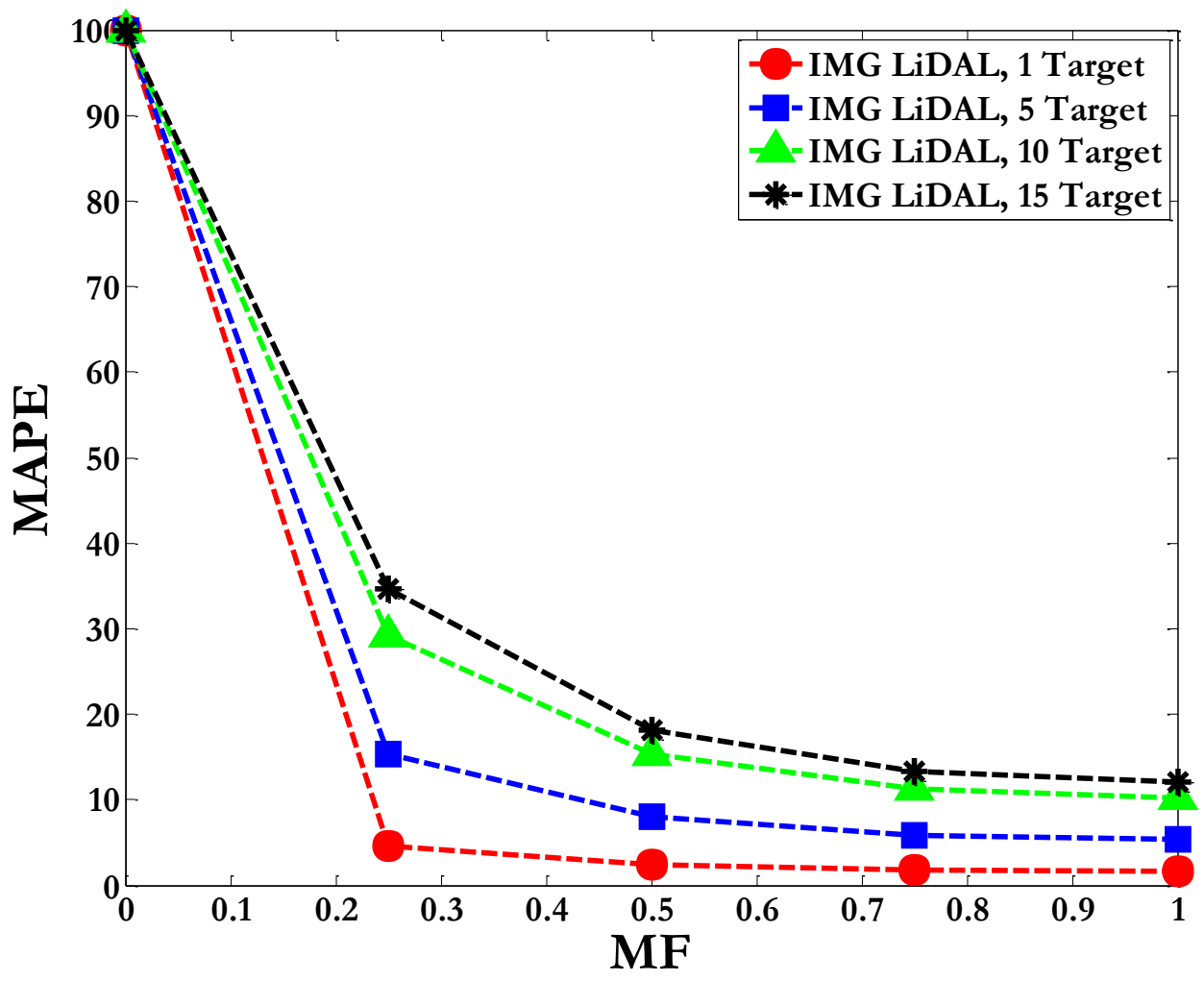

Fig.41: MISO-IMG LiDAL system MAPE CDF for nomadic targets with different MF.

Fig. 41 shows the MAPE CDF for MISO IMG LiDAL where nomadic behaviour is now considered. The MAPE decreases with increase in mobility, ie increase in the MF and also decreases with decrease in the number of targets that can cause clutter. The most important observation however is that the MAPE in the MISO IMG LiDAL system is much lower than that of the MIMO-LiDAL system. This is attributed mainly to the improved ability of the imaging receiver to resolve targets in space and subsequently track these targets as the targets move from pixel to pixel. This also means that a stationary target that was mobile at a previous point in time, continues to be marked as a target in a new pixel. This reduces the MAPE by correctly identifying targets from obstacles. For example, for MAPE of 20% or less a MF of 0.5 or higher is sufficient.

### E. *Case Study*

In this case study, we extend the cases we considered in the three scenarios. We build on scenario 3, namely, the case study considers obstacles and nomadic behaviour. The case study however extends scenario 3 in a number of ways. In particular, we consider (i) arrival and departure processes for human targets into and out of the office environment (not considered in scenario 3); (ii) obstacles (furniture as in scenario 3); (iii) challenging nomadic mobility behaviour, (nomadic pathway mobility (not considered in scenario 3) and random walk with nomadic behaviour (this was considered in scenario 3)); (iv) one hour evaluation period (new in the case study); (v) both MIMO-LiDAL and MISO-IMG-LiDAL systems with the better CCM for mobile target distinguishing.

The parameters used in the case study are shown in Table XI. The arrival and departure rates into and out of the office environment are 12 arrivals per hour and 14 departures per hour following a Poisson distribution as outlined in Section IV-C. This leads to an average of 30 minutes spent in the environment, with an average of 6 targets present in the environment as shown in Section IV-C and in Table XI.

The case study considers both pedestrian targets who move at 1m/s and nomadic targets who move at 0.5 m/s – 2 m/s when moving between locations of interest as shown in Table XI and in Section IV-C. We considered 9 locations of interest in the room where the nomadic user spends random and uniformly distributed times.

The simulation time was $T_{ob}$ equal to one hour. The LiDAL frame duration is 1ms as discussed in Section V-D, where at the start of the frame the LiDAL system, carries out its transmissions and measurements as discussed to determine the targets locations. LiDAL localization measurements are not carried out in each LiDAL frame, instead in this case study a LiDAL set of measurements is carried out every 200 frames, ie every 200ms, leading to 5 snapshot location measurements per second as shown in Table XI. This leads to a total of 18000 snapshot measurements in the one hour duration of the case study.

TABLE XI
MOBILITY SIMULATION PARAMETERS

| Parameters | Configurations |
|---|---|
| Simulation time $T_{Ob}$ | 60 min |
| LiDAL frame time $T$ | 1ms |
| Snapshots per second | 5 |
| Total number of snapshots $n$ per $T_{ob}$ | 18000 |
| Target destinations of interest $L_D$ | 9 |
| Buffering window | 1000 snapshots |
| Targets mobility behaviour | random walk and pathways |
| Nomadic target range speed | 0.5-2 m/s |
| Pedestrian target speed | 1 m/s |
| Targets arrival rate $\lambda$ | 12 arrivals per hour |
| Targets departure rate $\gamma$ | 14 departures per hour |
| Expected no. of target per $T_{Ob}$ | 6 |

The nomadic targets have 9 locations of interest in the room and spend 30 minutes on average in this office environment. The localization measurements are aggregated for the duration of a buffering window (see Table XI) and are processed in batch mode. This batch processing mode allows the localization process to consider a time span long enough for the nomadic user to start moving again. With 30 minutes on average in the office environment, 9 locations of interest, equally popular with random stay duration per location, and with 5 snapshot measurements per second, we considered a buffering window of duration equal to 1000 frames to capture the nomadic motion after stationary periods as shown in Table XI.

Fig. 42 presents the CDF of the MAPE associated with counting targets for the proposed LiDAL systems when the targets are either pedestrians or nomadic targets. Both types of targets move in Fig. 42 following a pathway model as described in Section IV-C-iii. Three key observations can be made on the results in Fig. 42. Firstly, nomadic target behaviour leads to higher MAPE when counting the number of targets regardless of the type of LiDAL system used. Secondly, the IMG-LIDAL system performs better than the MIMO-LiDAL system due to its improved spatial resolution. Finally, the difference in counting MAPE between cases when the targets are pedestrian and when they are mobile is smaller when the IMG-LiDAL system is considered compared to the MIMO-LiDAL system. This is due to the ability of the tracking algorithms to identify targets in pixels and track these targets, labelling them as targets even when they become stationary during their nomadic motion.

In Fig. 42, the counting error of targets with nomadic behaviour is more than the counting errors associated with pedestrian targets with an average (at CDF=0.5) MAPE of 28% and 15% for MIMO-LiDAL and MISO-IMG-LiDAL systems respectively under nomadic mobility. The average MAPE of pedestrian targets in MISO-IMG-LiDAL is 10% while for MIMO-LiDAL is 15%. For the 0.9 CDF interval, the MAPE of pedestrian targets is 14% and 11% for MIMO-LiDAL and MISO-IMG-LiDAL systems respectively. For nomadic targets, the MAPE is 33% and 18% for MIMO-LiDAL and MISO-IMG-LiDAL for the 0.9 CDF interval.

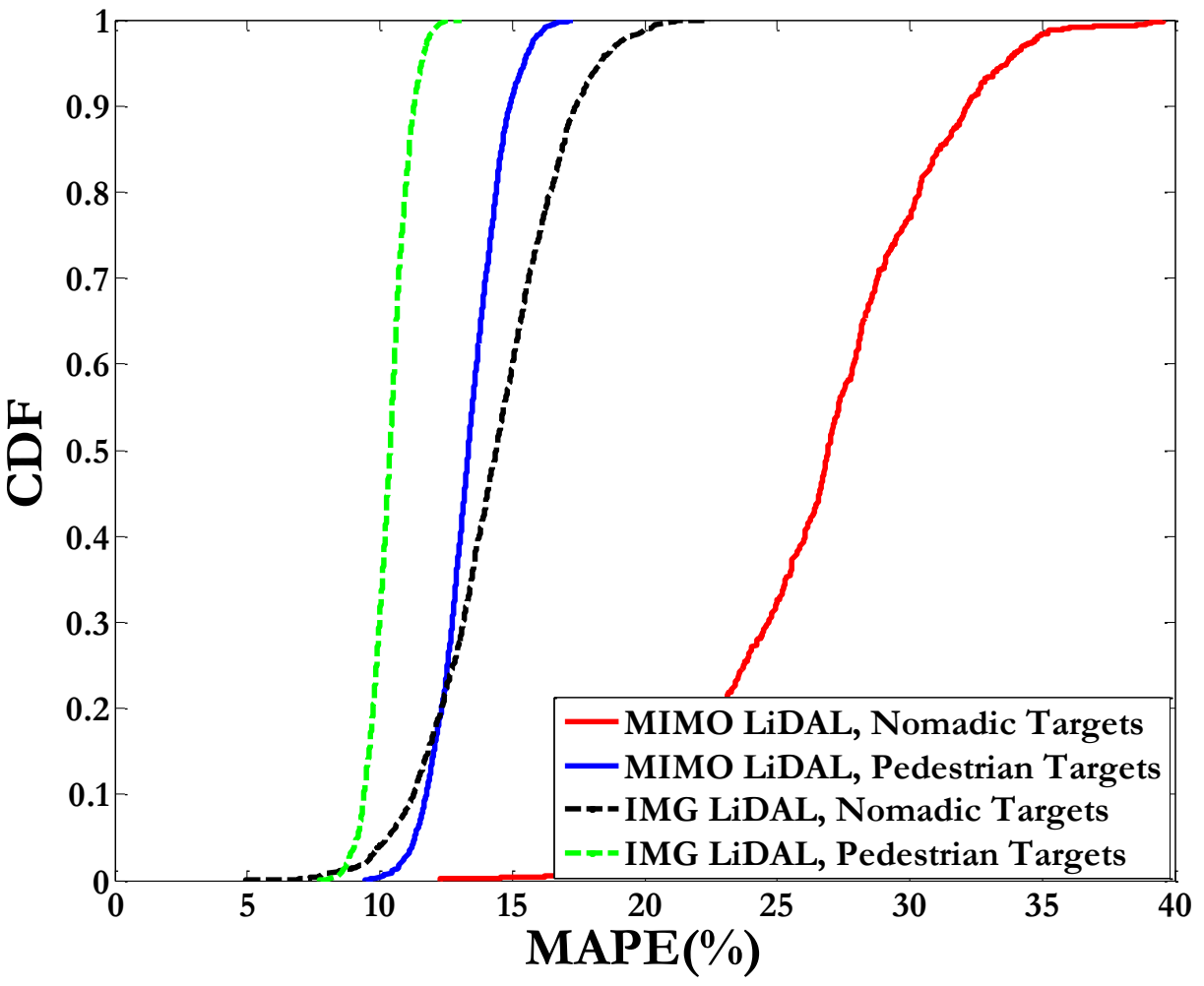

Fig. 42: CDF of counting MAPE of the targets, when the targets move along fixed pathways.

The same three observations we made in relation to Fig. 42, apply in Fig. 43 where the pedestrian targets and nomadic targets now follow a random walk pattern (see Section V-C-ii) instead of the pathway motion pattern used in Fig. 42.

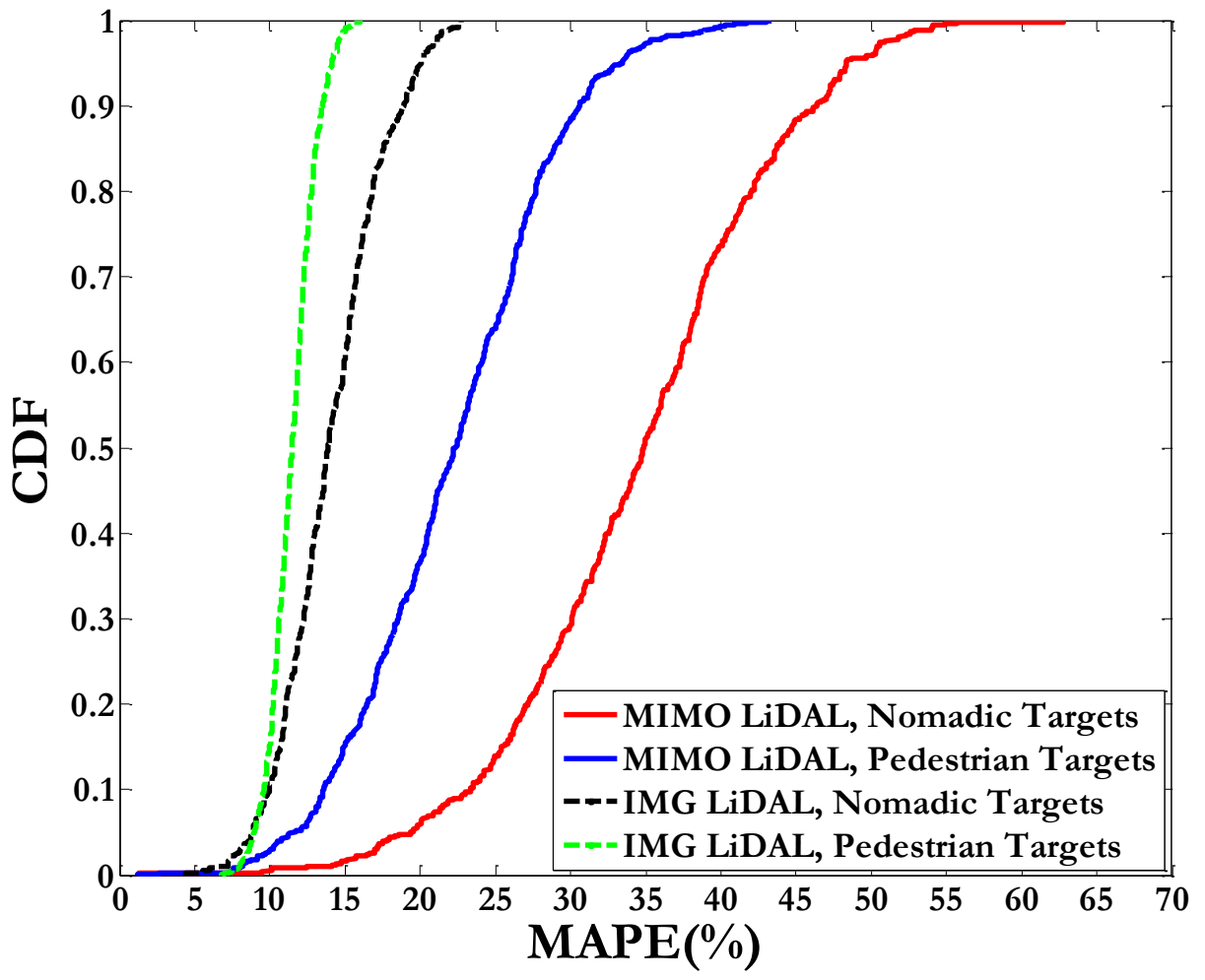

Fig.43: CDF of counting MAPE of the targets, when the targets move following a random walk model.

As can be seen in Fig. 43, the MAPE associated with the number of targets with nomadic behaviour detected by the MIMO-LiDAL system increased significantly in Fig. 43 (random walk) with average error of 38%, compared to 28% in Fig.42 (pathway mobility pattern). The increase in MAPE under random walk compared to pathway mobility is attributed to the nature of the random walk, where the random walk can result in (almost locked) mobility in a small geographic region, whereas the pathway mobility results in targets covering larger spans in the room and hence the detection of such "very" mobile targets improves.

For the MISO-IMG-LiDAL system with nomadic targets, the average MAPE in counting targets is 15% and 16% for pathway mobility and random walk mobility respectively. It should be noted that the increase in MAPE in the MISO-IMG-LiDAL system when mobility becomes a random walk rather than pathway based, is smaller compared to the corresponding increase in MAPE when the mobility pattern changes in the MIMO-LiDAL system. This is attributed to the ability of the imaging system to detect small movements on the detection floor, where each pixel corresponds to 0.5m × 0.5m whereas the MIMO-LiDAL coverage is within a circle of radius 1.25m.

## VII. Discussion and Conclusions

This paper presented the first study, to the best of our knowledge, of light used in a 'radar' fashion for people localization in indoor environments where visible light communication (VLC) and optical wireless communication may be present and in use. Our LiDAL systems can be used to count and localize people in indoor environments, and as such the LiDAL systems introduced can find application in a wide range of areas from security and safety to crowd management and marketing.

We introduced models for the indoor environment and for the human body, the materials used indoor and their reflection coefficients as well as the reflection coefficients of different forms of clothing taking into account colours and textures of clothing.

We introduced for the first time monostatic and bistatic optical indoor 'radar' configurations. Our resulting LiDAL systems provide coverage of the indoor environment through the use of multiple transmitters. The transmitters have broad beams for illumination, however we use relatively narrow FOV receivers to define optical target detection zones on the floor. This is very compatible with VLC systems where multiple light engines are used to illuminate the indoor environment. These light VLC sources can also act as our LiDAL transmitters. Humans located in the optical zones reflect the incident optical pulses, thus allowing optical receivers collocated with the transmitters in bistatic or monostatic configurations to detect the reflected pulses. Each optical zone is defined by the receiver FOV. We therefore developed models for the LiDAL systems range, namely the horizontal distance covered by each receiver / optical zone. We modelled the optical channel and estimated the receiver bandwidth needed and developed models for the spatial resolution that can be achieved with a given optical pulse duration. Based on indoor human occupancy, we concluded that the minimum human to human separation is typically more than 30cm even in meeting rooms and thus determined the LiDAL pulse duration needed as 2ns.

We identified the forms of target ambiguity that can occur in our LiDAL systems given that the target (human) has to be on the detection floor of the room and hence determined the number of anchors (light sources / light engines) needed concluding that three such anchors are needed for three dimensional localization.

We developed models for the sources of randomness in our LiDAL environment considering randomness due to the random nature of the reflection factor of humans (random colour and texture of clothing), the variable cross section of the target (human) which depends on human orientation with respect to the light source; and finally randomness due to receiver noise and background noise.

We derived optimum Bayes receiver structures based on the signal and noise models, considering and interpreting the priors associated with target presence and absence and the costs associated with correct decisions and the costs associated with wrong decisions together with the forms of decision errors. To simplify the receiver design, we derived a sub-optimum receiver structure that uses two thresholds for detection thus eliminating the need for exhaustive search and quantified the complexity reduction and the sacrifice in performance.

To distinguish reflections due to furniture from reflections attributed to the human targets, we used human mobility as the discriminator. We introduced two methods that use human motion to distinguish human targets from furniture; namely the background subtraction method (BSM) and the cross correlation method (CCM). We integrated both methods in the receiver designs developed.

To enable the evaluation of our LiDAL systems in a realistic environment, we furthermore developed models for human motion in the indoor environment of interest. In particular, we developed a directed random walk with obstacle avoidance mobility model and a pathway mobility model. Both models are based on Markov chains.

We introduced two LiDAL system configurations for target localization, a MIMO LiDAL system which has multiple transmitters (can be the same transmitters as the VLC transmitters, with MAC which we outlined) and multiple collocated receivers, with each receiver having a single photodiode. An improved alternative system design, MISO-IMG-LiDAL, was introduced making use of the spatial resolution afforded by the multiple pixels of an imaging receiver.

We studied the performance of our systems in three scenarios and in a case study which progressively test our LiDAL systems. The first scenario is a baseline system that produces the best performance possible. This scenario has an empty room with no obstacles (furniture) which reduces the localization errors and has continuous human (pedestrian) motion which helps distinguish humans. When the better target distinguishing method, ie CCM, is used the maximum target counting MAPE was reduced from 5.5% to 3.5% when the MISO-IMG-LiDAL system is used instead of the MIMO-LiDAL system. The maximum MAPE occurs at maximum number of targets, which was 15 human targets in our 8m×4m×3m room.

In the second scenario, obstacles (furniture) are introduced, however the environment has continuous pedestrian motion. Here the maximum target counting MAPE was reduced from 16% to 12% for the two systems respectively.

In terms of localization errors, in scenario 1, the average DRMSE was 0.28m and 0.16m for the MIMO LiDAL system and the MISO-IMG-LiDAL system respectively, while for scenario 2 the corresponding values were 0.38m and 0.19m respectively.

The third scenario is more challenging, with obstacles (furniture) present in the room and with targets moving in a nomadic fashion rendering the target distinguishing task harder. We defined a target mobility factor (MF), with MF=1 representing a fully mobile target and MF=0 being the extreme end of nomadic behaviour (fully stationary target). It is worth observing that a MPAE of 20% or less is only achieved in the MIMO-LiDAL system for mobility levels that correspond to MF approaching one. The MISO-IMG-LiDAL system offered improved performance in scenario 3 compared to the MIMO LiDAl system due to the ability of the imaging receiver to track a human target that then becomes stationary, but is still marked as a human target. For example, for MAPE of 20% or less a MF of 0.5 or higher is sufficient in MISO-IMG-LiDAL.

The case study added a number of additional realistic features to the environment including arrival rates and departure rates of targets (humans) and hence finite time spent per target in the environment, as well as more realistic directed pathways mobility with nomadic motion or pedestrian motion (continuous motion). This more challenging environment resulted in increased localization and counting errors. For example, the worst performance was observed in the MIMO-LiDAL system with nomadic random walk for the targets where the average MAPE associated with counting was 38%. In contrast the best system evaluated in this case study, ie the MISO-IMG-LiDAL system with nomadic random walk, reduced the counting MAPE from 38% to 16%. The best result for the MISO-IMG-LiDAL system was when the targets were pedestrian (continuous motion) pathway targets, and here the counting MAPE was 10%. In all three scenarios and case study, the presence of additional targets in the room increases the amount of reflections, hence the LiDAL clutter and hence leads to worse MAPE and DRMSE performance.

*Future areas of work can include* (i) consideration of MIMO-IMG-LiDAL where an imaging receiver is used with each light source instead of our MISO-IMG-LiDAL which uses a single imaging receiver in the entire room. This can lead to improved performance; (ii) angle diversity receivers can be evaluated with our systems; (iii) the artificial neural network (ANN) can be trained as an additional / alternative mobility distinguishing method instead of our BSM and CCM; (iv) the time domain can be introduced through pulses and snapshots and used with the spatial domain in the imaging receiver to determine the target location in the third dimension, ie not only the pixel or two dimensional location of the target on the floor, but also the height of the target; (v) passive LiDAL structures can be designed where the visible light *communications* (VLC) signals reflected from targets (humans) are observed and measured to determine the target locations (vi) the LiDAL localization information can be used to aid the VLC system, for example in terms of improved handovers through mobility direction and speed prediction (vii) LiDAL can be used for improved resource allocation in VLC systems by knowing the locations of users hence steering beams or allocating resources (wavelengths, time slots, transmitters etc) to reduce interference.

## ACKNOWLEDGEMENTS

Aubida A. Al-Hameed would like to thank the Higher Committee for Education Development in Iraq (HCED) and the University of Mosul for financial support and funding his PhD scholarship.

This work was supported by the Engineering and Physical Sciences Research Council (ESPRC), INTERNET (EP/H040536/1), and STAR (EP/K016873/1) projects. All data are provided in full in the results section of this paper.

**Aubida A. Al-Hameed** received a B.Sc. in electronic and electrical engineering from the University of Mosul, Iraq, in 2010 and an M.Sc. degree in communication systems from the University of Mosul, Iraq, in 2013. He is a Higher Committee for Education Developments in Iraq (HCED) Scholar and is currently working towards a Ph.D. degree in the school of Electronic and Electrical Engineering, University of Leeds, Leeds, UK.

**Safawn Hafeedh Younus** received the B.Sc. degree in electronic and electrical engineering, and the M.Sc. degree (Hons.) in communication systems from the University of Mosul, Iraq, in 2008 and 2010, respectively. He is currently pursuing the Ph.D. degree with the School of Electronic and Electrical Engineering,

University of Leeds, Leeds, U.K. He is a 'Ministry of Higher Education and Scientific Research (MOHESR) of Iraq' Scholar. Prior to his Ph.D. study, he was with the Ministry of Communication, Iraq, as a Maintenance and Support Engineer of the NGN Local Network, from 2010 to 2012. He also worked as a Lecturer with the Communication Department, College of Electronics, University of Mosul, from 2012 to 2014. His research interests include performance enhancement techniques for visible light communication systems, visible light communication system design, and indoor visible light communication networking.

**Ahmed Taha Hussein** received the B.Sc. degree (Hons.) in electronic and electrical engineering and the M.Sc. degree (Hons.) in communication systems from the University of Mosul, Iraq, in 2006 and 2011, respectively, and the Ph.D. degree in visible light communication systems from the University of Leeds, Leeds, U.K., in 2017. Prior to his Ph.D. study, he worked as a Communication Instructor with the Electronic and Electrical Engineering Department, College of Engineering, University of Mosul, from 2006 to 2009. He also worked as a Lecturer at the Electronic and Electrical Engineering Department, College of Engineering, University of Mosul, from 2011 to 2012. He published widely in the top IEEE communications conferences and journals. He has received the Carter award, University of Leeds, for the best Journal. His research interests include performance enhancement techniques for visible light communication systems, visible light communication.

**Mohammed Thamer Alresheed** received the B.Sc. degree (Hons.) in electrical engineering from King Saud University, Riyadh, Saudi Arabia, in 2006, and the M.Sc. degree (Hons.) in communication engineering, and the Ph.D. degree in electronic and electrical engineering, both from Leeds University, Leeds, U.K., in 2009 and 2014, respectively. He is currently an Assistant Professor with the Department of Electrical Engineering, King Saud University. His research interests include adaptive techniques for optical wireless (OW), OW systems design, indoor OW networking, and visible light communications.

**Prof. Jaafar M. H. Elmirghani** is the Director of the Institute of Integrated Information Systems within the School of Electronic and Electrical Engineering, University of Leeds, UK. He joined Leeds in 2007 and prior to that (2000–2007) as chair in optical communications at the University of Wales Swansea he founded, developed and directed the Institute of Advanced Telecommunications and the Technium Digital (TD), a technology incubator/spin-off hub. He has provided outstanding leadership in a number of large research projects at the IAT and TD. He received the Ph.D. in the synchronization of optical systems and optical receiver design from the University of Huddersfield UK in 1994 and the DSc in Communication Systems and Networks from University of Leeds, UK, in 2014. He has co-authored Photonic switching Technology: Systems and Networks, (Wiley) and has published over 500 papers. He has research interests in optical systems and networks. Prof. Elmirghani is Fellow of the IET, Fellow of the Institute of Physics and Senior Member of IEEE. He was Chairman of IEEE Comsoc Transmission Access and Optical Systems technical committee and was Chairman of IEEE Comsoc Signal Processing and Communications Electronics technical committee, and an editor of IEEE Communications Magazine. He was founding Chair of the Advanced Signal Processing for Communication Symposium which started at IEEE GLOBECOM'99 and has continued since at every ICC and GLOBECOM. Prof. Elmirghani was also founding Chair of the first IEEE ICC/GLOBECOM optical symposium at GLOBECOM'00, the Future Photonic Network Technologies, Architectures and Protocols Symposium. He chaired this Symposium, which continues to date under different names. He was the founding chair of the first Green Track at ICC/GLOBECOM at GLOBECOM 2011, and is Chair of the IEEE Green ICT committee within the IEEE Technical Activities Board (TAB) Future Directions Committee (FDC), a pan IEEE Societies committee responsible for Green ICT activities across IEEE, 2012-present. He is and has been on the technical program committee of 34 IEEE ICC/GLOBECOM conferences between 1995 and 2015 including 15 times as Symposium Chair. He received the IEEE Communications Society Hal Sobol award, the IEEE Comsoc Chapter Achievement award for excellence in chapter activities (both in 2005), the University of Wales Swansea Outstanding Research Achievement Award, 2006, the IEEE Communications Society Signal Processing and Communication Electronics outstanding service award, 2009, a best paper award at IEEE ICC'2013, the IEEE Comsoc Transmission Access and Optical Systems outstanding Service award 2015 in recognition of "Leadership and Contributions to the Area of Green Communications" and received the GreenTouch 1000x award in 2015 for "pioneering research contributions to the field of energy efficiency in telecommunications". He is currently an editor of: IET Optoelectronics, Journal of Optical Communications, IEEE Communications Surveys and Tutorials and IEEE Journal on Selected Areas in Communications series on Green Communications and Networking. He is Co-Chair of the GreenTouch Wired, Core and Access Networks Working Group, an adviser to the Commonwealth Scholarship Commission, member of the Royal Society International Joint Projects Panel and member of the Engineering and Physical Sciences Research Council (EPSRC) College. He has been awarded in excess of £22 million in grants to date from EPSRC, the EU and industry and has held prestigious fellowships funded by the Royal Society and by BT. He is an IEEE Comsoc Distinguished Lecturer 2013-2016.